\PassOptionsToPackage{table}{xcolor}

\documentclass{article}

\usepackage{promptcard}
\usepackage{iclr2027_conference,times}
\usepackage{microtype}
\usepackage{graphicx}
\usepackage{booktabs}
\usepackage{tabularx}
\usepackage{array}
\usepackage{xcolor}
\usepackage{placeins}
\usepackage{booktabs}
\usepackage{slotcatalog}
\newcolumntype{Y}{>{\raggedright\arraybackslash}X}

\definecolor{petct}{HTML}{D96C8C}
\definecolor{vimed}{HTML}{8C6BB1}
\definecolor{merlin}{HTML}{5B9BD5}
\definecolor{bdmap}{HTML}{D89A5B}
\definecolor{datasetgray}{RGB}{242,242,245}
\definecolor{oursblue}{RGB}{235,240,255}

\newcommand{\tabbest}[1]{\textbf{#1}}
\newcommand{\tabsecond}[1]{\underline{#1}}
\providecommand{\methodname}{\textsc{TEMPLAR}}

\usepackage{amsmath,amsfonts,bm}

\def\eqref#1{equation~\ref{#1}}

\def\1{\bm{1}}

\DeclareMathAlphabet{\mathsfit}{\encodingdefault}{\sfdefault}{m}{sl}
\SetMathAlphabet{\mathsfit}{bold}{\encodingdefault}{\sfdefault}{bx}{n}

\usepackage{hyperref}
\hypersetup{
  pdftitle={TEMPLAR: Agentic Induction and Evolution of Standardized Radiology Reporting Templates from Large-Scale Clinical Corpora},
  pdfauthor={Xiaotian Hu, Mingxuan Liu, Zhonghan Wang, Xinfeng Zhang, Yiming Huang, Ziang Wang, Kasidit Anmahapong, Yijin Li, Yifei Chen, Hongjia Yang, Zihan Li, Qiyuan Tian},
  pdfsubject={Public preprint},
  hidelinks
}
\usepackage{url}

\title{TEMPLAR: Agentic Induction and Evolution of Standardized Radiology Reporting Templates from Large-Scale Clinical Corpora}

\author{%
\parbox{\dimexpr\textwidth-2\tabcolsep\relax}{%
\raggedright\normalfont\small
\mbox{\textbf{Xiaotian Hu}$^{1}$\thanks{Equal contribution.}},
\mbox{\textbf{Mingxuan Liu}$^{1}$\footnotemark[1]},
\mbox{\textbf{Zhonghan Wang}$^{1}$},
\mbox{\textbf{Xinfeng Zhang}$^{2}$},
\mbox{\textbf{Yiming Huang}$^{3}$},
\mbox{\textbf{Ziang Wang}$^{1}$},
\mbox{\textbf{Kasidit Anmahapong}$^{1}$},
\mbox{\textbf{Yijin Li}$^{1}$},
\mbox{\textbf{Yifei Chen}$^{1}$},
\mbox{\textbf{Hongjia Yang}$^{1}$},
\mbox{\textbf{Zihan Li}$^{1}$},
\mbox{\textbf{Qiyuan Tian}$^{1}$\thanks{Corresponding author. E-mail: \texttt{qiyuantian@tsinghua.edu.cn}}}\\[0.45em]
{\normalfont\footnotesize
\mbox{$^{1}$Tsinghua University}\hfill
\mbox{$^{2}$University of Hong Kong}\hfill
\mbox{$^{3}$University of California, San Diego}}%
}%
}

\iclrfinalcopy
\begin{document}
\microtypesetup{activate=false}

\maketitle
\lhead{}

\raggedbottom
\AddToHook{shipout/after}[restore-template-bottom]{%
  \ifnum\value{page}=2\relax
    \begingroup\globaldefs=1\flushbottom\endgroup
    \RemoveFromHook{shipout/after}[restore-template-bottom]%
  \fi
}

\begin{abstract}
Structured radiology reporting mitigates the heterogeneity of free-text reports, yet its benefits depend on high-quality reporting templates. In practice, such templates are conventionally built through labor-intensive expert consensus and therefore vary across institutions and lag behind evolving clinical practice. Large language models (LLMs) enable automated template induction, but existing approaches remain limited: single-LLM induction is constrained by context length, and the corpus-scale method ASTAR produces a static, closed-corpus template without external grounding or downstream adaptation. To address these limitations, we propose \methodname{}, a TEMPLate-centric Agentic framework for inducing and evolving standardized Radiology reporting templates from large-scale clinical corpora. \methodname{} treats the template as a persistent central state maintained alongside two provenance-aware knowledge graphs, namely an anatomical graph that constrains template construction and a diagnostic graph that supports finding-to-diagnosis reasoning. Three agents operate on this state. The Induction Agent derives canonical clinical slots from anatomy-constrained Span-Triple atoms via dual-view similarity clustering; the Evolution Agent then assembles these slots into a hierarchical template and revises it under consistency constraints, external clinical evidence, and downstream structuring feedback; and the Clinical Agent applies the evolved template to report structuring, reconstruction, and diagnostic reasoning. Across four datasets (PETCT, ViMed, MERLIN, and BDMAP), \methodname{} outperforms ASTAR, three medical LLMs, and six general-purpose LLMs in coverage, information fidelity, and diagnostic fidelity, while achieving the highest or tied-highest LLM-rated template quality. Its fidelity advantages over ASTAR persist under cross-dataset transfer, and cumulative ablations support complementary contributions of its key components.
\end{abstract}

\suppressfloats[t]
\begin{figure}[!t]
    \centering
    \includegraphics[width=0.95\textwidth]{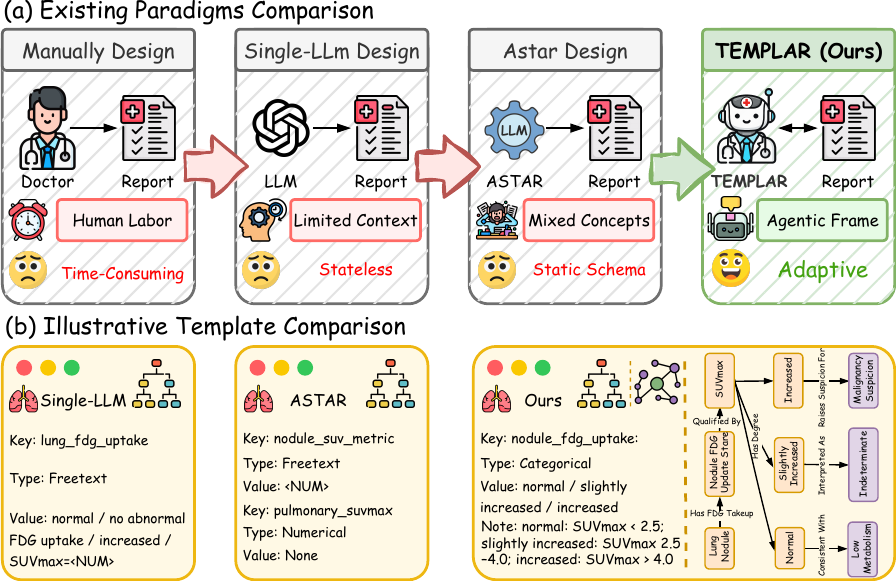}
\caption{
\textbf{Comparison of reporting-template construction paradigms.}
(a) Labor-intensive manual design, context-limited single-LLM induction,
static corpus-scale induction with ASTAR, and adaptive agentic evolution
with \methodname{}.
(b) Single-LLM templates may conflate clinical concepts and ASTAR templates
retain redundant keys, whereas \methodname{} couples an evidence-grounded,
evolvable template with complementary knowledge graphs.
}
    \label{fig:teaser}
\end{figure}

\section{Introduction}
Radiology reports communicate imaging findings for clinical decisions~\citep{Ganeshan2018Structured,pons2016natural} and provide training data for medical artificial intelligence (AI)~\citep{irvin2019aaai-chexpert,pereira2024automated}. Institutional conventions and individual styles produce inconsistent terminology, organization, and granularity in free-text reports~\citep{naik2001radiology,esr2018structured}, complicating evidence extraction and normalization for clinical interpretation and AI development~\citep{vosshenrich2023investigating,datta2021radlex,huesch2018evaluating}.

Structured reporting first defines a template of clinical fields and value specifications, then populates it from free-text reports~\citep{zhang2026astar,11219055,fytas2024radprompt}. Although LLMs increasingly automate field population~\citep{10.1145/3688868.3689200,10.1007/978-3-032-05997-0_11}, template construction remains largely manual, relying on institutional committees or professional-society consensus~\citep{larson2013improving,dhondt2026standardised}. Expert involvement preserves clinical relevance, but repeated design, review, and adaptation require substantial labor~\citep{busch2025large}, limiting deployment across tasks, institutions, and evolving practice.

LLMs offer a path toward automated template construction~\citep{adams2023leveraging,busch2025large,woznicki2025automatic}, but context limits prevent direct processing of large reporting corpora~\citep{pmlr-v267-modarressi25a}. Truncation and sampling can reduce input length at the cost of incomplete coverage and uneven granularity~\citep{zhang2026astar}. ASTAR scales induction by consolidating report-level concepts into canonical slots and a refined anatomical hierarchy~\citep{zhang2026astar}. Its closed-corpus templates nevertheless remain static, without systematic external grounding or downstream feedback to detect and address information loss.

We propose \methodname{}, a template-centric agentic framework that induces templates from clinical corpora and maintains them as persistent states. Two provenance-aware knowledge graphs preserve clinical relations: an anatomical graph constrains construction, and a diagnostic graph supports finding-to-diagnosis reasoning (Fig.~\ref{fig:teaser}b). The Induction Agent extracts anatomy-constrained Span--Triple atoms, induces slots through dual-view similarity and Leiden clustering~\citep{traag2019louvain}, and consolidates both graphs. The Evolution Agent assembles the hierarchy, revises it under consistency constraints and external evidence, and performs one round of feedback-guided granularity calibration. The Clinical Agent applies the template and graphs to structuring, reconstruction, and diagnostic reasoning; the Evaluation Agent assesses template coverage, template quality, information fidelity, and diagnostic fidelity.

Our contributions are threefold: (1) We formulate template construction as agentic induction and evolution, treating the template as a persistent central state coupled with provenance-aware knowledge graphs (Fig.~\ref{fig:teaser}a). (2) We design an Induction Agent for anatomy-constrained slot induction, an Evolution Agent that refines the template under external evidence and downstream feedback, and a Clinical Agent that applies it to structuring and diagnostic reasoning. (3) On four datasets (PETCT, ViMed, MERLIN, and BDMAP), \methodname{} achieves higher template coverage, information fidelity, and diagnostic fidelity than ASTAR and nine LLMs.

\section{Related Work}
\label{sec:related_work}
\textbf{Structured Representations of Medical Reports.} Medical report structuring converts free-text narratives into standardized, machine-readable representations, ranging from template-based clinical fields to normalized concepts and entity--relation graphs~\citep{kreimeyer2017natural,jain2021radgraph}.
In radiology, the target structure is often specified through expert-defined reporting systems. For example, PI-RADS, O-RADS, and LI-RADS standardize terminology, categorization, and reporting for prostate MRI, ovarian and adnexal ultrasound, and liver imaging, respectively~\citep{turkbey2019prostate,andreotti2020orads,chernyak2018lirads}.
The European Society of Paediatric Radiology (ESPR) Fetal Task Force has also developed structured reporting recommendations for fetal MRI~\citep{sofia2024standardised}.
With the target fields or templates specified in advance, Transformer- and LLM-based approaches primarily automate their population from free-text reports~\citep{adams2023leveraging,woznicki2025automatic}.
Beyond field-based representations, a complementary line of work explicitly captures relations among clinical concepts through entity extraction, terminology normalization, and relation extraction~\citep{datta2021radlex,jain2021radgraph}.
For knowledge-graph construction from unstructured text more broadly, KGGen combines entity--relation extraction with cross-source aggregation and iterative clustering of entities and edges~\citep{mo2025kggen}.
Shifting the focus from report content extraction to template induction, ASTAR derives clinical fields and their hierarchical organization from report corpora, reducing reliance on manual template design~\citep{zhang2026astar}.
However, the resulting template remains a static representation derived from a closed corpus, without external knowledge grounding or adaptation to downstream structuring feedback.
\methodname{} extends this corpus-driven paradigm by incorporating external clinical evidence into template revision and using downstream structuring feedback to calibrate descriptive granularity.

\textbf{Stateful Agentic Systems.} Agentic systems combine tools, specialized roles, and persistent state in interactive workflows~\citep{xi2025rise,10.24963/ijcai.2024/890}. InnoEval retrieves heterogeneous evidence for research-idea evaluation~\citep{qiao2026innoeval}; EvolveR distills interaction trajectories into strategies updated through subsequent feedback~\citep{wu2026from}; and MedKGent assigns extraction, evidence integration, and graph updates to specialized agents~\citep{zhang2026medkgent}. These systems motivate persistent representations revised through evidence and feedback. \methodname{} applies this principle to reporting templates: external clinical evidence grounds semantics, structuring feedback calibrates granularity, and structural constraints and provenance support controlled updates.

\section{Method}
\label{sec:method}

\begin{figure}[h]
    \centering
    \includegraphics[width=\textwidth]{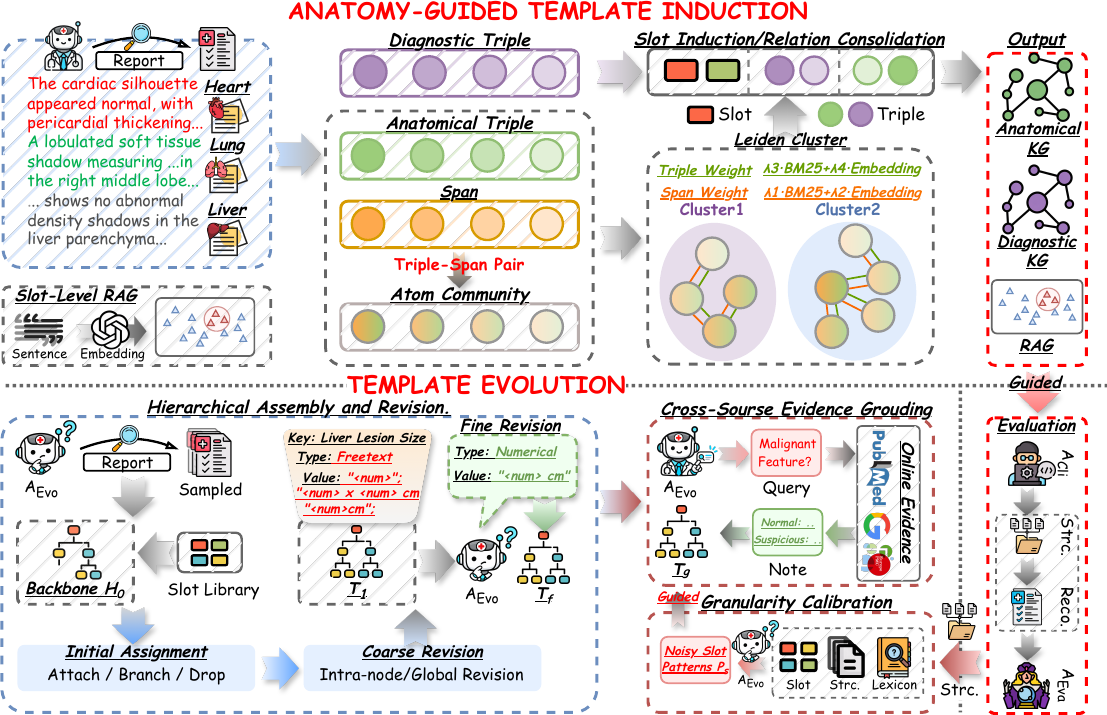}
    \caption{\textbf{\methodname{} overview.}
Anatomy-guided induction derives clinical slots, which evolve into a template through assembly, revision, external evidence grounding, and structuring feedback.}
    \label{fig:pipeline}
\end{figure}

\textbf{Overview.} Given a corpus of free-text radiology reports
$\mathcal{D}=\{x_i\}_{i=1}^{N}$,
the goal of \methodname{} (Fig. \ref{fig:pipeline}) is to induce and evolve a hierarchical
reporting template $T^\star$, where anatomical branches organize
clinical slots $\phi=(k,\tau,\Omega)$ specifying a canonical field
name $k$, a value type $\tau$, and a candidate-value set $\Omega$.
\methodname{} obtains $T^\star$ through
$\mathcal{D}\xrightarrow{\mathcal{A}_{\mathrm{ind}}}
\Phi_{\mathrm{keep}}\xrightarrow{\mathcal{A}_{\mathrm{evo}}}T^\star$,
where the Induction Agent $\mathcal{A}_{\mathrm{ind}}$ derives
the retained canonical slot library $\Phi_{\mathrm{keep}}$ from
anatomy-constrained Span--Triple atoms, and the Evolution Agent
$\mathcal{A}_{\mathrm{evo}}$ assembles and refines the template
using consistency checks, external evidence, and structuring feedback.
Two accompanying graphs, $G_{\mathrm{anat}}$ and
$G_{\mathrm{diag}}$, encode anatomical relations and
finding--diagnosis associations, respectively, supporting
template construction and downstream clinical applications
while preserving source provenance for each canonical relation.
Detailed procedures and agent prompts are provided (Appendices~\ref{app:method-details} and~\ref{app:prompts}).

\subsection{Anatomy-Guided Template Induction}


\textbf{Dual-View Similarity Graph.} To prevent cross-region concept conflation, $\mathcal{A}_{\mathrm{ind}}$
segments reports at changes in anatomical focus and groups segments by
canonical region. Within each segment, it separately extracts a clinical
span $u^{\mathrm{span}}=(r,\ell,c,\tau,v,\rho)$ and a relation
$u^{\mathrm{kg}}=(e^{\mathrm{subj}},p,e^{\mathrm{obj}})$.
The span records region $r$, finer location $\ell$, attribute $c$, type
$\tau$, value $v$, and clinical role $\rho$. The relation specifies subject,
predicate, and object. Segment-local alignment forms atoms
$a=(u^{\mathrm{span}},u^{\mathrm{kg}},\xi)$ with source provenance $\xi$
(Appendices~\ref{app:ordered_segmentation} and~\ref{app:span_relation_alignment}).
The full induction procedure is detailed separately (Appendix~\ref{app:template-induction}).

For atom $a_i$, the two views are
$q_i^{\mathrm{text}}=\operatorname{concat}(r_i,\ell_i,c_i,v_i)$ and
$q_i^{\mathrm{graph}}=\operatorname{linearize}
(e_i^{\mathrm{subj}},p_i,e_i^{\mathrm{obj}})$.
Within each region, fused similarities and sparse edge weights are:
\begin{equation}
\begin{aligned}
S_{ij}&=\sum_{m\in\{\mathrm{text},\mathrm{graph}\}}\lambda_m
\left(\lambda_{\mathrm{bm25}}B_{ij}^{m}
+\lambda_{\mathrm{emb}}E_{ij}^{m}\right),\\
A_{ij}&=S_{ij}\,\mathbb{I}\!\left[S_{ij}\geq\delta\ \land\
\bigl(j\in\mathcal{N}_k(i)\lor i\in\mathcal{N}_k(j)\bigr)\right].
\end{aligned}
\label{eq:dual_view_graph}
\end{equation}
Here, $B^m$ is row-normalized, symmetrized BM25 similarity~\citep{BM25},
$E^m$ is cosine similarity between normalized embeddings, the $\lambda$
terms are fusion weights, and $\mathcal{N}_k(i)$ contains the top-$k$
neighbors of $i$. Cross-region edges are excluded. Weighted Leiden clustering~\citep{traag2019louvain} yields anatomically
consistent communities of related atoms
(Appendix~\ref{app:leiden_construction}).

\textbf{Slot and Relation Consolidation.} 
Each community is partitioned into semantic subgroups, with one slot
$\phi=(k,\tau,\Omega)$ created for each subgroup, where $k$ denotes the
canonical slot key, $\tau\in\{\texttt{categorical},\texttt{numerical},
\texttt{freetext}\}$ denotes its value type, and $\Omega$ is its
candidate-value set. Proposals sharing a normalized key are merged while
retaining their evidence. Let $A(\phi)$ contain the distinct identifiers
of the atoms supporting $\phi$. To suppress incidental concepts, we retain
only slots whose distinct-atom support exceeds $\gamma_s$:
\begin{equation}
\Phi_{\mathrm{keep}}
=\{\phi\in\Phi:|A(\phi)|>\gamma_s\},
\label{eq:slot_support_filter}
\end{equation}
where $\Phi$ is the merged slot library. Regional relation consolidation
produces $G_{\mathrm{anat}}$ while preserving measurement semantics,
laterality, and polarity. Separately, paired training findings and impressions
yield $G_{\mathrm{diag}}$ by normalizing diagnostic targets, including
disease and clinical-status concepts, and consolidating finding--target
relations with anatomical scope and provenance. Both graphs retain source evidence. Slot induction and relation consolidation are detailed separately (Appendices~\ref{app:slot_triple_induction} and~\ref{app:relation-consolidation}); representative subgraphs illustrate the resulting relations (Appendix~\ref{app:knowledge-graph-examples}).

\textbf{Sentence Retrieval Corpus.} To provide linguistic exemplars for report reconstruction,
$\mathcal{A}_{\mathrm{ind}}$ indexes normalized, deduplicated training
sentences as $\mathcal{M}_{\mathrm{sent}}=\{(u_j,\mathbf{h}_j)\}_{j=1}^{M}$,
where $\mathbf{h}_j$ is the normalized embedding of sentence $u_j$.
The retrieval procedure is described in
Sec.~\ref{sec:structuring_reasoning}.

\subsection{Template Evolution}
\label{sec:template_grounding}

To improve template consistency and preserve clinical detail,
$\mathcal{A}_{\mathrm{evo}}$ assembles and refines the retained slots (Appendix~\ref{app:template-evolution}):
\begin{equation}
\Phi_{\mathrm{keep}}\xrightarrow{\mathrm{assembly}}T_1
\xrightarrow{\mathrm{revision}}T_f
\xrightarrow{\mathrm{grounding}}T_g
\xrightarrow{\mathrm{calibration}}T^\star.
\end{equation}
The resulting $T^\star$ integrates an anatomical hierarchy,
evidence-grounded slot annotations, and extraction guidance
for preserving clinical detail when structuring individual
free-text radiology reports.

\textbf{Hierarchical Assembly and Revision.} From representative reports, $\mathcal{A}_{\mathrm{evo}}$ induces a backbone
$H_0$. The first assignment round attaches slots or proposes branches;
the second reassigns retained slots against the completed hierarchy.
Coarse revision through within-node consolidation and cross-path concept harmonization yields $T_1$
(Appendix~\ref{app:hierarchical_template_construction}). Fine revision corrects ambiguous
keys, incompatible types, fragmented vocabularies, and numerical or unit
inconsistencies. A deterministic reducer rejects edits violating structural
or unit checks. Empty vocabularies are supplemented only when indexed
training sentences support coherent values, producing $T_f$.

\textbf{Cross-Source Evidence Grounding.}
Slots induced from the local report corpus may remain
semantically underspecified or exhibit dataset-specific biases.
Accordingly, $\mathcal{A}_{\mathrm{evo}}$ queries PubMed, Europe PMC,
and Google using each slot's key, type, vocabulary, and anatomical
path in $T_f$. It deduplicates documents by identifiers and metadata,
ranks them by authority and information richness, and attaches
 supported evidence as slot-level reference notes. These notes clarify
categorical or free-text semantics and provide units, reference
intervals, or thresholds, yielding $T_g$ (Appendix~\ref{app:evidence-grounding}).

\textbf{Behavior-Aware Granularity Calibration.} The Clinical Agent $\mathcal{A}_{\mathrm{cli}}$
(Sec.~\ref{sec:structuring_reasoning}) structures sampled training
findings with $T_g$. For each slot $s$, $\mathcal{A}_{\mathrm{evo}}$
audits its specification $\phi_s$, extracted values $Y_s$, aligned
source contexts $C_s$, and a template-derived generic-value lexicon
$V_{\mathrm{gen}}$. If detailed descriptions repeatedly collapse to
generic values (e.g., \texttt{present}), the agent derives slot-specific
linguistic patterns $P_s$. These patterns augment $T_g$ with extraction
guidance to form $T^\star$, without changing $\Omega_s$.
For example, mapping ``a 2.1-cm spiculated lesion in the right upper
lobe'' to \texttt{present} discards its size, morphology, and location.
The corresponding $P_s$ instructs subsequent extraction to preserve
these explicitly reported attributes (Appendix~\ref{app:granularity-calibration}).
Representative final templates span PETCT and ViMed (Appendices~\ref{app:template-examples-petct} and~\ref{app:template-examples-vimed}) and MERLIN and BDMAP (Appendices~\ref{app:template-examples-merlin} and~\ref{app:template-examples-bdmap}); their catalog documents the resulting specifications (Appendix~\ref{app:template_examples}).


\subsection{Clinical Structuring and Reasoning}
\label{sec:structuring_reasoning}

The Clinical Agent $\mathcal{A}_{\mathrm{cli}}$ uses $T^\star$ to
structure reports, then supports report reconstruction and
diagnostic reasoning from the resulting structured records.

\textbf{Template-Conditioned Structuring.}
For a new report $x$, $\mathcal{A}_{\mathrm{cli}}$ fills $T^\star$
by anatomical region to obtain a hierarchical record $z$.
Highly supported, region-relevant triples retrieved from
$G_{\mathrm{anat}}$ provide priors. Notably, $\mathcal{A}_{\mathrm{cli}}$
extracts values only from $x$, leaving unsupported slots unfilled.

\textbf{Slot-Conditioned Reconstruction.} For each non-empty slot--value pair $(k_s,v_s)\in z$,
$\mathcal{A}_{\mathrm{cli}}$ constructs a query and retrieves
the closest training sentence:
\begin{equation}
q_s=\operatorname{concat}(k_s,\texttt{:},v_s),\qquad
j_s=\arg\max_j\cos\!\left(\operatorname{Enc}(q_s),\mathbf{h}_j\right).
\label{eq:sentence_retrieval}
\end{equation}
After deduplication, the retrieved sentences are ranked by similarity
score, and the top-$K$ sentences form the reference set $R(z)$.
$\mathcal{A}_{\mathrm{cli}}$ reconstructs the report $\hat{x}$ using
the structured record $z$ and these reference sentences.
The references guide wording and anatomical organization, while
all case-specific findings and measurements are drawn from $z$.

\begingroup
\microtypesetup{activate=true}
\textbf{Graph-Assisted Diagnostic Reasoning.}
$\mathcal{A}_{\mathrm{cli}}$ uses the structured record $z$ to retrieve
relevant finding--target relations from the frozen diagnostic graph
$G_{\mathrm{diag}}$. These relations provide training-derived priors
for interpreting the findings, inferring a diagnostic conclusion, and
identifying the supporting clinical findings.
During reasoning, $\mathcal{A}_{\mathrm{cli}}$ accesses only $z$ and
$G_{\mathrm{diag}}$, with $z$ supplying all case-specific evidence.
The original report and reference diagnosis are withheld and used
only~for~evaluation.\par
\endgroup

\subsection{Multi-Dimensional Evaluation}
\label{sec:quality_evaluation}

The Evaluation Agent $\mathcal{A}_{\mathrm{eval}}$ assesses template
quality and downstream performance (Appendix~\ref{app:evaluation}). \textbf{Template quality} is quantified
by averaging the scores for design quality (13 dimensions) and clinical
utility (6 dimensions), computed separately from
LLM-as-judge~\citep{llmasjudge} and expert ratings (Appendix~\ref{app:template-quality-evaluation}).
\textbf{Template coverage} measures how well template slots represent
clinical concepts in held-out reports through LLM-based semantic
matching, with coverage aggregated at the case-macro and key-micro
levels (Appendix~\ref{app:template-coverage}). \textbf{Information fidelity} quantifies the agreement between
reconstructed reports $\hat{x}$ and the original held-out findings
using ROUGE-1/2/L~\citep{rouge}, chrF/chrF++~\citep{chrf,chrf++},
and BERTScore~\citep{BERTScore}, all computed directly against the
references (Appendix~\ref{app:information-fidelity}). \textbf{Diagnostic fidelity} assesses the preservation of
diagnostically relevant information through LLM ratings of primary
diagnosis accuracy (PDA), key finding preservation (KFP), and clinical
actionability (CA), using the held-out diagnoses and findings as
references. All three metrics are reported on a
$[0,1]$ scale (Appendix~\ref{app:diagnostic-fidelity}).

\section{Experiments and Results}
\label{sec:experiments}

\begin{table*}[t]
\caption{
Performance comparison on PETCT and ViMed in template coverage,
information fidelity, and diagnostic fidelity. Best in \textbf{bold} and
runner-up in \underline{underlined}. Case = case-macro coverage; Key = key-micro coverage;
Avg. = average; R-1/R-2/R-L = ROUGE-1/ROUGE-2/ROUGE-L;
chrF = character n-gram F-score; chrF++ = character and word n-gram F-score;
BERT = BERTScore; PDA = Primary Diagnosis Accuracy;
KFP = Key Finding Preservation; CA = Clinical Actionability.
}
\label{tab:comparison_petct_vimed}
\centering

\setlength{\tabcolsep}{2.8pt}
\renewcommand{\arraystretch}{0.96}
\scriptsize

\resizebox{\textwidth}{!}{%
\begin{tabular}{
@{}l
rrr
@{\hspace{4pt}}
rrrrrrr
@{\hspace{4pt}}
rrrr
@{}
}
\toprule

& \multicolumn{3}{c}{\textbf{Template Coverage}}
& \multicolumn{7}{c}{\textbf{Information Fidelity}}
& \multicolumn{4}{c}{\textbf{Diagnostic Fidelity}} \\

\cmidrule(lr){2-4}
\cmidrule(lr){5-11}
\cmidrule(lr){12-15}

\textbf{Method}
& \textbf{Case}
& \textbf{Key}
& \textbf{Avg.}
& \textbf{R-1}
& \textbf{R-2}
& \textbf{R-L}
& \textbf{chrF}
& \textbf{chrF++}
& \textbf{BERT}
& \textbf{Avg.}
& \textbf{PDA}
& \textbf{KFP}
& \textbf{CA}
& \textbf{Avg.} \\
\midrule

\rowcolor{datasetgray}
\multicolumn{15}{@{}l@{}}{\textbf{PETCT} \hspace{2pt} ($n=49$)} \\

\multicolumn{15}{@{}l@{}}{\textit{Medical LLM baselines}} \\

lingshu-32b
& 0.6678 & 0.7192 & 0.6935
& 0.5920 & 0.3089 & 0.3937 & 0.4916 & 0.4687 & 0.8865 & 0.5236
& 0.4939 & 0.5310 & 0.6000 & 0.5416 \\

medgemma-27b
& 0.6744 & 0.7027 & 0.6886
& \tabsecond{0.7495} & 0.5163 & 0.5727
& \tabsecond{0.6214} & \tabsecond{0.6058}
& \tabsecond{0.9069} & \tabsecond{0.6621}
& 0.5633 & 0.6051 & 0.6571 & 0.6085 \\

clinfusion-32b
& 0.4710 & 0.4910 & 0.4810
& 0.1419 & 0.0580 & 0.1089 & 0.0598 & 0.0556 & 0.7380 & 0.1937
& 0.5102 & 0.5320 & 0.6204 & 0.5542 \\

\rowcolor{black!4}
\textbf{Medical LLM Avg.}
& 0.6044 & 0.6376 & 0.6210
& 0.4945 & 0.2944 & 0.3584 & 0.3909 & 0.3767 & 0.8438 & 0.4598
& 0.5225 & 0.5560 & 0.6258 & 0.5681 \\

\addlinespace[1pt]
\multicolumn{15}{@{}l@{}}{\textit{General LLM baselines}} \\

deepseek-v4-flash
& 0.7632 & 0.8079 & 0.7856
& 0.6736 & 0.3825 & 0.4418 & 0.5555 & 0.5327 & 0.8991 & 0.5809
& 0.6245 & 0.6473 & 0.7265 & 0.6661 \\

deepseek-v4-pro
& 0.7565 & 0.7652 & 0.7609
& 0.6863 & 0.3939 & 0.4692 & 0.5748 & 0.5524 & 0.9029 & 0.5966
& 0.5959 & 0.6271 & 0.6857 & 0.6362 \\

qwen3.6-flash
& 0.7679 & 0.7948 & 0.7814
& 0.5713 & 0.2861 & 0.3607 & 0.4307 & 0.4101 & 0.8718 & 0.4885
& 0.6327 & 0.6739 & 0.7265 & 0.6777 \\

gemini-3.6-flash
& 0.5967 & 0.5908 & 0.5938
& 0.7228 & \tabsecond{0.5350} & \tabsecond{0.5834}
& 0.5812 & 0.5701 & 0.8954 & 0.6480
& 0.6408 & 0.6876 & 0.7265 & 0.6850 \\

gpt-5.4
& 0.7683 & 0.8144 & 0.7914
& 0.7117 & 0.4825 & 0.5293 & 0.5831 & 0.5688 & 0.8973 & 0.6288
& 0.5878 & 0.6380 & 0.6816 & 0.6358 \\

gemini-3.5-flash-lite
& 0.6330 & 0.6450 & 0.6390
& 0.5577 & 0.3449 & 0.4086 & 0.4140 & 0.3994 & 0.8808 & 0.5009
& 0.6449 & 0.6654 & \tabsecond{0.7510} & 0.6871 \\

\rowcolor{black!4}
\textbf{General LLM Avg.}
& 0.7143 & 0.7363 & 0.7253
& 0.6539 & 0.4042 & 0.4655 & 0.5232 & 0.5056 & 0.8912 & 0.5739
& 0.6211 & 0.6566 & 0.7163 & 0.6647 \\

\addlinespace[1pt]
\multicolumn{15}{@{}l@{}}{\textit{Template induction baseline}} \\

ASTAR
& \tabsecond{0.8074} & \tabsecond{0.8674} & \tabsecond{0.8374}
& 0.6473 & 0.3397 & 0.3863 & 0.5978 & 0.5655 & 0.8911 & 0.5713
& \tabsecond{0.6735} & \tabsecond{0.7086}
& \tabsecond{0.7510} & \tabsecond{0.7110} \\

\rowcolor{oursblue}
\textbf{Ours}
& \tabbest{0.8783} & \tabbest{0.9041} & \tabbest{0.8912}
& \tabbest{0.8151} & \tabbest{0.6094} & \tabbest{0.5840}
& \tabbest{0.7756} & \tabbest{0.7570} & \tabbest{0.9268}
& \tabbest{0.7447}
& \tabbest{0.7265} & \tabbest{0.7533} & \tabbest{0.8041}
& \tabbest{0.7613} \\

\midrule

\rowcolor{datasetgray}
\multicolumn{15}{@{}l@{}}{\textbf{VIMED} \hspace{2pt} ($n=277$)} \\

\multicolumn{15}{@{}l@{}}{\textit{Medical LLM baselines}} \\

lingshu-32b
& 0.5595 & 0.5501 & 0.5548
& 0.5179 & 0.2648 & 0.3881 & 0.3754 & 0.3580 & 0.8774 & 0.4636
& 0.3502 & 0.4091 & 0.4188 & 0.3927 \\

medgemma-27b
& 0.5122 & 0.5086 & 0.5104
& 0.5485 & 0.2944 & 0.4038 & 0.4372 & 0.4143 & 0.8866 & 0.4975
& 0.3690 & 0.4341 & 0.4390 & 0.4140 \\

clinfusion-32b
& 0.4557 & 0.4634 & 0.4596
& 0.2440 & 0.0987 & 0.1986 & 0.0849 & 0.0785 & 0.7542 & 0.2432
& 0.5588 & 0.6049 & 0.6426 & 0.6021 \\

\rowcolor{black!4}
\textbf{Medical LLM Avg.}
& 0.5091 & 0.5074 & 0.5083
& 0.4368 & 0.2193 & 0.3302 & 0.2992 & 0.2836 & 0.8394 & 0.4014
& 0.4260 & 0.4827 & 0.5001 & 0.4696 \\

\addlinespace[1pt]
\multicolumn{15}{@{}l@{}}{\textit{General LLM baselines}} \\

deepseek-v4-pro
& 0.5692 & 0.5693 & 0.5693
& 0.5634 & 0.2798 & 0.4381 & 0.4247 & 0.3997 & 0.8856 & 0.4986
& 0.3863 & 0.4600 & 0.4751 & 0.4405 \\

qwen3.6-flash
& 0.6522 & 0.6513 & 0.6518
& 0.7101 & 0.4610 & 0.5310 & 0.6182 & 0.5877 & 0.9112 & 0.6365
& 0.5935 & 0.6588 & 0.6534 & 0.6352 \\

gemini-3.6-flash
& 0.6384 & 0.6381 & 0.6383
& 0.6044 & 0.3774 & 0.4881 & 0.4946 & 0.4722 & 0.9039 & 0.5568
& 0.4498 & 0.5348 & 0.5249 & 0.5032 \\

gpt-5.4
& 0.5750 & 0.5795 & 0.5773
& 0.6070 & 0.3938 & 0.4993 & 0.5002 & 0.4804 & 0.9049 & 0.5643
& 0.3458 & 0.4376 & 0.4231 & 0.4022 \\

gemini-3.5-flash-lite
& 0.7283 & 0.7335 & 0.7309
& \tabsecond{0.7720} & \tabsecond{0.5921} & \tabsecond{0.6462}
& \tabsecond{0.7378} & \tabsecond{0.7183} & \tabsecond{0.9300}
& \tabsecond{0.7327}
& 0.6325 & 0.6928 & \tabsecond{0.7032} & 0.6762 \\

deepseek-v4-flash
& 0.7293 & 0.7325 & 0.7309
& 0.5648 & 0.2829 & 0.4326 & 0.4465 & 0.4273 & 0.8883 & 0.5071
& 0.5025 & 0.5495 & 0.5841 & 0.5454 \\

\rowcolor{black!4}
\textbf{General LLM Avg.}
& 0.6487 & 0.6507 & 0.6497
& 0.6370 & 0.3978 & 0.5059 & 0.5370 & 0.5143 & 0.9040 & 0.5827
& 0.4851 & 0.5556 & 0.5606 & 0.5338 \\

\addlinespace[1pt]
\multicolumn{15}{@{}l@{}}{\textit{Template induction baseline}} \\

ASTAR
& \tabsecond{0.8489} & \tabsecond{0.8518} & \tabsecond{0.8504}
& 0.6339 & 0.3147 & 0.4180 & 0.5666 & 0.5332 & 0.8955 & 0.5603
& \tabsecond{0.6469} & \tabsecond{0.7033} & 0.6931
& \tabsecond{0.6811} \\

\rowcolor{oursblue}
\textbf{Ours}
& \tabbest{0.8815} & \tabbest{0.8758} & \tabbest{0.8787}
& \tabbest{0.9066} & \tabbest{0.7861} & \tabbest{0.7539}
& \tabbest{0.8742} & \tabbest{0.8680} & \tabbest{0.9609}
& \tabbest{0.8583}
& \tabbest{0.6578} & \tabbest{0.7318} & \tabbest{0.7148}
& \tabbest{0.7015} \\

\bottomrule
\end{tabular}%
}
\end{table*}

\begin{table*}[t]
\caption{Performance comparison on MERLIN and BDMAP in template coverage,
information fidelity, and diagnostic fidelity. Best in \textbf{bold} and
runner-up in \underline{underlined}.
}
\label{tab:comparison_merlin_bdmap}
\centering

\setlength{\tabcolsep}{2.8pt}
\renewcommand{\arraystretch}{0.96}
\scriptsize

\resizebox{\textwidth}{!}{%
\begin{tabular}{
@{}l
rrr
@{\hspace{4pt}}
rrrrrrr
@{\hspace{4pt}}
rrrr
@{}
}
\toprule

& \multicolumn{3}{c}{\textbf{Template Coverage}}
& \multicolumn{7}{c}{\textbf{Information Fidelity}}
& \multicolumn{4}{c}{\textbf{Diagnostic Fidelity}} \\

\cmidrule(lr){2-4}
\cmidrule(lr){5-11}
\cmidrule(lr){12-15}

\textbf{Method}
& \textbf{Case}
& \textbf{Key}
& \textbf{Avg.}
& \textbf{R-1}
& \textbf{R-2}
& \textbf{R-L}
& \textbf{chrF}
& \textbf{chrF++}
& \textbf{BERT}
& \textbf{Avg.}
& \textbf{PDA}
& \textbf{KFP}
& \textbf{CA}
& \textbf{Avg.} \\
\midrule

\rowcolor{datasetgray}
\multicolumn{15}{@{}l@{}}{\textbf{MERLIN} \hspace{2pt} ($n=408$)} \\

\multicolumn{15}{@{}l@{}}{\textit{Medical LLM baselines}} \\

lingshu-32b
& 0.5214 & 0.5347 & 0.5281
& 0.5677 & 0.3047 & 0.3765 & 0.4965 & 0.4701 & 0.8521 & 0.5113
& \tabsecond{0.8662} & \tabsecond{0.8975}
& \tabsecond{0.8637} & \tabsecond{0.8758} \\

medgemma-27b
& 0.3230 & 0.3801 & 0.3516
& 0.6072 & 0.3891 & 0.3860 & 0.4705 & 0.4538 & 0.8766 & 0.5305
& 0.7137 & 0.8057 & 0.7456 & 0.7550 \\

clinfusion-32b
& 0.3293 & 0.3435 & 0.3364
& 0.2769 & 0.1159 & 0.1589 & 0.2084 & 0.1947 & 0.7833 & 0.2897
& 0.5853 & 0.6650 & 0.6333 & 0.6279 \\

\rowcolor{black!4}
\textbf{Medical LLM Avg.}
& 0.3912 & 0.4194 & 0.4053
& 0.4839 & 0.2699 & 0.3071 & 0.3918 & 0.3729 & 0.8373 & 0.4438
& 0.7217 & 0.7894 & 0.7475 & 0.7529 \\

\addlinespace[1pt]
\multicolumn{15}{@{}l@{}}{\textit{General LLM baselines}} \\

deepseek-v4-flash
& 0.4111 & 0.4556 & 0.4334
& 0.4970 & 0.2116 & 0.2791 & 0.3566 & 0.3384 & 0.8553 & 0.4230
& 0.6956 & 0.7632 & 0.7225 & 0.7271 \\

deepseek-v4-pro
& 0.5563 & 0.5854 & 0.5709
& 0.4962 & 0.2143 & 0.2855 & 0.4492 & 0.4179 & 0.8466 & 0.4516
& 0.7348 & 0.8016 & 0.7637 & 0.7667 \\

qwen3.6-flash
& 0.5505 & 0.6082 & 0.5794
& 0.5745 & 0.3059 & 0.3596 & 0.4484 & 0.4273 & 0.8657 & 0.4969
& 0.7480 & 0.8157 & 0.7657 & 0.7765 \\

gemini-3.6-flash
& \tabsecond{0.6724} & \tabsecond{0.7002} & \tabsecond{0.6863}
& 0.6750 & 0.4555 & 0.4379 & 0.5344 & 0.5152 & 0.8819 & 0.5833
& 0.7583 & 0.8413 & 0.7868 & 0.7955 \\

gpt-5.4
& 0.5738 & 0.6334 & 0.6036
& \tabsecond{0.6979} & \tabsecond{0.5033} & \tabsecond{0.4799}
& \tabsecond{0.5663} & \tabsecond{0.5498} & \tabsecond{0.8877}
& \tabsecond{0.6142}
& 0.7426 & 0.8152 & 0.7681 & 0.7753 \\

gemini-3.5-flash-lite
& 0.5411 & 0.6103 & 0.5757
& 0.5751 & 0.3497 & 0.3849 & 0.4334 & 0.4175 & 0.8711 & 0.5053
& 0.6892 & 0.7695 & 0.7162 & 0.7250 \\

\rowcolor{black!4}
\textbf{General LLM Avg.}
& 0.5509 & 0.5989 & 0.5749
& 0.5860 & 0.3400 & 0.3711 & 0.4647 & 0.4444 & 0.8681 & 0.5124
& 0.7281 & 0.8011 & 0.7538 & 0.7610 \\

\addlinespace[1pt]
\multicolumn{15}{@{}l@{}}{\textit{Template induction baseline}} \\

ASTAR
& 0.6546 & 0.6918 & 0.6732
& 0.5866 & 0.2967 & 0.3346 & 0.4847 & 0.4594 & 0.8641 & 0.5044
& 0.8431 & 0.8825 & 0.8520 & 0.8592 \\

\rowcolor{oursblue}
\textbf{Ours}
& \tabbest{0.6833} & \tabbest{0.7168} & \tabbest{0.7001}
& \tabbest{0.7919} & \tabbest{0.6342} & \tabbest{0.5387}
& \tabbest{0.6872} & \tabbest{0.6760} & \tabbest{0.8982}
& \tabbest{0.7044}
& \tabbest{0.8931} & \tabbest{0.9409} & \tabbest{0.8966}
& \tabbest{0.9102} \\

\midrule

\rowcolor{datasetgray}
\multicolumn{15}{@{}l@{}}{\textbf{BDMAP} \hspace{2pt} ($n=200$)} \\

\multicolumn{15}{@{}l@{}}{\textit{Medical LLM baselines}} \\

lingshu-32b
& 0.3364 & 0.3316 & 0.3340
& 0.3909 & 0.2214 & 0.3091 & 0.1973 & 0.1998 & 0.8920 & 0.3684
& 0.2340 & 0.2310 & 0.2390 & 0.2347 \\

medgemma-27b
& 0.6456 & 0.6278 & 0.6367
& 0.6477 & 0.4416 & 0.5244 & 0.4199 & 0.4307 & 0.9178 & 0.5637
& 0.6380 & 0.6861 & 0.6510 & 0.6584 \\

clinfusion-32b
& 0.7014 & 0.6742 & 0.6878
& 0.7041 & 0.4668 & 0.4965 & 0.5188 & 0.5213 & 0.9226 & 0.6050
& 0.8400 & 0.8778 & 0.8530 & 0.8569 \\

\rowcolor{black!4}
\textbf{Medical LLM Avg.}
& 0.5611 & 0.5445 & 0.5528
& 0.5809 & 0.3766 & 0.4433 & 0.3787 & 0.3839 & 0.9108 & 0.5124
& 0.5707 & 0.5983 & 0.5810 & 0.5833 \\

\addlinespace[1pt]
\multicolumn{15}{@{}l@{}}{\textit{General LLM baselines}} \\

deepseek-v4-flash
& 0.6730 & 0.6543 & 0.6637
& \tabsecond{0.7557} & \tabsecond{0.5204} & \tabsecond{0.5769}
& 0.5190 & 0.5298 & \tabsecond{0.9307} & \tabsecond{0.6388}
& 0.8440 & 0.8796 & 0.8660 & 0.8632 \\

deepseek-v4-pro
& 0.6857 & 0.6714 & 0.6786
& 0.7060 & 0.5128 & 0.5677 & 0.5281 & 0.5334 & 0.9268 & 0.6291
& 0.7980 & 0.8229 & 0.8200 & 0.8136 \\

qwen3.6-flash
& 0.3756 & 0.3664 & 0.3710
& 0.2308 & 0.1351 & 0.1884 & 0.1101 & 0.1126 & 0.8712 & 0.2747
& 0.1870 & 0.1870 & 0.2090 & 0.1943 \\

gemini-3.6-flash
& 0.6374 & 0.6091 & 0.6233
& 0.6665 & 0.4643 & 0.5488 & 0.4442 & 0.4535 & 0.9215 & 0.5831
& 0.6070 & 0.6540 & 0.6260 & 0.6290 \\

gpt-5.4
& 0.4338 & 0.4368 & 0.4353
& 0.4412 & 0.3090 & 0.3452 & 0.2471 & 0.2511 & 0.9039 & 0.4163
& 0.2770 & 0.2707 & 0.2860 & 0.2779 \\

gemini-3.5-flash-lite
& 0.6742 & 0.6535 & 0.6639
& 0.6823 & 0.4718 & 0.5463 & 0.4832 & 0.4879 & 0.9242 & 0.5993
& 0.6300 & 0.6693 & 0.6570 & 0.6521 \\

\rowcolor{black!4}
\textbf{General LLM Avg.}
& 0.5800 & 0.5652 & 0.5726
& 0.5804 & 0.4022 & 0.4622 & 0.3886 & 0.3947 & 0.9131 & 0.5235
& 0.5572 & 0.5806 & 0.5773 & 0.5717 \\

\addlinespace[1pt]
\multicolumn{15}{@{}l@{}}{\textit{Template induction baseline}} \\

ASTAR
& \tabsecond{0.8641} & \tabsecond{0.8885} & \tabsecond{0.8763}
& 0.7333 & 0.4609 & 0.4861
& \tabsecond{0.5582} & \tabsecond{0.5552} & 0.9129 & 0.6178
& \tabsecond{0.8690} & \tabsecond{0.9194}
& \tabsecond{0.8810} & \tabsecond{0.8898} \\

\rowcolor{oursblue}
\textbf{Ours}
& \tabbest{0.9011} & \tabbest{0.9083} & \tabbest{0.9047}
& \tabbest{0.9202} & \tabbest{0.8508} & \tabbest{0.8399}
& \tabbest{0.8731} & \tabbest{0.8740} & \tabbest{0.9705}
& \tabbest{0.8881}
& \tabbest{0.9810} & \tabbest{0.9870} & \tabbest{0.9100}
& \tabbest{0.9593} \\

\bottomrule
\end{tabular}%
}
\end{table*}

\subsection{Experimental Setup}

\textbf{Datasets.} We evaluate \methodname{} on four English-language radiology report datasets covering two distinct imaging settings, namely whole-body $^{18}\mathrm{F}$-FDG PET/CT and abdominal CT. The PET/CT datasets comprise PETCT~\citep{PETCT} with 490 reports (441/49 train/test) and ViMed~\citep{ViMed} with 2,761 reports (2,484/277). The abdominal CT datasets comprise MERLIN~\citep{merlin} with 4,080 reports (3,672/408) and BDMAP~\citep{BDMAP} with 2,000 reports (1,800/200). Template induction, knowledge-graph construction, and method-specific adaptation use only the training split, with the test split held out for evaluation. We use findings for template construction, structuring, and reconstruction, reserving training impressions for diagnostic-graph construction and test impressions for diagnostic evaluation (Appendix~\ref{app:datasets}).

\textbf{Baselines.} We compare \methodname{} with ASTAR~\citep{zhang2026astar} and nine LLMs. (1) general-purpose LLMs: \texttt{DeepSeek-V4-Flash}, \texttt{DeepSeek-V4-Pro}~\citep{deepseekv4}, \texttt{Qwen3.6-Flash}~\citep{qwen36flash}, \texttt{Gemini-3.6-Flash}~\citep{gemini36flash}, \texttt{GPT-5.4}~\citep{gpt54}, and \texttt{Gemini-3.5-Flash-Lite}~\citep{google2026gemini35flashlite}; and (2) medical LLMs: \texttt{Lingshu-32B}~\citep{Lingshu}, \texttt{MedGemma-27B}~\citep{Medgemma}, and \texttt{ClinFusion-32B}~\citep{ClinFusion}. To accommodate context limits, each LLM baseline processes the training reports in batches and merges valid local templates into a single template (Appendix~\ref{app:llm-baseline-construction}).
Implementation settings are specified separately (Appendix~\ref{app:implementation_details}). Additional model screening and construction failures (Appendices~\ref{app:failure-analysis} and~\ref{app:baseline-screening}), parsing and recovery (Appendices~\ref{app:json-failures} and~\ref{app:fallback-parser}), and repetition (Appendix~\ref{app:degeneration}) are documented.

\begin{figure}[h]
    \centering
    \includegraphics[width=\textwidth]{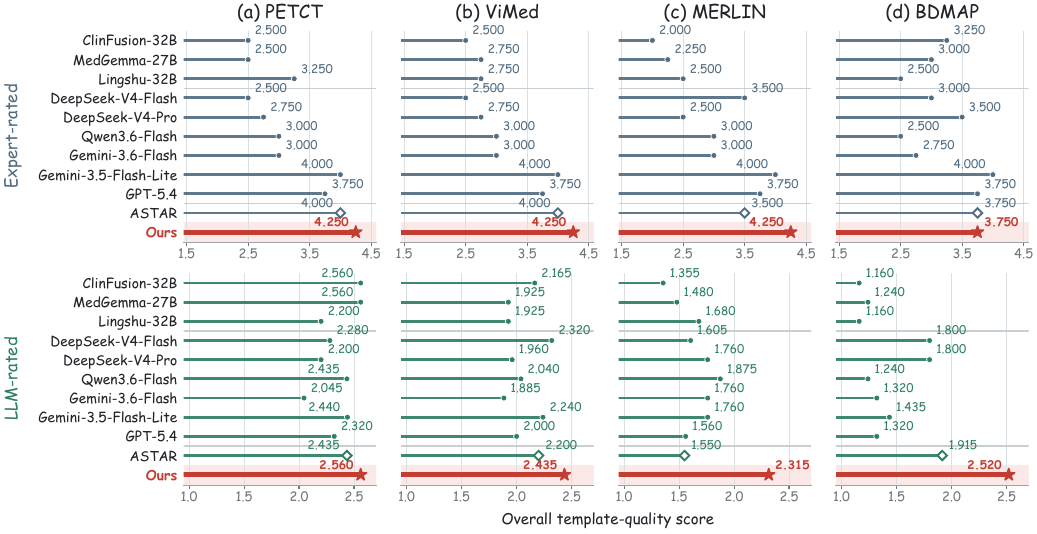}
    \caption{\textbf{Template quality across four datasets.}
Expert ratings (top) and LLM ratings (bottom) are shown for
(a) PETCT, (b) ViMed, (c) MERLIN, and (d) BDMAP.}
    \label{fig:template-quality-llm}
\end{figure}




\subsection{Comparison Results}
\textbf{Template quality.} \methodname{} consistently ranked among the best-performing methods in both expert and LLM assessments of overall template quality (Fig.~\ref{fig:template-quality-llm}). Under expert evaluation, it ranked first on PETCT, ViMed, and MERLIN (4.250 each), while tying with ASTAR and GPT-5.4 on BDMAP (3.750), below Gemini-3.5-Flash-Lite (4.000). Under LLM evaluation, it tied with MedGemma-27B and ClinFusion-32B on PETCT (2.560) and ranked first on ViMed (2.435), MERLIN (2.315), and BDMAP (2.520). Its margins over the strongest baseline on the latter three datasets were 0.115, 0.440, and 0.605, respectively.

\textbf{Template coverage.} \methodname{} achieved broader coverage of clinical concepts in held-out reports across all four datasets (Tables~\ref{tab:comparison_petct_vimed} and~\ref{tab:comparison_merlin_bdmap}). It ranked first at both the case and key levels, reaching a macro-average coverage of 0.8437, representing a gain of 3.44 percentage points over ASTAR (0.8093), the strongest baseline on this aggregate (Table~\ref{tab:macro_average}). Notably, \methodname{} showed its largest coverage advantage on the PETCT dataset, scoring 0.8912 versus 0.8374 for ASTAR.

\textbf{Information fidelity.} \methodname{} reconstructed reports more faithfully across all four datasets, with gains spanning lexical, character-level, and semantic measures (Tables~\ref{tab:comparison_petct_vimed} and~\ref{tab:comparison_merlin_bdmap}). It ranked first on all six reconstruction metrics and achieved a macro-average information fidelity of 0.7988, exceeding Gemini-3.6-Flash, the strongest baseline (0.5928), by 34.8\% (Table~\ref{tab:macro_average}).

\textbf{Diagnostic fidelity.} \methodname{} achieved the highest diagnostic fidelity across all four datasets (Tables~\ref{tab:comparison_petct_vimed} and~\ref{tab:comparison_merlin_bdmap}). It ranked first in primary diagnosis accuracy, key finding preservation, and clinical actionability, reaching a macro-average score of 0.8331 compared with 0.7853 for ASTAR, the strongest baseline on this aggregate (Table~\ref{tab:macro_average}). Notably, \methodname{} showed its largest diagnostic-fidelity advantage on BDMAP, scoring 0.9593 versus 0.8898 for ASTAR (Table~\ref{tab:comparison_merlin_bdmap}).

Supplementary results include cross-dataset aggregates and detailed template-quality comparisons (Appendices~\ref{app:result-details}, \ref{app:aggregate-results}, and~\ref{app:detailed-template-quality-results}).

\begin{figure}[h]
    \centering
    \includegraphics[width=\textwidth]{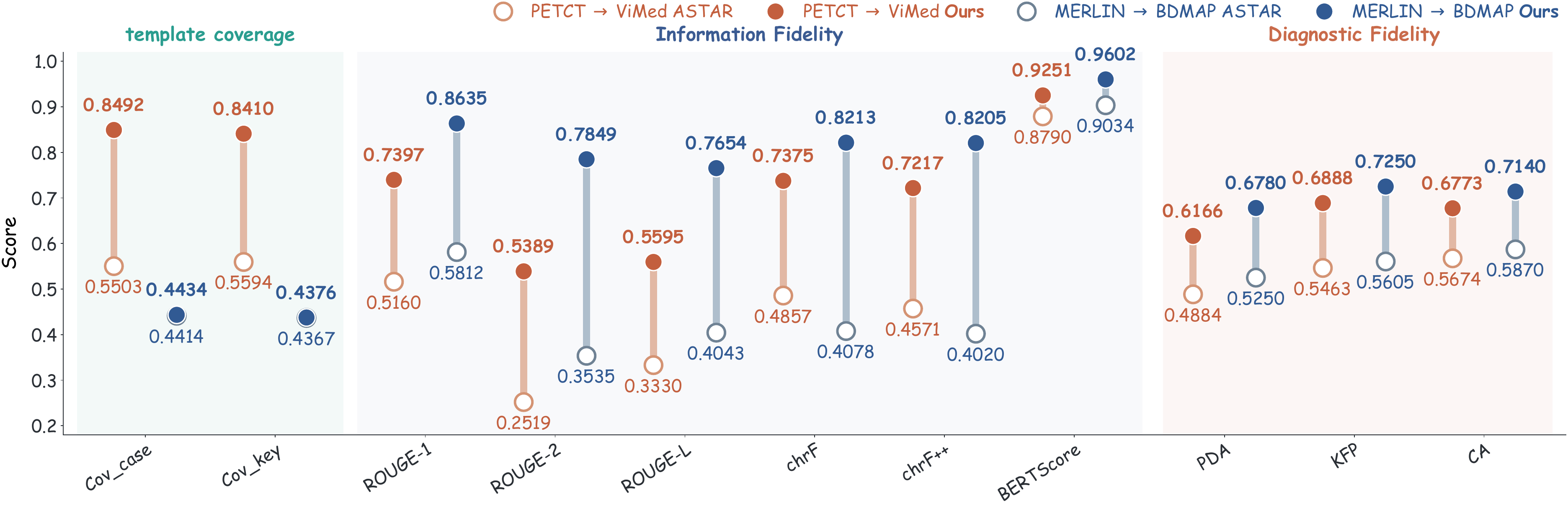}
    \caption{
\textbf{Cross-dataset generalization.}
Template coverage, information fidelity, and diagnostic fidelity under
\textcolor[HTML]{C35F3E}{PETCT$\rightarrow$ViMed} and
\textcolor[HTML]{315A93}{MERLIN$\rightarrow$BDMAP} transfer, without target-domain template re-induction. Open and filled circles denote ASTAR and \methodname{}, respectively.
}
    \label{fig:ood-transfer}
\end{figure}

\subsection{Cross-Dataset Generalization}

We compared the cross-dataset generalization of \methodname{} and ASTAR in two transfer settings: PETCT$\rightarrow$ViMed and MERLIN$\rightarrow$BDMAP (Fig. \ref{fig:ood-transfer}). Specifically, templates induced on each source dataset were applied directly to held-out reports from the corresponding target dataset, without target-domain template re-induction. For PETCT$\rightarrow$ViMed, \methodname{} achieved case/key coverage of 0.8492/0.8410, exceeding ASTAR's 0.5503/0.5594. It also outperformed ASTAR in reconstruction and diagnostic fidelity, with ROUGE-1 of 0.7397 versus 0.5160 and PDA of 0.6166 versus 0.4884. For MERLIN$\rightarrow$BDMAP, \methodname{} achieved comparable case/key coverage to ASTAR (0.4414/0.4367 vs. 0.4434/0.4376), while attaining higher ROUGE-2 (0.7849 vs. 0.3535), BERTScore (0.9602 vs. 0.9034), and PDA (0.6780 vs. 0.5250), highlighting benefits beyond broader coverage alone. Together, these results support the cross-dataset reuse of \methodname{} templates, with higher reconstruction and diagnostic fidelity observed both alongside broader target-domain coverage (PETCT$\rightarrow$ViMed) and at coverage comparable to ASTAR (MERLIN$\rightarrow$BDMAP).

\subsection{Ablation Studies}

\begin{figure}[h]
    \centering
    \includegraphics[width=\textwidth]{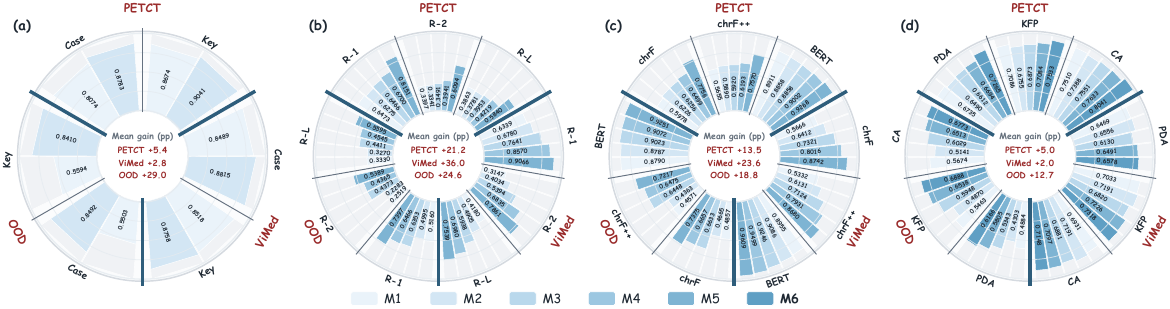}
    \caption{\textbf{Ablation study results.} Template coverage (a), information fidelity (b,c), and diagnostic fidelity (d) on PETCT, ViMed, and PETCT$\rightarrow$ViMed transfer (OOD). Starting from the baseline (M1), M2--M6 successively add the Dual-View Similarity Graph, Anatomical KG Structuring, Slot-Conditioned RAG, Template Evolution, and Diagnostic KG Reasoning.
    }
    \label{fig:figure4}
\end{figure}

We assessed component contributions of \methodname{} on PETCT, ViMed, and PETCT$\rightarrow$ViMed transfer (Fig.~\ref{fig:figure4}). Starting from the baseline (M1, i.e., ASTAR), we successively added the Dual-View Similarity Graph (M2), Anatomical KG Structuring (M3), Slot-Conditioned RAG (M4), Template Evolution (M5), and Diagnostic KG Reasoning (M6), with each configuration retaining all preceding components. Coverage is reported for M1--M2 and reconstruction for M1--M5, as subsequent additions leave the respective metrics unchanged. Diagnostic fidelity includes M1--M3, M5, and M6; M4 is omitted because its retrieval step affects reconstruction alone. 

The induction stage expanded concept coverage, most markedly under PETCT$\rightarrow$ViMed transfer, where M2 achieved case/key coverage of 0.8492/0.8410 versus 0.5503/0.5594 for M1. Yet M2 had lower ROUGE-1 and PDA than M1 on both PETCT and PETCT$\rightarrow$ViMed, indicating that broader coverage alone did not ensure higher reconstruction or diagnostic fidelity. Subsequent stages supported more faithful reconstruction. On PETCT, ROUGE-1 reached 0.6466 with anatomical guidance (M3) and 0.6700 with retrieval (M4), compared with 0.6275 for M2. Template Evolution (M5) yielded the largest incremental ROUGE-1 gain in this setting, reaching 0.8151. M5 also attained ROUGE-1/BERTScore of 0.9066/0.9609 on ViMed and 0.7397/0.9251 under PETCT$\rightarrow$ViMed transfer, with the latter exceeding M4's 0.6466/0.9072. Finally, Diagnostic KG Reasoning (M6) achieved higher PDA than M5 on PETCT (0.7265 vs. 0.6694), ViMed (0.6578 vs. 0.6491), and PETCT$\rightarrow$ViMed (0.6166 vs. 0.5805). These findings support complementary roles for concept coverage, information preservation, and diagnostic reasoning in carrying clinical evidence from free-text reports into useful structured representations. Additional ablations examine template revision and evidence notes in terms of template quality and clinical utility (Appendix~\ref{app:template-quality-ablation}).

\section{Conclusion}
We presented \methodname{}, a template-centric agentic framework for inducing and evolving standardized radiology reporting templates from clinical corpora. It integrates anatomy-guided induction, external evidence grounding, and structuring feedback with anatomical and diagnostic knowledge graphs for report structuring and reasoning. Across four datasets, \methodname{} achieved broader template coverage and higher reconstruction and diagnostic fidelity than ASTAR and nine LLM baselines. Its fidelity advantages over ASTAR persisted under cross-dataset transfer, while cumulative ablations supported complementary contributions from its components. By enabling scalable template construction, \methodname{} could support clearer communication of diagnostic evidence for clinical decision-making and facilitate downstream clinical AI model development.

\subsection*{AI use statement}
We used generative AI tools for language polishing, manuscript formatting checks, and reference sketches that
informed the design of some figures. The authors made all final
decisions regarding the manuscript, figures, and scientific content
and take full responsibility for the accuracy and integrity of
this paper, including its experimental results, numerical values,
and bibliographic references.

\subsection*{Ethics Statement}
This work investigates automated template construction and
report processing using existing radiology report datasets,
as described in Appendix~\ref{app:datasets}.
Differences in data provenance, imaging modality, reporting
practice, and language may introduce biases and limit
generalization. In particular, translated reports and
automatically generated reports may contain artifacts that
affect both template construction and evaluation.

The framework is intended for research purposes. Generated
templates, clinical reference notes, reconstructed reports,
and diagnostic outputs may omit relevant information or
contain unsupported interpretations. Automated metrics and
expert assessments in this study do not establish clinical
safety or readiness for deployment. Any clinical application
would require independent validation in the intended setting
and oversight by qualified clinicians. Generated outputs
should not be used as a substitute for professional clinical
judgment.

\subsection*{Reproducibility Statement}
The appendices document the experimental setup and methodological
details to support reproducibility.
Appendix~\ref{app:datasets} describes the datasets, report
selection, and training/test split sizes.
Appendix~\ref{app:implementation_details} specifies the models,
computing environment, and implementation hyperparameters.
Appendix~\ref{app:method-details} details the framework,
and Appendix~\ref{app:llm-baseline-construction} describes
batched template construction for the LLM baselines.
Evaluation procedures are described in
Appendix~\ref{app:evaluation}, and the provided prompt
specifications appear in Appendix~\ref{app:prompts}.
Structured-output failures and fallback recovery are discussed
in Appendices~\ref{app:json-failures}
and~\ref{app:fallback-parser}.
Representative template specifications are presented
in Appendix~\ref{app:template_examples}.

\bibliography{iclr2027_conference}
\bibliographystyle{iclr2027_conference}

\appendix


\newpage

\section{Implementation Details and Hyperparameter Settings}
\label{app:implementation_details}

\textbf{Models and computing environment.} Unless otherwise specified, the Induction Agent and Evolution Agent use Qwen3.6-Flash for anatomy segmentation, semantic-atom induction, slot construction, and template refinement. The Clinical Agent uses Gemini-3.5-Flash for template-conditioned report structuring and slot-conditioned report reconstruction. The Evaluation Agent uses DeepSeek-V4-Flash for clinical-concept extraction and semantic concept-to-slot matching. GPT-5.5 serves as the LLM judge for the automated assessment of Template Design Quality and Clinical Utility. Diagnostic outputs were generated using Gemini-3.5-Flash and evaluated using
deepseek-v4-flash. All model calls use a temperature of 0. Experiments were conducted on a server equipped with NVIDIA A800 GPUs (80GB). The GPUs were used primarily for BERTScore evaluation.

\textbf{Embeddings and multi-view graph construction.} We use \texttt{text-embedding-3-small} to encode clinical spans, relation triples, and training sentences. The embeddings are 1,536-dimensional, and corpus embeddings are computed in batches of 128. Within each text or relation view, row-normalized BM25 similarity and cosine embedding similarity are combined with equal weights of 0.5. BM25 uses $k_1=1.5$, $b=0.75$, and $\epsilon=0.25$. The text and relation views are then fused with weights of 0.4 and 0.6. For each anatomical region, we construct a symmetric $k$-nearest-neighbor graph using $k=20$ and an edge threshold of $\delta=0.25$. Leiden partitioning uses the Reichardt--Bornholdt configuration model, a resolution of $\gamma=1.5$, and random seed 0, and is run until convergence. Regions containing fewer than 30 atoms are retained as single communities.

\textbf{Slot induction and external evidence grounding.} A slot is retained when it is supported by more than $\gamma_s=3$ distinct atoms, corresponding to at least four supporting atoms. Candidate categorical values are ranked by support frequency, and at most 20 values are retained for each slot. The anatomical backbone is induced from 100 sampled training reports. For evidence grounding, the Evolution Agent queries PubMed, Europe PMC, and Google. It retrieves up to $K_r=8$ records from each source, deduplicates the combined pool, and provides the six highest-ranked records for evidence-note synthesis. Retrieved documents are truncated to 6,000 characters, the evidence text supplied from each selected record is truncated to 600 characters, and each slot-level reference note is limited to 200 characters.

\textbf{Sentence retrieval and metrics.} The sentence corpus is constructed from normalized and deduplicated training sentences. Sentences shorter than 15 characters or longer than 400 characters are discarded. For each non-empty slot-value pair, the Clinical Agent retrieves its nearest training sentence by cosine similarity. Retrieved sentences are deduplicated and ranked, and the top 80 sentences are supplied as linguistic and organizational references for reconstruction. ROUGE-1, ROUGE-2, and ROUGE-L are reported using F-measure with stemming. chrF and chrF++ are computed at the sentence level, with word order 2 for chrF++. BERTScore uses \texttt{roberta-large}, layer 17, batch size 16, and a maximum sequence length of 512, and reports the F1 score.

\section{Method Details}
\label{app:method-details}
\subsection{Template Induction}
\label{app:template-induction}
\subsubsection{Anatomical-Order Prior and Ordered Segmentation}

\label{app:ordered_segmentation}

Medical reports often exhibit a weak anatomical-order prior, whereby findings from the same anatomical region tend to appear contiguously. We use this prior to prevent lexically similar findings from distinct anatomical regions from being incorrectly linked during subsequent global processing.

For each report $x_i$, the Induction Agent $\mathcal{A}_{\mathrm{ind}}$ partitions the text into contiguous segments according to their anatomical regions (Fig.~\ref{fig:prompt-anatomy-segmentation}):
\begin{equation}
\mathcal{G}_i
=
\mathcal{A}_{\mathrm{ind}}^{\mathrm{seg}}(x_i)
=
\{g_{ij}\}_{j=1}^{M_i},
\qquad
g_{ij}=(\widetilde{b}_{ij},x_{ij}),
\label{eq:app_region_segmentation}
\end{equation}
where $x_{ij}$ denotes a contiguous excerpt from the original report and $\widetilde{b}_{ij}$ is its anatomical region identified by the agent. Each segment retains its original text and position in the report.

Because reports are segmented independently, the same anatomical region may receive different labels across reports. For example, \textit{chest}, \textit{thorax}, and \textit{thoracic region} may all refer to the same region. We therefore standardize variations in capitalization, spacing, formatting, and terminology by mapping equivalent labels to a shared English snake-case name, such as \texttt{chest}.

After label standardization, semantic atoms associated with the same anatomical region are pooled across reports. Similarity computation, neighborhood construction, and community detection are then performed independently within each region. Semantic atoms from different anatomical regions are not matched, and their cross-region affinities are fixed to zero.

\subsubsection{Segment-Local Span--Relation Alignment}
\label{app:span_relation_alignment}
Clinical spans encode fine-grained attributes and values, whereas relation triples capture the relational context among clinical entities. Since the two views are extracted independently (Fig.~\ref{fig:prompt-span-kg-extraction}), \(\mathcal{A}_{\mathrm{ind}}\) aligns them to construct unified semantic atoms for subsequent dual-view graph construction. For a segment \(g\), let
\begin{equation}
U_g = \left\{u_s^{\mathrm{span}}\right\}_{s=1}^{n_g},
\qquad
R_g = \left\{u_t^{\mathrm{kg}}\right\}_{t=1}^{m_g}
\label{eq:segment_span_relation_sets}
\end{equation}
denote its ordered span collection and relation set, respectively. For each span, the Induction Agent selects the most semantically relevant relation from the same segment according to the span fields and the relation's subject, predicate, and object (Fig.~\ref{fig:prompt-span-relation-alignment}):
\begin{equation}
t_s
=
\mathcal{A}_{\mathrm{ind}}^{\mathrm{align}}
\left(
u_s^{\mathrm{span}}, R_g
\right),
\qquad
a_s
=
\left(
u_s^{\mathrm{span}},
u_{t_s}^{\mathrm{kg}},
\xi_s
\right).
\label{eq:span_relation_alignment}
\end{equation}
The resulting atoms preserve the number and order of the input spans, with \(\xi_s\) retaining source provenance. A relation may be assigned to multiple spans, while the alignment encourages coverage of distinct segment-level relations before reuse whenever they are semantically compatible.

\subsubsection{Anatomy-Constrained Multi-View Graph and Leiden Partition}
\label{app:leiden_construction}

ASTAR applies K-means to globally cluster dense embeddings of clinical spans by minimizing the squared distance between each span and its assigned cluster centroid:
\begin{equation}
\mathcal{J}_{\mathrm{KM}}
=
\sum_i
\left\|
\mathbf{h}_i-\boldsymbol{\mu}_{c_i}
\right\|_2^2,
\label{eq:app_kmeans_objective}
\end{equation}
where \(\mathbf{h}_i\) is the embedding of span \(i\) and \(\boldsymbol{\mu}_{c_i}\) is the centroid of its assigned cluster. This objective represents each clinical concept cluster by a single prototype and measures cluster coherence through proximity to that prototype. It is therefore better suited to concept groups that form compact distributions around representative centroids in a shared vector space, but can be poorly matched to clinical concepts with heterogeneous local densities. In contrast, our atom similarity is not derived from a single vector distance. It integrates sparse lexical similarity and dense semantic similarity from both the text and relation views, producing heterogeneous pairwise affinities that cannot necessarily be reduced to Euclidean distances from a single centroid without losing information. To preserve the multi-view information encoded by these pairwise affinities, we directly represent the fused affinity \(S_{ij}\) as a weighted graph.

Let \(b_i\in\mathcal{B}\) denote the normalized anatomical region of atom \(a_i\), and let
\begin{equation}
\mathcal{V}_b
=
\left\{
i\mid b_i=b
\right\}
\label{eq:app_anatomical_bucket}
\end{equation}
denote the atoms assigned to region \(b\). Graph construction is restricted to atoms within the same region, such that \(A_{ij}=0\) whenever \(b_i\neq b_j\). For each \(i\in\mathcal{V}_b\), we define its regional \(k\)-nearest-neighbor set according to the fused text--relation affinity in Eq.~(\ref{eq:dual_view_graph}):
\begin{equation}
N_k^{(b)}(i)
=
\operatorname{TopK}_{j\in\mathcal{V}_b\setminus\{i\}}
S_{ij}.
\label{eq:app_regional_knn}
\end{equation}
The directed neighborhoods are then converted into a sparse undirected graph:
\begin{equation}
A_{ij}^{(b)}
=
S_{ij}
\mathbb{I}
\left[
S_{ij}\geq\delta
\right]
\mathbb{I}
\left[
j\in N_k^{(b)}(i)
\;\lor\;
i\in N_k^{(b)}(j)
\right],
\qquad
i,j\in\mathcal{V}_b.
\label{eq:app_symmetric_knn}
\end{equation}
The neighborhood-union rule preserves a strong one-sided nearest-neighbor relation when local concept densities are asymmetric, while the threshold \(\delta\) removes weak edges that could otherwise connect unrelated concepts. We use \(k=20\) and \(\delta=0.25\).

For the resulting regional graph, let
\begin{equation}
d_i^{(b)}
=
\sum_{j\in\mathcal{V}_b}
A_{ij}^{(b)},
\qquad
2w_b
=
\sum_{i\in\mathcal{V}_b}
d_i^{(b)}
\label{eq:app_regional_degree}
\end{equation}
denote the weighted degree of node \(i\) and twice the total undirected edge weight, respectively. For \(w_b>0\), we define the degree-corrected excess affinity between two atoms as
\begin{equation}
\Delta_{ij}^{(b)}
=
A_{ij}^{(b)}
-
\gamma
\frac{
d_i^{(b)}d_j^{(b)}
}{
2w_b
},
\label{eq:app_excess_affinity}
\end{equation}
where the second term is their expected connection strength under a degree-preserving null model. Leiden community detection~\citep{traag2019louvain} then approximately maximizes the Reichardt--Bornholdt objective:
\begin{equation}
\widehat{\mathbf{c}}^{(b)}
\approx
\arg\max_{\mathbf{c}^{(b)}}
\frac{1}{2w_b}
\sum_{i,j\in\mathcal{V}_b}
\Delta_{ij}^{(b)}
\mathbb{I}
\left[
c_i^{(b)}=c_j^{(b)}
\right].
\label{eq:app_rb_objective}
\end{equation}
This objective identifies atom communities whose observed internal affinities exceed those expected from their weighted degrees. We set the resolution parameter to \(\gamma=1.5\). Anatomical regions containing fewer than \(n_{\min}=30\) atoms bypass graph partitioning and are retained as single communities because the available evidence is insufficient for a stable finer partition. For regions with at least \(n_{\min}\) atoms, the Leiden objective applies when \(w_b>0\), with isolated atoms retained as singleton communities. If \(w_b=0\), every atom is isolated; the stated singleton convention therefore defines the partition without evaluating the objective. Letting \(\widehat{\mathcal{C}}_b\) denote the resulting partition of region \(b\), the complete partition is
\begin{equation}
\widehat{\mathcal{C}}
=
\bigcup_{b\in\mathcal{B}}
\widehat{\mathcal{C}}_b.
\label{eq:app_global_partition}
\end{equation}
Each community retains its member atom identifiers and source provenance for subsequent slot induction. Unlike K-means, which groups atoms according to their proximity to a shared centroid and requires the selection of a global cluster count \(C\), our construction uses anatomical labels to restrict the admissible comparison scope and multi-view evidence to determine local connections. Leiden then groups atoms according to whether their observed affinities exceed the expectations of a degree-preserving null model, allowing the number and structure of communities to adapt to each anatomical region. This combination reduces the risk that anatomically distinct or semantically heterogeneous clinical concepts are absorbed into the same global centroid-based cluster.

\subsubsection{Community-to-Slot Transition}
\label{app:slot_triple_induction} 

Let $\mathcal{C}=\{a_i\}_{i=1}^{|\mathcal{C}|}$ 
denote a Leiden community, where \(a_i\) is its \(i\)-th semantic atom and
\(|\mathcal{C}|\) is the number of atoms in the community. Although atoms
within the same community are similar in anatomical location, textual
semantics, and relational structure, they may encode distinct clinical
attributes. The slot-induction action therefore identifies clinically
meaningful semantic distinctions within \(\mathcal{C}\) and divides it into
\(Q_{\mathcal{C}}\) semantically coherent subgroups, where
\(q\in\{1,\ldots,Q_{\mathcal{C}}\}\) indexes the \(q\)-th subgroup. One
canonical slot is induced from each subgroup (Fig.~\ref{fig:prompt-community-slot-induction}):
\begin{equation}
\phi_{\mathcal{C},q}
=
\left(
k_{\mathcal{C},q},
\tau_{\mathcal{C},q},
\Omega_{\mathcal{C},q}
\right),
\label{eq:app_community_slot}
\end{equation}
where \(k_{\mathcal{C},q}\) is the canonical English slot key,
\(\tau_{\mathcal{C},q}
\in
\{\texttt{categorical},\texttt{numerical},\texttt{freetext}\}\)
is its value type, and \(\Omega_{\mathcal{C},q}\) is the candidate-value
set supported by the subgroup evidence.

For example, one community in the induced graph contains 65
adrenal-gland atoms, including 44 atoms describing SUVmax, 13 describing
size, and two describing diameter. Taking the SUVmax subgroup as an
example, its corresponding slot is
\begin{equation}
\phi_{\mathcal{C},2}
=
\left(
\texttt{adrenal\_suvmax},
\texttt{numerical},
\left\{
\langle\mathrm{NUM}\rangle
\right\}
\right).
\label{eq:app_adrenal_suvmax_example}
\end{equation}
This example illustrates that atoms from the same anatomical region may
belong to a common Leiden community while still requiring separate slots
for anatomically related but clinically distinct measurements, such as
structural dimensions and metabolic activity. A single slot is induced by
default when a community concentrates on one clinical attribute; multiple
slots are produced only when the community contains a clinically meaningful
semantic separation.

After community-level induction, slot proposals with the same canonical key
are merged globally. Candidate values are first deduplicated and then ranked
by the number of supporting proposals, with the most frequently supported
values retained. When proposals disagree on value type, the majority type is
used as the initial merged type. For keys supported by multiple proposals,
an LLM may further harmonize the canonical name, description, value type,
and candidate-value set without changing the underlying atom-to-slot
evidence assignments (Fig.~\ref{fig:prompt-slot-consolidation}).

Let \(A(\phi)\) denote the set of distinct atoms associated with slot
\(\phi\). Slot support is defined as \(|A(\phi)|\), and only slots satisfying
the support threshold are retained:
\begin{equation}
\Phi_{\mathrm{keep}}
=
\left\{
\phi\in\Phi
\;\middle|\;
|A(\phi)|>\gamma_s
\right\}.
\label{eq:app_slot_support_filter}
\end{equation}
We set \(\gamma_s=3\), requiring each retained slot to be supported by at
least four distinct atoms. Only the filtered slot library
\(\Phi_{\mathrm{keep}}\) is passed to the hierarchical construction stage
described in Appendix~\ref{app:hierarchical_template_construction}.

\subsubsection{Provenance-Preserving Relation Consolidation}
\label{app:relation-consolidation}

Relation normalization is performed in parallel with slot induction. Within each community \(C\), the subject, predicate, and object of each relation triple are normalized, and semantically equivalent relations are mapped to a canonical triple (Fig.~\ref{fig:prompt-relation-normalization})
\[
\bar{g} = (\bar{h}, \bar{p}, \bar{t}).
\]

Across communities, whitespace, capitalization, and delimiters are first standardized. Triples with identical normalized forms are then merged deterministically according to their complete signatures:
\begin{equation}
\sigma(g)
=
\left(
\operatorname{norm}(h),
\operatorname{norm}(p),
\operatorname{norm}(t)
\right).
\label{eq:relation_signature}
\end{equation}

Semantic comparison is subsequently restricted to triples associated with the same normalized anatomical region. Subject to predefined clinical constraints, the LLM merges relations that use different entity or predicate expressions but convey equivalent clinical semantics (Fig.~\ref{fig:prompt-anatomical-relation-consolidation}). These constraints preserve clinically meaningful distinctions. Measurement relations retain their measurement targets and quantitative semantics and are not collapsed into generic attribute relations. Relations involving left- and right-sided anatomical structures remain separate to preserve laterality. Relations representing positive findings, negative findings, and normal states are also kept distinct to prevent changes in clinical polarity during consolidation.

After consolidation, each semantic-equivalence class is represented by a canonical triple \(\bar{g} = (\bar{h}, \bar{p}, \bar{t})\). The canonical triple retains its normalized anatomical region, source atom identifiers, and source community identifiers, thereby preserving its correspondence to the original evidence. During clinical structuring, these canonical triples provide region-specific relational priors over recurrent anatomy--finding--attribute patterns.

\subsection{Template Evolution}
\label{app:template-evolution}

\subsubsection{Detailed Hierarchical Template Construction}
\label{app:hierarchical_template_construction}

\textbf{Anatomical Backbone Induction.} To construct the initial hierarchy, the Evolution Agent \(\mathcal{A}_{\mathrm{evo}}\) samples a fixed number of training reports and provides their description fields to the LLM. The LLM is instructed to generate a nested JSON tree with \texttt{desc} as its unique root (Fig.~\ref{fig:prompt-anatomical-backbone-induction}). Internal nodes organize the reports by major anatomical regions and organ-level structures, whereas leaf nodes remain empty for subsequent slot attachment. The resulting anatomical backbone \(H_0\) defines the candidate paths used during the first round of slot assignment.

\textbf{First-Round Slot Assignment.} For each retained slot \(\phi_j=(k_j,\tau_j,\Omega_j)\), the first-round assignment jointly considers its normalized key, semantic description, value type, and candidate vocabulary (Fig.~\ref{fig:prompt-first-round-slot-assignment}). The LLM returns
\begin{equation}
a_j^{(1)}
=
\left(
d_j^{(1)},
t_j^{(1)},
p_j^{(1)},
b_j^{(1)}
\right),
\label{eq:first_round_assignment}
\end{equation}
where \(d_j^{(1)}\in\{\texttt{attach},\texttt{propose\_new},\texttt{drop}\}\) specifies whether the slot is attached to an existing path, requires a new branch, or is discarded. The tier \(t_j^{(1)}\in\{\texttt{core},\texttt{extended},\texttt{drop}\}\) indicates whether the slot belongs to the core template, the extended template, or neither. The candidate path is denoted by \(p_j^{(1)}\), and \(b_j^{(1)}\) optionally specifies a proposed branch through its parent path and branch name.

A proposed branch is accepted only if its parent path exists and its branch name is valid. Duplicate branches are ignored. Accepted branches are incorporated into \(H_0\) to produce the updated backbone \(H_1\). Because all first-round assignments are performed independently against the initial path set, a slot cannot use a branch proposed by another slot in the same round. We therefore introduce a second global reassignment stage.

\textbf{Global Path Reassignment.}
All non-dropped slots are re-evaluated against the complete path set of \(H_1\) (Fig.~\ref{fig:prompt-global-path-reassignment}). This stage does not create additional branches. Instead, it selects or corrects each slot's path within the completed hierarchy. Slots assigned to the \texttt{core} or \texttt{extended} tier are attached to their selected nodes, whereas dropped slots and assignments without valid paths are excluded. The resulting hierarchy and slot assignments form the preliminary template \(T_{\mathrm{pre}}\).

\textbf{Semantic Reorganization and Concept Harmonization.}
After path assignment, \(\mathcal{A}_{\mathrm{evo}}\) refines the template in two steps. First, slots within the same anatomical node are jointly examined to identify semantically redundant fields and strongly overlapping candidate vocabularies. Their candidate values are consolidated and their value types are reconciled while the original anatomical path is preserved (Fig.~\ref{fig:prompt-intra-node-reorganization}). The published intra-node prompt uses substantial candidate-value overlap as its merging criterion and majority voting to reconcile types. Value overlap alone does not establish semantic identity. Second, slots with the same normalized key across different anatomical paths are aligned in naming, candidate vocabulary, and value type while retaining their anatomical locations. These operations preserve the assigned anatomical paths and produce the coarsely revised template \(T_1\).

\textbf{Fine Revision.} Starting from \(T_1\), fine revision addresses ambiguous slot keys, incompatible value types, fragmented vocabularies, and numerical or unit inconsistencies. Candidate edits that fail the structural or unit checks of the deterministic reducer are rejected. Empty vocabularies are supplemented only when indexed training sentences support coherent values. The accepted revisions produce \(T_f\), which is passed to evidence grounding.

\subsubsection{Cross-Source Evidence Grounding}
\label{app:evidence-grounding}

For each slot $s \in T_f$ in the refined template, the Evolution Agent 
$\mathcal{A}_{\mathrm{evo}}$ constructs a type-aware retrieval query from the 
slot specification and its anatomical context (Fig.~\ref{fig:prompt-evidence-query-generation}):
\begin{equation}
q_s =
\operatorname{QueryGen}
\left(k_s, \tau_s, \Omega_s, \pi_s\right),
\end{equation}
where $k_s$, $\tau_s$, $\Omega_s$, and $\pi_s$ denote the slot key, value type, 
current vocabulary, and anatomical path, respectively. A bilingual slot label 
is also provided when available.

The retrieval objective depends on the slot type. For numerical slots, $\mathcal{A}_{\mathrm{evo}}$ searches for accepted units, reference intervals, and abnormality thresholds. For categorical and free-text slots, it searches for terminology and descriptions that distinguish normal, benign, and pathological states within the anatomical context. If query generation returns an empty result, the system constructs a fallback query by combining the slot key with the phrase \texttt{normal range clinical}.

Let $\mathcal{B} =
\{\text{PubMed}, \text{Europe PMC}, \text{Google}\}$ denote the external evidence sources. The initial evidence pool is defined as
\begin{equation}
\widetilde{R}_s =
\bigcup_{b \in \mathcal{B}}
\operatorname{Retrieve}_{b}(q_s, K_r),
\qquad K_r = 8,
\end{equation}
where up to eight records are retrieved from each source. Available abstracts 
and full texts are retained, with each document truncated to at most 
$6{,}000$ characters. Records returned by multiple sources are deduplicated 
using DOI, PMID, PMCID, normalized URL, and normalized title. When two records 
refer to the same document, their identifiers are merged, and the version with 
richer textual evidence is retained. In this subsection, $q_s$ denotes the external-evidence search
query for slot $s$, rather than the slot-value query used for
report reconstruction.

The deduplicated documents are ranked first by source authority and then by 
information richness. The source priority is PubMed $>$ Europe PMC $>$ Google. 
For a document $d$, its information-richness score is defined as
\begin{equation}
\rho(d) =
2\mathbb{I}_{\mathrm{fulltext}}(d)
+
\mathbb{I}_{\mathrm{abstract}}(d)
+
\min\left(
\left\lfloor
\frac{\lvert b_d \rvert}{500}
\right\rfloor,
5
\right),
\end{equation}
where $b_d$ denotes the textual evidence contained in the document. This 
ranking prioritizes authoritative sources and, within the same source, favors 
records containing full text, an abstract, or a richer evidence body. The 
resulting ranked records constitute the candidate evidence set $R_s$ for slot 
$s$. Notably, $\rho(d)$ denotes document information richness,
rather than the clinical-role field $\rho$ of a span.

The top six records are provided to $\mathcal{A}_{\mathrm{evo}}$, with the 
evidence body of each record truncated to $600$ characters. The agent then 
condenses the retrieval query and candidate evidence into a slot-level 
reference note (Fig.~\ref{fig:prompt-evidence-note-synthesis}). Each note is written in English and limited to $200$ 
characters. For numerical slots, the note summarizes evidence-informed units, 
reference intervals, and abnormality thresholds, prioritizing values explicitly 
reported in the retrieved snippets. For categorical and free-text slots, it 
summarizes the main clinical distinctions among normal, benign, and pathological 
states.

The generated reference note is attached to the slot as auxiliary metadata 
without replacing its key, value type, or vocabulary. The grounded template is 
therefore constructed as
\begin{equation}
T_g =
T_f \oplus
\left\{
s.\mathrm{note} \leftarrow n_s
\right\}_{s \in T_f},
\end{equation}
where $n_s$ denotes the reference note associated with slot $s$. If no usable evidence is retrieved or note generation fails, the original slot is retained without a note. The system also records the retrieval query, evidence provenance, generation rationale, and execution status for auditing.

\subsubsection{Behavior-Aware Granularity Calibration.}
\label{app:granularity-calibration}
Although compact slot vocabularies facilitate standardized extraction, they may collapse detailed clinical descriptions into generic states. After the Clinical Agent $\mathcal{A}_{\mathrm{cli}}$ applies $T_g$ to sampled training reports, the Evolution Agent $\mathcal{A}_{\mathrm{evo}}$ audits the resulting structured records $Z=\{z_i\}$. For each slot $s$, it examines the slot specification $(k_s,\tau_s,\Omega_s)$, the extracted values $Y_s$, their aligned source contexts $C_s$, and a template-induced generic-value lexicon $V_{\mathrm{gen}}$, which contains generic categorical values such as \textit{present} and \textit{absent}. A slot is flagged when fine-grained source descriptions are repeatedly mapped to such generic values.

For each flagged slot, $\mathcal{A}_{\mathrm{evo}}$ summarizes the fine-grained information lost during structuring into a compact set of slot-specific linguistic patterns $P_s$. For example, if descriptions containing explicit measurements are repeatedly mapped to \textit{present}, $P_s$ guides subsequent structuring to preserve the exact measurement and unit, as well as explicitly stated anatomical location, lesion morphology, and other clinically meaningful attributes. These patterns provide slot-level guidance for source-grounded structuring without modifying the candidate-value set $\Omega_s$.

\subsection{Multi-Dimensional Evaluation Details}
\label{app:evaluation}

\textit{Template Design Quality} and \textit{Clinical Utility} are assessed separately by LLM-based and expert evaluators, whereas \textbf{Template Coverage} and \textbf{Diagnostic Fidelity} use LLM judgments anchored to held-out reports and diagnostic references. \textbf{Information Fidelity} further provides a deterministic reference-based assessment without LLM involvement. Together, these complementary evaluations provide multiple perspectives on template quality and downstream performance, reducing reliance on any single LLM-as-judge assessment.

\subsubsection{Template Coverage}
\label{app:template-coverage}

For the $i$-th held-out report, let $C_i$ denote the set of extracted clinical 
concept keys (Fig.~\ref{fig:prompt-clinical-concept-extraction}), and let $M(c,T)\in\{0,1\}$ indicate whether concept $c$ receives 
lexical or semantic coverage in template $T$ (Fig.~\ref{fig:prompt-concept-slot-matching}). Given an evaluation set of 
$N_{\mathrm{eval}}$ reports, case-macro coverage is defined as
\begin{equation}
\mathrm{Cov}_{\mathrm{case}}
=
\frac{1}{N_{\mathrm{eval}}}
\sum_{i=1}^{N_{\mathrm{eval}}}
\frac{
\sum_{c\in C_i}M(c,T)
}{
|C_i|
}.
\end{equation}
This metric first computes the coverage ratio for each report and then averages 
the resulting ratios, thereby assigning equal weight to each held-out case.

Concept-micro coverage is defined as
\begin{equation}
\mathrm{Cov}_{\mathrm{key}}
=
\frac{
\sum_{i=1}^{N_{\mathrm{eval}}}
\sum_{c\in C_i}M(c,T)
}{
\sum_{i=1}^{N_{\mathrm{eval}}}|C_i|
}.
\end{equation}
Unlike case-macro coverage, $\mathrm{Cov}_{\mathrm{key}}$ assigns equal weight 
to each concept occurrence across the complete evaluation set.

\subsubsection{Template Quality and Clinical Utility Evaluation}
\label{app:template-quality-evaluation}

$\mathcal{A}_{\mathrm{eval}}$ evaluates each template from
two complementary perspectives: \emph{Template Design Quality} and
\emph{Clinical Utility}. Both the LLM-as-judge and domain experts assess the
same 19 dimensions, comprising 13 dimensions of template design quality and
six dimensions of clinical utility. The complete evaluation criteria specify both aspects (Tables~\ref{tab:template-design-quality} and~\ref{tab:clinical-utility}). The identifiers TD1--TD13 and DQ1--DQ13 refer to the same
13 template-design dimensions in the same order.

The same 19 evaluation dimensions and scoring scale were retained across datasets, while the evaluator prompts and expert forms were instantiated with dataset-specific modality and task descriptions: for abdominal CT, PET/CT-specific wording was replaced with abdominal-CT terminology, including the imaging modality (TD1), oncology-related abnormality examples such as pancreatic tumors, vascular invasion, and tumor staging (TD2), and examination ranges such as single-organ, upper-abdominal, and whole-abdominal CT (TD10); although the source reports were in English, TD5 evaluated both the English slot names and their required Chinese glosses ($k_{\mathrm{zh}}$).

\begin{table}[!htbp]
\centering
\caption{\textbf{Dimensions of Template Design Quality.}
The 13 dimensions assess the structural design, clinical safety,
adaptability, and overall usability of a radiology reporting template.
The criteria shown below correspond to the PET/CT instantiation.}
\label{tab:template-design-quality}
\small
\setlength{\tabcolsep}{5pt}
\renewcommand{\arraystretch}{1.15}
\begin{tabularx}{\linewidth}{
@{}
>{\raggedright\arraybackslash}p{3.25cm}
>{\centering\arraybackslash}p{1.05cm}
Y
@{}
}
\toprule
\textbf{Category} & \textbf{ID} & \textbf{Evaluation criterion} \\
\midrule

Coverage and completeness
& \texttt{TD1}
& Coverage of the essential structures and measurements required in
PET/CT reports. \\

Coverage and completeness
& \texttt{TD2}
& Explicit support for common abnormalities, including primary tumors,
lymph-node metastases, and distant metastases involving the bone,
liver, or lung. \\

Structure and granularity
& \texttt{TD3}
& Hierarchical organization consistent with radiological reasoning,
proceeding from findings to organ systems and key anatomical
structures. \\

Structure and granularity
& \texttt{TD4}
& Appropriate field granularity that preserves clinically relevant
information without making the template unnecessarily fragmented or
burdensome to complete. \\

Terminology and clarity
& \texttt{TD5}
& Clear field names and Chinese labels that conform to commonly used
clinical terminology. \\

Terminology and clarity
& \texttt{TD6}
& Natural and usable permissible values, such as \emph{Normal},
\emph{Abnormal}, \emph{Uncertain}, and \emph{NotAssessed}. \\

Safety and error prevention
& \texttt{TD7}
& Ability of the template to reduce the risk of omitting clinically
important findings. \\

Safety and error prevention
& \texttt{TD8}
& Protection against incorrectly recording an unassessed structure as
normal, including natural support for the \emph{NotAssessed} state. \\

Safety and error prevention
& \texttt{TD9}
& Appropriate representation of diagnostic uncertainty, including
expressions such as \emph{consider}, \emph{possible}, and
\emph{cannot be excluded}. \\

Adaptability and extensibility
& \texttt{TD10}
& Adaptability to different examination ranges, including regional,
torso, and whole-body PET/CT examinations. \\

Deployment feasibility
& \texttt{TD11}
& Feasibility of using the template in routine clinical reporting. \\

Sign-off safety
& \texttt{TD12}
& Whether the template is sufficiently safe to support clinical report
sign-off from a template-design perspective. \\

Overall assessment
& \texttt{TD13}
& Overall clinical usability of the template after considering all
preceding dimensions. \\

\bottomrule
\end{tabularx}
\end{table}

\begin{table}[!htbp]
\centering
\caption{\textbf{Dimensions of Clinical Utility.}
The 6 dimensions assess the expected value of the template in routine
radiological interpretation and reporting workflows.
The criteria shown below correspond to the PET/CT instantiation.}
\label{tab:clinical-utility}
\small
\setlength{\tabcolsep}{5pt}
\renewcommand{\arraystretch}{1.15}
\begin{tabularx}{\linewidth}{
@{}
>{\raggedright\arraybackslash}p{3.25cm}
>{\centering\arraybackslash}p{1.05cm}
Y
@{}
}
\toprule
\textbf{Category} & \textbf{ID} & \textbf{Evaluation criterion} \\
\midrule

Omission reduction
& \texttt{CU1}
& The template reduces omissions of clinically important findings. \\

Reporting consistency
& \texttt{CU2}
& The template improves consistency across PET/CT reports. \\

Workflow alignment
& \texttt{CU3}
& The ordering of template fields is consistent with the clinician's
image-reading workflow. \\

Unassessed-state support
& \texttt{CU4}
& The template enables unassessed structures or findings to be
represented naturally using the \emph{NotAssessed} state. \\

Communication efficiency
& \texttt{CU5}
& The template improves communication efficiency between radiologists
and referring clinicians. \\

Overall satisfaction
& \texttt{CU6}
& Overall evaluator satisfaction with the template. \\

\bottomrule
\end{tabularx}
\end{table}

\FloatBarrier

\textbf{Score aggregation.}
For each evaluator type
$e \in \{\mathrm{LLM},\mathrm{Expert}\}$, we first average the corresponding
dimension-level ratings to obtain the Template Design Quality score
$S_{\mathrm{design}}^{(e)}$ and the Clinical Utility score
$S_{\mathrm{utility}}^{(e)}$:
\begin{equation}
S_{\mathrm{design}}^{(e)}
=
\frac{1}{13}\sum_{i=1}^{13}s_{\mathrm{DQ},i}^{(e)},
\qquad
S_{\mathrm{utility}}^{(e)}
=
\frac{1}{6}\sum_{j=1}^{6}s_{\mathrm{CU},j}^{(e)}.
\end{equation}
Here $s_{\mathrm{DQ},i}^{(e)}$ and $s_{\mathrm{CU},j}^{(e)}$ denote the ratings of design-quality dimension $i$ (DQ1--DQ13) and clinical-utility dimension $j$ (CU1--CU6), respectively. The overall template-quality score for evaluator type $e$ is then computed as
\begin{equation}
S_{\mathrm{template}}^{(e)}
=
\frac{1}{2}
\left(
S_{\mathrm{design}}^{(e)}
+
S_{\mathrm{utility}}^{(e)}
\right).
\end{equation}

\subsubsection{Information Fidelity}
\label{app:information-fidelity}

Information fidelity measures the preservation of patient-specific information throughout the complete \emph{text $\rightarrow$ structure $\rightarrow$ text} cycle. Specifically, given an original report $x$, the Clinical Agent first converts it into a hierarchical structured record $z$ using the final template $T^\star$ (Fig.~\ref{fig:prompt-region-conditioned-structuring}) and then generates a reconstructed report $\hat{x}$ from $z$ (Fig.~\ref{fig:prompt-report-reconstruction}). The reconstructed report $\hat{x}$ is compared with the original report $x$ using ROUGE-1, ROUGE-2, ROUGE-L, chrF, chrF++, and BERTScore. These metrics characterize reconstruction fidelity from lexical, character-level, and contextual-semantic perspectives. The structured record $z$ remains the sole source of patient-specific facts during reconstruction. Retrieved training sentences provide only linguistic and organizational priors and must not introduce patient-specific findings absent from $z$.

\subsubsection{Diagnostic Fidelity}
\label{app:diagnostic-fidelity}

Diagnostic fidelity is evaluated under an \emph{information-closure} setting.
During diagnosis generation (Fig.~\ref{fig:prompt-diagnostic-reasoning}), the Clinical Agent may access only the
structured representation $z$ of the current case and the frozen diagnostic
graph $G_{\mathrm{diag}}$ constructed exclusively from the training set.
The reference report and diagnosis annotation are withheld during reasoning
and used only for final evaluation. The Evaluation Agent acts as an LLM judge, scoring
the generated diagnostic output against the withheld reference report and
diagnosis annotation.
The judging prompt specifies diagnostic-output evaluation (Fig.~\ref{fig:prompt-diagnostic-fidelity-evaluation}).
We assess three complementary criteria:

PDA is rated on a 0--5 scale: 0, absent or wholly incorrect;
1, nonspecific but compatible with the reference;
2, partially correct information but an incorrect primary diagnosis;
3, correct primary diagnosis with a clinically important qualifier
missing or incorrect;
4, correct diagnosis with only minor discrepancies;
5, clinically equivalent to the reference, including essential
qualifiers and diagnostic uncertainty. Important qualifiers include
anatomical site, laterality, and severity or stage when specified
in the reference.

KFP is the proportion of reference key findings correctly preserved:
$\mathrm{KFP}=M/K$, where $K$ is the number of distinct
diagnostically relevant reference findings and $M$ is the number
correctly retained. The reference findings and their essential
qualifiers are fixed before inspecting any method's output and
shared across methods. Semantically equivalent expressions count
as matches; omissions, contradictions, and missing essential
qualifiers do not. Unsupported additions receive no credit.
Cases with $K=0$ are excluded from KFP aggregation, with the
effective sample size reported.

CA measures whether the output provides sufficient supported
information for a reference-consistent clinical interpretation:
0, uninterpretable;
1, misleading;
2, usable only after major correction;
3, useful for a broad interpretation but missing important context;
4, adequate with minor omissions;
5, clear, sufficiently specific, and fully supported.
Fabricated findings or unsupported recommendations that could
materially change clinical interpretation or management cap CA
at 1. Recommendations are not required when the references
provide no basis for them.

\textbf{Score normalization and aggregation.}
For each evaluated case $i$, the judge returns raw scores
$r_i^{\mathrm{PDA}}, r_i^{\mathrm{CA}} \in \{0,1,2,3,4,5\}$
and $r_i^{\mathrm{KFP}} \in [0,1]$.
We linearly normalize PDA and CA by dividing their raw scores by 5,
while retaining KFP on its original scale.
All three metrics are therefore reported on a common $[0,1]$ scale,
with higher values indicating better diagnostic fidelity.
PDA represents a normalized ordinal rating of diagnostic correctness.

Let $N_{\mathrm{eval}}$ denote the number of evaluated cases in a dataset.
The dataset-level scores are computed as
\begin{equation}
\begin{aligned}
\mathrm{PDA}
&= \frac{1}{N_{\mathrm{eval}}}
   \sum_{i=1}^{N_{\mathrm{eval}}}
   \frac{r_i^{\mathrm{PDA}}}{5}, \\
\mathrm{KFP}
&= \frac{1}{N_{\mathrm{eval}}}
   \sum_{i=1}^{N_{\mathrm{eval}}}
   r_i^{\mathrm{KFP}}, \\
\mathrm{CA}
&= \frac{1}{N_{\mathrm{eval}}}
   \sum_{i=1}^{N_{\mathrm{eval}}}
   \frac{r_i^{\mathrm{CA}}}{5}.
\end{aligned}
\end{equation}
The Diagnostic Fidelity Avg.\ reported in the result tables
is the unweighted mean of these three normalized metrics:
\begin{equation}
\mathrm{Avg.}_{\mathrm{DF}}
= \frac{\mathrm{PDA}+\mathrm{KFP}+\mathrm{CA}}{3}.
\end{equation}

\section{Result Details}
\label{app:result-details}

\subsection{Dataset}
\label{app:datasets}

(1) \textbf{PETCT}, our shorthand for PETWB-REP~\citep{PETCT}, comprises 490 whole-body \(^{18}\mathrm{F}\)-FDG PET/CT examinations, each accompanied by de-identified bilingual imaging reports in Chinese and English and structured clinical information. The dataset covers various malignancies, including lung, liver, breast, prostate, and ovarian cancers. The reports describe both CT morphological findings and PET metabolic findings, including lesion location, size, density, FDG uptake, SUV/SUVmax, lymph-node involvement, and distant metastases. We used all 490 English-language reports, of which 441 were assigned to the training set and 49 to the test set. (2) \textbf{ViMed} is a whole-body PET/CT dataset collected at a tertiary hospital in Vietnam. The official dataset contains PET/CT studies and their corresponding complete Vietnamese clinical reports collected in 2017, 2018, 2019, and 2023. The reports are organized into head and neck, chest, and abdominopelvic sections and include the examination indication, clinical history, demographic information, regional imaging findings, and diagnostic summary. We used the publicly available English version, in which the Vietnamese reports were translated into English using Gemma-4-26B-A4B-it. After excluding reports with empty imaging findings, 2,761 reports remained. We then performed a stratified split by year and month, yielding 2,484 training reports and 277 test reports. (3) \textbf{MERLIN} is a large-scale abdominal CT dataset covering normal and abnormal findings involving the hepatobiliary system, spleen, pancreas, kidneys, gastrointestinal tract, vasculature, lymph nodes, and other abdominopelvic structures. We selected 4,080 reports containing valid content in both the FINDINGS and SUMMARY sections. The main text of the FINDINGS section was used as the imaging description, whereas the SUMMARY section was retained as the diagnostic reference. For the current experiments, the reports were divided into 3,672 training reports and 408 test reports. (4) \textbf{BDMAP} contains abdominal CT volumes, paired imaging reports, and voxel-level annotations of structures such as liver, kidney, and pancreatic tumors. The reports were generated by RadGPT from organ and tumor segmentation results and contain quantitative and qualitative information, including organ volumes, CT attenuation values, lesion sizes, tumor burden, and organ status. After quality filtering and narrative-text deduplication, we randomly selected 2,000 reports to construct a BDMAP-derived text subset. Of these, 1,800 reports were used for training and 200 for testing.

\subsection{Batched Template Construction for LLM Baselines}
\label{app:llm-baseline-construction}

Due to the context-length limitation of LLMs, the complete training corpus
cannot generally be processed within a single model context. Therefore, for
LLM baselines in the template construction task, a two-stage procedure is
adopted. The complete training corpus is first divided into multiple smaller
batches. In the first stage, each batch of reports is summarized into a local
template. In the second stage, these local templates are consolidated while
preserving their combined clinical coverage as much as possible.

\textbf{Batching Strategy.}
Let
$\mathcal{D}_{\mathrm{train}}=\{x_i\}_{i=1}^{N}$
denote the training report set. A fixed random seed is first used to shuffle
the reports, which are then partitioned into disjoint batches:
\begin{equation}
\mathcal{D}_{\mathrm{train}}
=
\biguplus_{b=1}^{B}\mathcal{B}_b.
\end{equation}
The same batch partition $\mathcal{B}_b$ is reused across all LLM baselines to
ensure a fair comparison. Since report lengths vary substantially across the
four datasets, the number of reports per batch is determined separately for
each dataset so that the serialized input payloads remain approximately
comparable. Specifically, the batch sizes are set to 20, 80, 40, and 150 for
PETCT, ViMed, MERLIN, and BDMAP, respectively. If the final remainder contains
no more than half of the nominal batch size, it is merged into the preceding
batch; otherwise, it is retained as a separate batch.

\textbf{Batch-Level Template Induction.}
For each batch $\mathcal{B}_b$, model $f_\theta$ receives the same system
instruction, a dataset-specific modality description, a shared one-shot
template example, and the report descriptions within the batch, producing a
local template $T_b^{(\theta)}$.

Each template is represented as a JSON object organized by anatomical regions
and organs. Generated outputs undergo strict JSON parsing and schema
validation. When strict parsing fails, the conservative fallback described
below attempts to recover a complete template object already present in the
output. Only parsed or recovered objects that satisfy the template schema
are admitted to the merging stage; outputs without such an object remain
failures.

\textbf{Local Template Merging.}
All valid local templates are serialized and passed again to the same model.
A merging strategy $\mathcal{M}_\theta$ is then used to consolidate them into
the final template:
\begin{equation}
T^{(\theta)}
=
\mathcal{M}_{\theta}
\left(
T_1^{(\theta)},\ldots,T_B^{(\theta)}
\right).
\end{equation}
During this process, the original anatomical organization and clinically
distinct information are preserved.

This procedure separates \textit{evidence acquisition} from
\textit{global normalization}: the induction stage summarizes the information
contained in each batch, whereas the merging stage reconciles naming and
semantic granularity across batches.

When the serialized local templates still exceed the context-length or
generation budget of a single model call, the same merging operator is applied
hierarchically.

At level $\ell$, the current templates are divided into groups
$\{\mathcal{G}^{(\ell)}_j\}_j$, and each group is reduced independently:
\begin{equation}
T_j^{(\ell+1)}
=
\mathcal{M}_{\theta}
\left(
\left\{
T_i^{(\ell)}
:
i\in\mathcal{G}^{(\ell)}_j
\right\}
\right).
\end{equation}
This process is repeated until a single unified template remains.

If the final merge group is too small, it is merged into the preceding group
so that the intermediate templates summarize approximately comparable amounts
of information.

\subsection{Aggregate Results Across Datasets}
\label{app:aggregate-results}

The unweighted macro-average results across PETCT, ViMed, MERLIN, and BDMAP (Table~\ref{tab:macro_average}) complement the dataset-specific comparisons (Tables~\ref{tab:comparison_petct_vimed} and~\ref{tab:comparison_merlin_bdmap}). Overall, ASTAR provides the strongest
baseline performance when considering Template Coverage, Information Fidelity,
and Diagnostic Fidelity jointly, although Gemini-3.6-Flash achieves the
highest baseline score on Information Fidelity alone. Our method achieves the
best overall performance across all three criteria, with aggregate scores of
0.8437, 0.7988, and 0.8331, respectively, and obtains the highest score on
every individual metric reported in the table.

\begin{table*}[t]
\caption{
Performance comparison across PETCT, ViMed, MERLIN, and BDMAP
in template coverage, information fidelity, and diagnostic fidelity,
reported as unweighted macro-averages across the four datasets.
Higher is better for all metrics.
Among individual methods, best in \textbf{bold} and
runner-up in \underline{underlined}.
Case = case-macro coverage; Key = key-micro coverage;
Avg. = average; R-1/R-2/R-L = ROUGE-1/ROUGE-2/ROUGE-L;
chrF = character n-gram F-score; chrF++ = character and word n-gram F-score;
BERT = BERTScore; PDA = Primary Diagnosis Accuracy;
KFP = Key Finding Preservation; CA = Clinical Actionability.
}
\label{tab:macro_average}
\centering

\providecommand{\tabbest}[1]{\textbf{#1}}
\providecommand{\tabsecond}[1]{\underline{#1}}

\setlength{\tabcolsep}{2.8pt}
\renewcommand{\arraystretch}{0.96}
\scriptsize

\resizebox{\textwidth}{!}{%
\begin{tabular}{
@{}l
rrr
@{\hspace{4pt}}
rrrrrrr
@{\hspace{4pt}}
rrrr
@{}
}
\toprule

& \multicolumn{3}{c}{\textbf{Template Coverage}}
& \multicolumn{7}{c}{\textbf{Information Fidelity}}
& \multicolumn{4}{c}{\textbf{Diagnostic Fidelity}} \\

\cmidrule(lr){2-4}
\cmidrule(lr){5-11}
\cmidrule(lr){12-15}

\textbf{Method}
& \textbf{Case}
& \textbf{Key}
& \textbf{Avg.}
& \textbf{R-1}
& \textbf{R-2}
& \textbf{R-L}
& \textbf{chrF}
& \textbf{chrF++}
& \textbf{BERT}
& \textbf{Avg.}
& \textbf{PDA}
& \textbf{KFP}
& \textbf{CA}
& \textbf{Avg.} \\
\midrule

\multicolumn{15}{@{}l@{}}{\textit{Medical LLM baselines}} \\

lingshu-32b
& 0.5213 & 0.5339 & 0.5276
& 0.5171 & 0.2750 & 0.3669 & 0.3902 & 0.3742 & 0.8770 & 0.4667
& 0.4861 & 0.5172 & 0.5304 & 0.5112 \\

medgemma-27b
& 0.5388 & 0.5548 & 0.5468
& 0.6382 & 0.4104 & 0.4717 & 0.4873 & 0.4762 & 0.8970 & 0.5634
& 0.5710 & 0.6328 & 0.6232 & 0.6090 \\

clinfusion-32b
& 0.4894 & 0.4930 & 0.4912
& 0.3417 & 0.1849 & 0.2407 & 0.2180 & 0.2125 & 0.7995 & 0.3329
& 0.6236 & 0.6699 & 0.6873 & 0.6603 \\

\rowcolor{black!4}
\textbf{Medical LLM Avg.}
& 0.5165 & 0.5272 & 0.5219
& 0.4990 & 0.2901 & 0.3598 & 0.3651 & 0.3543 & 0.8578 & 0.4543
& 0.5602 & 0.6066 & 0.6136 & 0.5935 \\

\addlinespace[2pt]
\multicolumn{15}{@{}l@{}}{\textit{General LLM baselines}} \\

deepseek-v4-flash
& 0.6442 & 0.6626 & 0.6534
& 0.6228 & 0.3494 & 0.4326 & 0.4694 & 0.4571 & 0.8934 & 0.5374
& 0.6667 & 0.7099 & 0.7248 & 0.7004 \\

deepseek-v4-pro
& 0.6419 & 0.6478 & 0.6449
& 0.6130 & 0.3502 & 0.4401 & 0.4942 & 0.4759 & 0.8905 & 0.5440
& 0.6288 & 0.6779 & 0.6861 & 0.6643 \\

qwen3.6-flash
& 0.5866 & 0.6052 & 0.5959
& 0.5217 & 0.2970 & 0.3599 & 0.4019 & 0.3844 & 0.8800 & 0.4741
& 0.5403 & 0.5839 & 0.5887 & 0.5709 \\

gemini-3.6-flash
& 0.6362 & 0.6346 & 0.6354
& \tabsecond{0.6672}
& \tabsecond{0.4581}
& \tabsecond{0.5146}
& 0.5136 & 0.5028 & 0.9007
& \tabsecond{0.5928}
& 0.6140 & 0.6794 & 0.6661 & 0.6532 \\

gpt-5.4
& 0.5877 & 0.6160 & 0.6019
& 0.6145 & 0.4222 & 0.4634 & 0.4742 & 0.4625 & 0.8985 & 0.5559
& 0.4883 & 0.5404 & 0.5397 & 0.5228 \\

gemini-3.5-flash-lite
& 0.6442 & 0.6606 & 0.6524
& 0.6468 & 0.4396 & 0.4965 & 0.5171 & 0.5058
& \tabsecond{0.9015} & 0.5846
& 0.6492 & 0.6993 & 0.7069 & 0.6851 \\

\rowcolor{black!4}
\textbf{General LLM Avg.}
& 0.6235 & 0.6378 & 0.6306
& 0.6143 & 0.3861 & 0.4512 & 0.4784 & 0.4647 & 0.8941 & 0.5481
& 0.5979 & 0.6485 & 0.6520 & 0.6328 \\

\addlinespace[2pt]
\multicolumn{15}{@{}l@{}}{\textit{Template induction baseline}} \\

ASTAR
& \tabsecond{0.7938}
& \tabsecond{0.8249}
& \tabsecond{0.8093}
& 0.6503 & 0.3530 & 0.4063
& \tabsecond{0.5518}
& \tabsecond{0.5283}
& 0.8909 & 0.5634
& \tabsecond{0.7581}
& \tabsecond{0.8035}
& \tabsecond{0.7943}
& \tabsecond{0.7853} \\

\rowcolor{oursblue}
\textbf{Ours}
& \tabbest{0.8361}
& \tabbest{0.8513}
& \tabbest{0.8437}
& \tabbest{0.8585}
& \tabbest{0.7201}
& \tabbest{0.6791}
& \tabbest{0.8025}
& \tabbest{0.7938}
& \tabbest{0.9391}
& \tabbest{0.7988}
& \tabbest{0.8146}
& \tabbest{0.8533}
& \tabbest{0.8314}
& \tabbest{0.8331} \\

\bottomrule
\end{tabular}%
}
\end{table*}


\subsection{Ablation Analysis of Revision and Note.}
\label{app:template-quality-ablation}
To further disentangle the contributions of \textit{Revision} and \textit{Note} to template quality and clinical utility, we compare ASTAR with three sequential variants of our framework: \texttt{Ours w/o (Note + Rev.)}, in which both Revision and Note are removed; \texttt{Ours w/o Note}, which retains Revision but removes Note; and the full model, \texttt{Ours}, which incorporates both components.

Because Note influences template usage by supplementing guidance on observation priorities, abnormal manifestations, interpretation criteria, and slot completion rather than by introducing additional structural fields, we focus on 13 evaluation dimensions that are closely aligned with its functional pathway. These comprise eight Template Design dimensions (TD1, TD2, TD7, TD8, TD9, TD11, TD12, and TD13) and five Clinical Utility dimensions (CU1, CU2, CU4, CU5, and CU6). Collectively, these dimensions assess coverage of key structures, representation of abnormal findings, omission prevention, prevention of erroneous normal-state completion, uncertainty representation, clinical feasibility, sign-off safety, overall usability, reduction of missing information, reporting consistency, support for not-assessed states, communication efficiency, and overall satisfaction.

For each dataset, we use an LLM judge to conduct exhaustive pairwise comparisons among the four candidate templates: ASTAR, \texttt{Ours w/o (Note + Rev.)}, \texttt{Ours w/o Note}, and \texttt{Ours}. The four templates form six unordered pairs. Each pair is evaluated twice with the A/B presentation order reversed, yielding 12 directed comparisons in total to mitigate potential positional bias. In each comparison, the judge is provided with the complete slot schemas of both templates and, for each of the 13 dimensions, outputs a winner (\texttt{A}, \texttt{B}, or \texttt{tie}) together with an advantage magnitude $m \in \{0,1,2,3\}$, where $0$ denotes no distinguishable difference and $3$ denotes a decisive advantage.

For dimension $d$, the directional advantage of the first-presented template over the second is defined as:
\begin{equation}
a_d(A,B)=
\begin{cases}
m/3, & \text{if A is preferred},\\
-m/3, & \text{if B is preferred},\\
0, & \text{if the judgment is a tie}.
\end{cases}
\end{equation}
For a pair of templates $X$ and $Y$, the two presentation orders are then symmetrized as:
\begin{equation}
s_d(X,Y)
=
\frac{a_d(X,Y)-a_d(Y,X)}{2},
\end{equation}
so that judgments from both A/B orderings jointly determine the relative advantage of $X$ over $Y$ on dimension $d$.

Because each template is compared against the other three candidates, its score on dimension $d$ is computed as:
\begin{equation}
S_d(X)
=
0.5
+
0.5
\times
\frac{1}{3}
\sum_{Y \neq X}
s_d(X,Y),
\end{equation}
which maps the resulting relative-preference score to the interval $[0,1]$. A score of $0.5$ indicates that the template is, on average, comparable to the other candidates in the current comparison set; values approaching $1$ indicate consistent preference over competing templates, whereas values approaching $0$ indicate consistent disadvantage.

Finally, \textit{Selected Design} is obtained by averaging $S_d(X)$ over the eight selected Template Design dimensions, while \textit{Selected Utility} is obtained by averaging over the five selected Clinical Utility dimensions. \textit{Selected Overall} is computed directly as the mean over all 13 selected dimensions. These scores therefore quantify the relative preference of each template within the current candidate set rather than an absolute assessment of clinical quality independent of the competing templates.

\begin{figure}[!htbp]
    \centering
    \includegraphics[width=\textwidth]{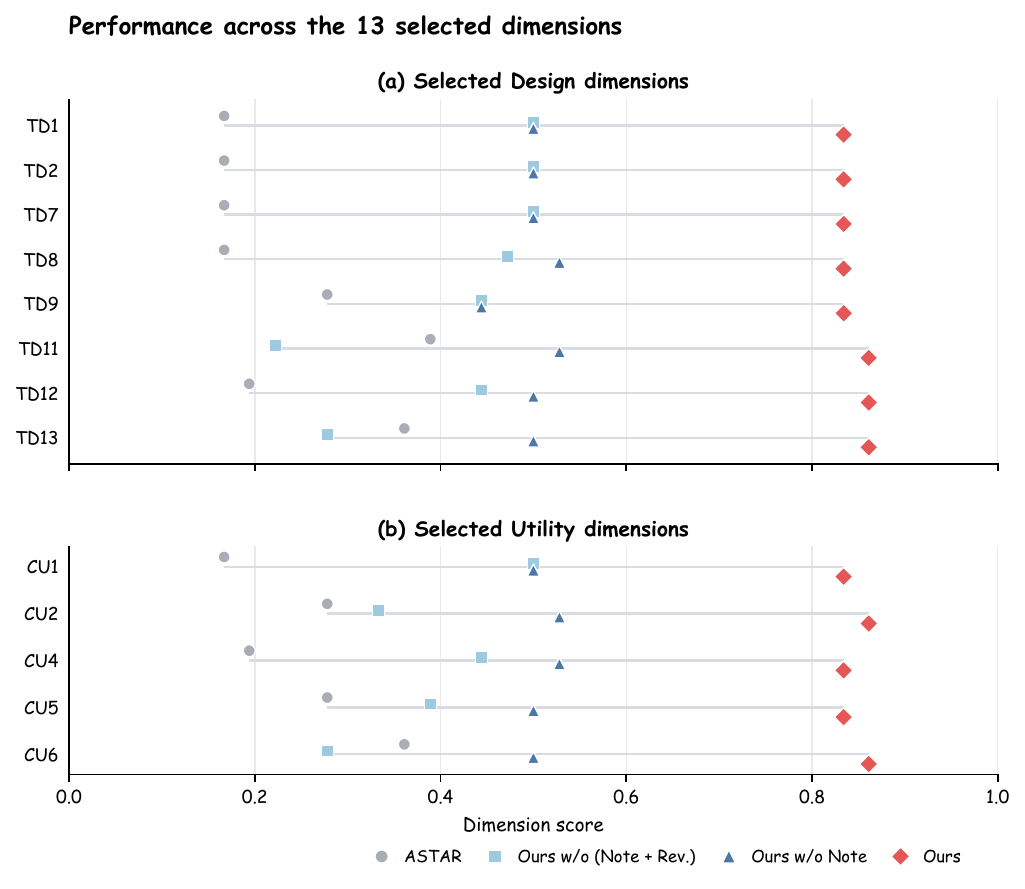}
\caption{
\textbf{Dimension-wise ablation analysis on the PETCT dataset.}
Results are reported for (a) eight selected Design Quality dimensions and (b) five selected Clinical Utility dimensions, comparing ASTAR, Ours w/o (Note + Rev.), Ours w/o Note, and the full model. Higher scores indicate better performance on the corresponding dimension.
}
    \label{fig:petct-4}
\end{figure}

\begin{figure}[!htbp]
    \centering
    \includegraphics[width=\textwidth]{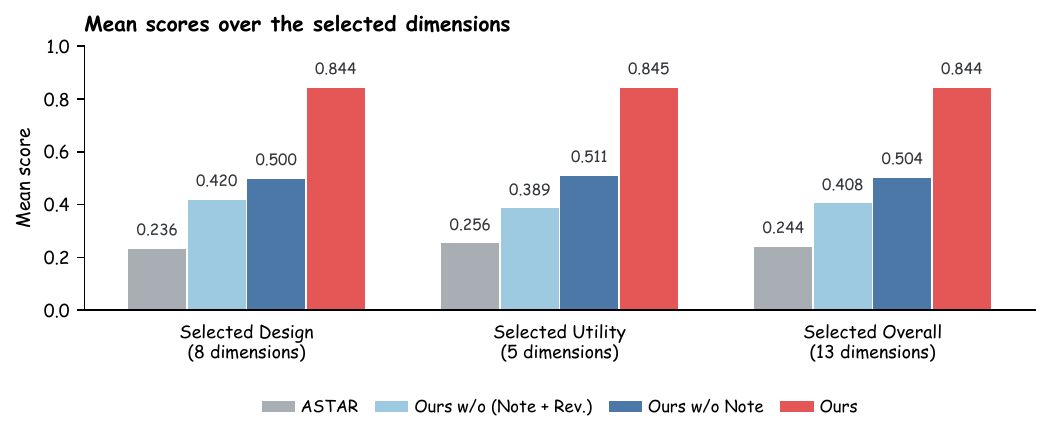}
\caption{
\textbf{Aggregated ablation results on the PETCT dataset.}
Mean scores are reported over the eight selected Design Quality dimensions, five selected Clinical Utility dimensions, and all 13 selected dimensions for ASTAR, Ours w/o (Note + Rev.), Ours w/o Note, and the full model.
}
    \label{fig:petct-5}
\end{figure}

\begin{figure}[!htbp]
    \centering
    \includegraphics[width=\textwidth]{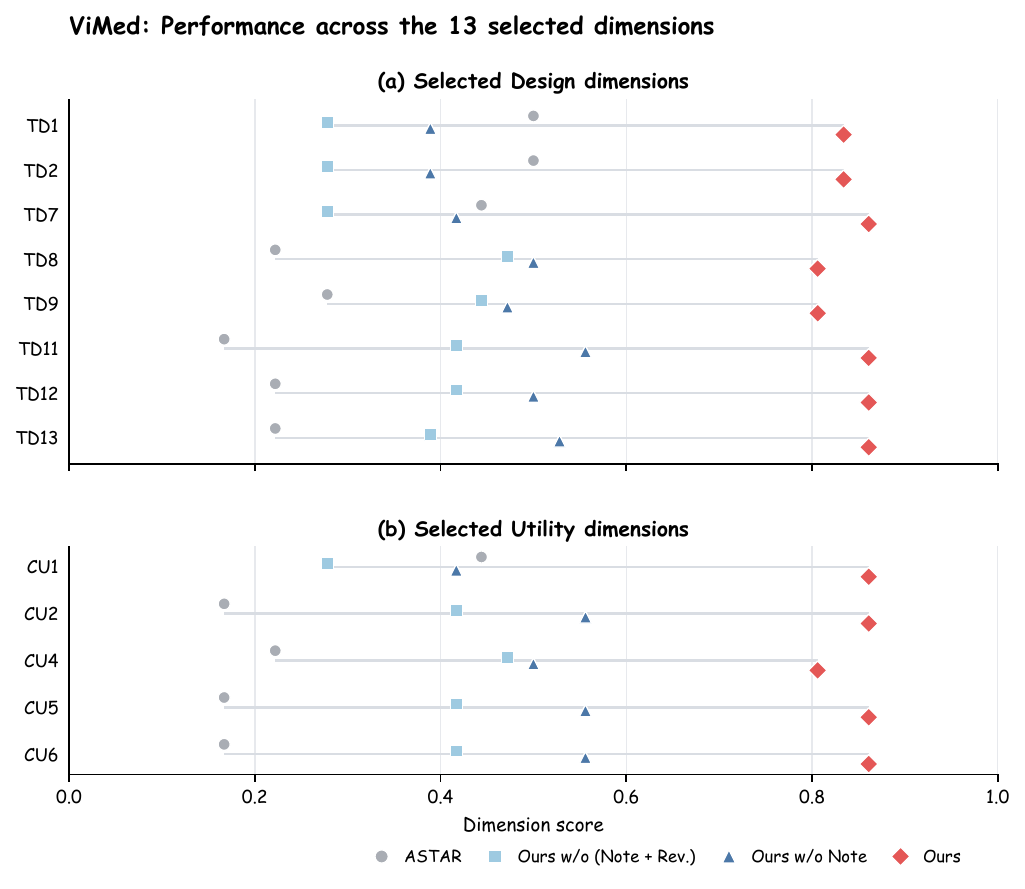}
\caption{
\textbf{Dimension-wise ablation analysis on the ViMed dataset.}
Results are reported for (a) eight selected Design Quality dimensions and (b) five selected Clinical Utility dimensions, comparing ASTAR, Ours w/o (Note + Rev.), Ours w/o Note, and the full model. Higher scores indicate better performance on the corresponding dimension.
}
    \label{fig:vimed-4}
\end{figure}

\begin{figure}[!htbp]
    \centering
    \includegraphics[width=\textwidth]{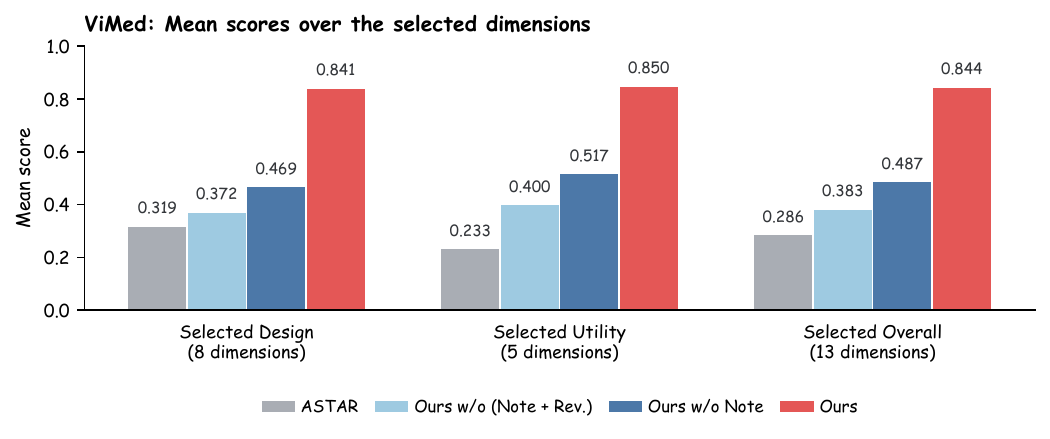}
\caption{
\textbf{Aggregated ablation results on the ViMed dataset.}
Mean scores are reported over the eight selected Design Quality dimensions, five selected Clinical Utility dimensions, and all 13 selected dimensions for ASTAR, Ours w/o (Note + Rev.), Ours w/o Note, and the full model.
}
    \label{fig:vimed-5}
\end{figure}

\begin{figure}[!htbp]
    \centering
    \includegraphics[width=\textwidth]{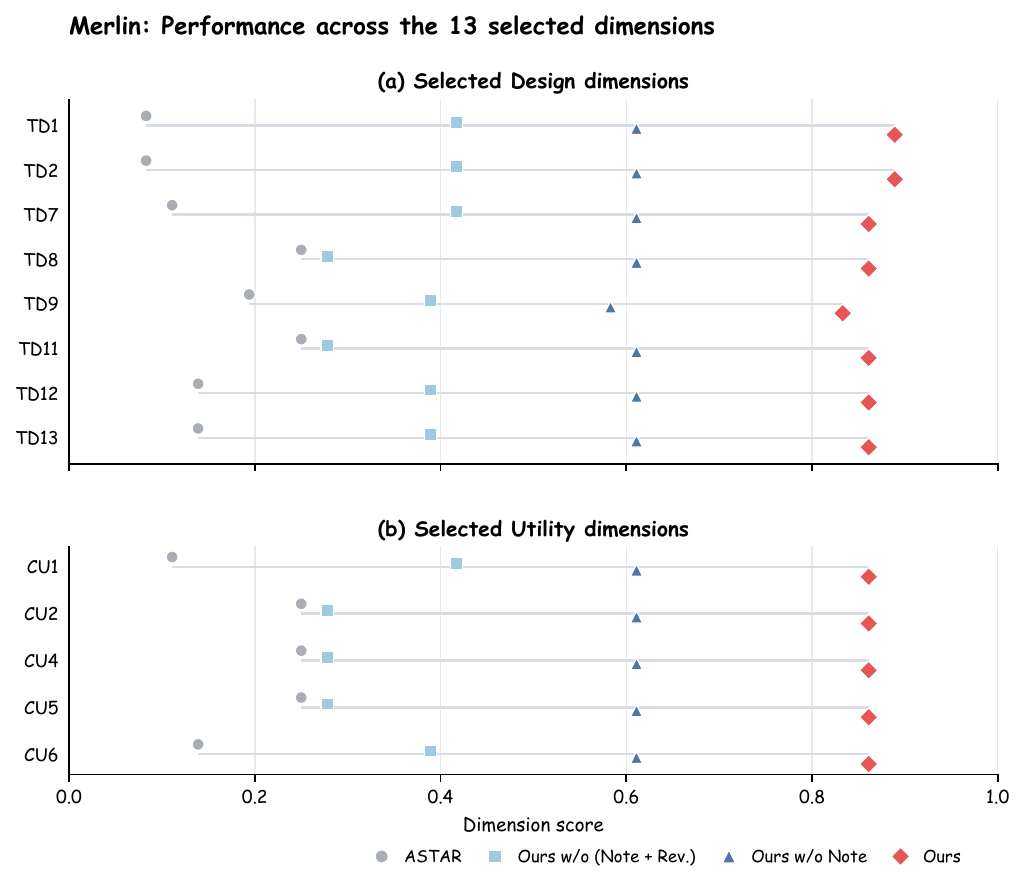}
\caption{
\textbf{Dimension-wise ablation analysis on the MERLIN dataset.}
Results are reported for (a) eight selected Design Quality dimensions and (b) five selected Clinical Utility dimensions, comparing ASTAR, Ours w/o (Note + Rev.), Ours w/o Note, and the full model. Higher scores indicate better performance on the corresponding dimension.
}
    \label{fig:merlin-4}
\end{figure}

\begin{figure}[!htbp]
    \centering
    \includegraphics[width=\textwidth]{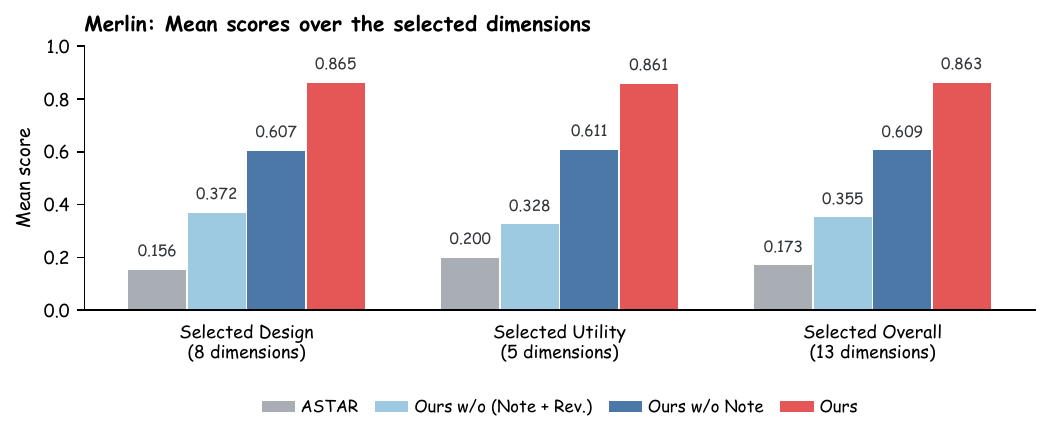}
\caption{
\textbf{Aggregated ablation results on the MERLIN dataset.}
Mean scores are reported over the eight selected Design Quality dimensions, five selected Clinical Utility dimensions, and all 13 selected dimensions for ASTAR, Ours w/o (Note + Rev.), Ours w/o Note, and the full model.
}
    \label{fig:merlin-5}
\end{figure}

\begin{figure}[!htbp]
    \centering
    \includegraphics[width=\textwidth]{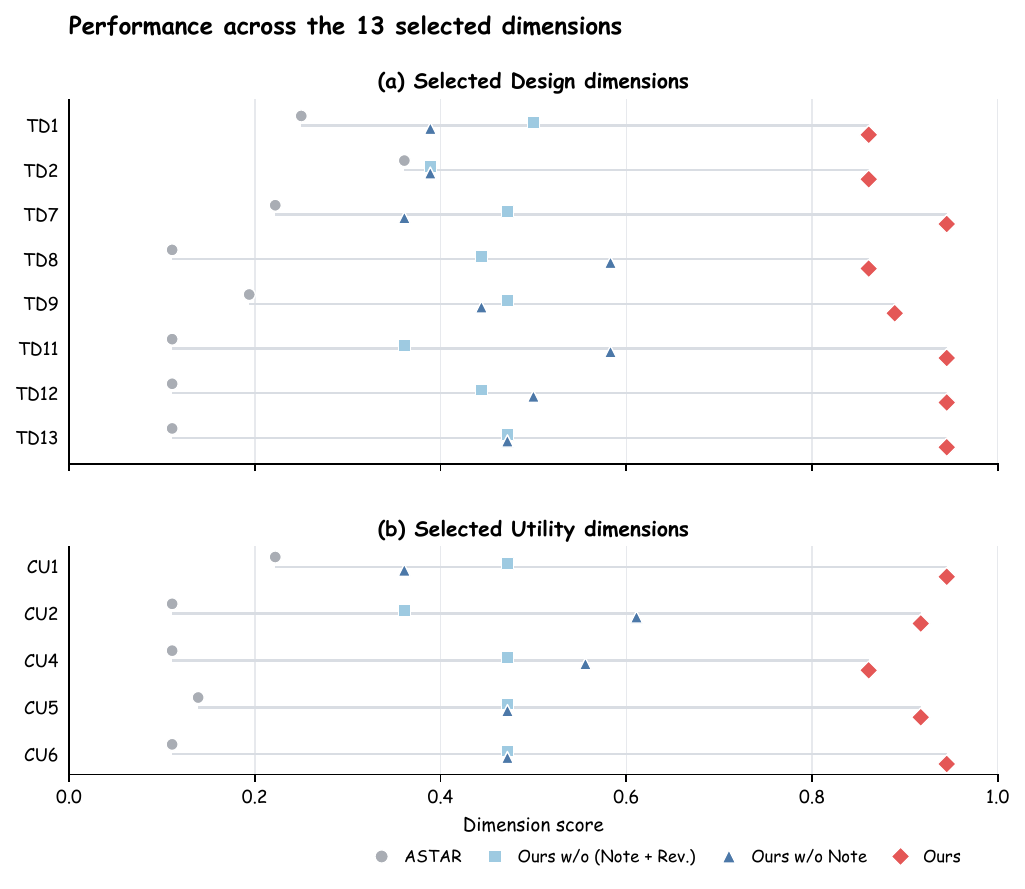}
\caption{
\textbf{Dimension-wise ablation analysis on the BDMAP dataset.}
Results are reported for (a) eight selected Design Quality dimensions and (b) five selected Clinical Utility dimensions, comparing ASTAR, Ours w/o (Note + Rev.), Ours w/o Note, and the full model. Higher scores indicate better performance on the corresponding dimension.
}
    \label{fig:bdmap-4}
\end{figure}

\begin{figure}[!htbp]
    \centering
    \includegraphics[width=\textwidth]{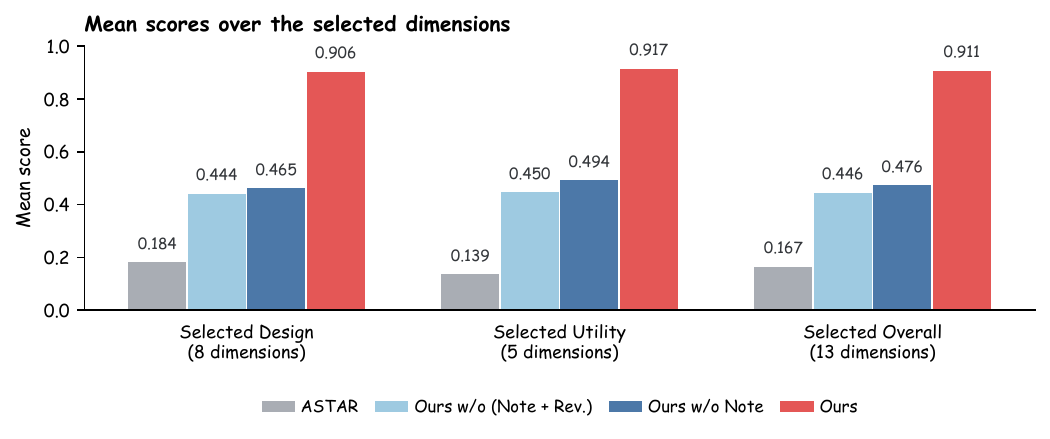}
\caption{
\textbf{Aggregated ablation results on the BDMAP dataset.}
Mean scores are reported over the eight selected Design Quality dimensions, five selected Clinical Utility dimensions, and all 13 selected dimensions for ASTAR, Ours w/o (Note + Rev.), Ours w/o Note, and the full model.
}
    \label{fig:bdmap-5}
\end{figure}

\textbf{Results.}
Across all four datasets, Revision and Note provide consistent cumulative
improvements over the corresponding ablated variants
(Figs.~\ref{fig:petct-4}--\ref{fig:bdmap-5}). Adding Revision increases the
Selected Overall score from 0.408 to 0.504 on PETCT, from 0.383 to 0.487 on
ViMed, and from 0.355 to 0.609 on MERLIN, with a smaller increase from 0.446
to 0.476 on BDMAP. Incorporating Note yields a further and more pronounced
improvement, raising the corresponding scores to 0.844, 0.844, 0.863, and
0.911, respectively. The dimension-wise results exhibit the same general
pattern, with the full model consistently receiving stronger preferences
across the selected Design Quality and Clinical Utility criteria. Overall,
these results support complementary roles for Revision and Note, with Note
contributing particularly strongly to the dimensions associated with
clinically grounded template guidance and downstream usability.

\FloatBarrier


\FloatBarrier

\subsection{Detailed Template Quality Results}
\label{app:detailed-template-quality-results}

\begin{figure}[h]
    \centering
    \includegraphics[width=\textwidth]{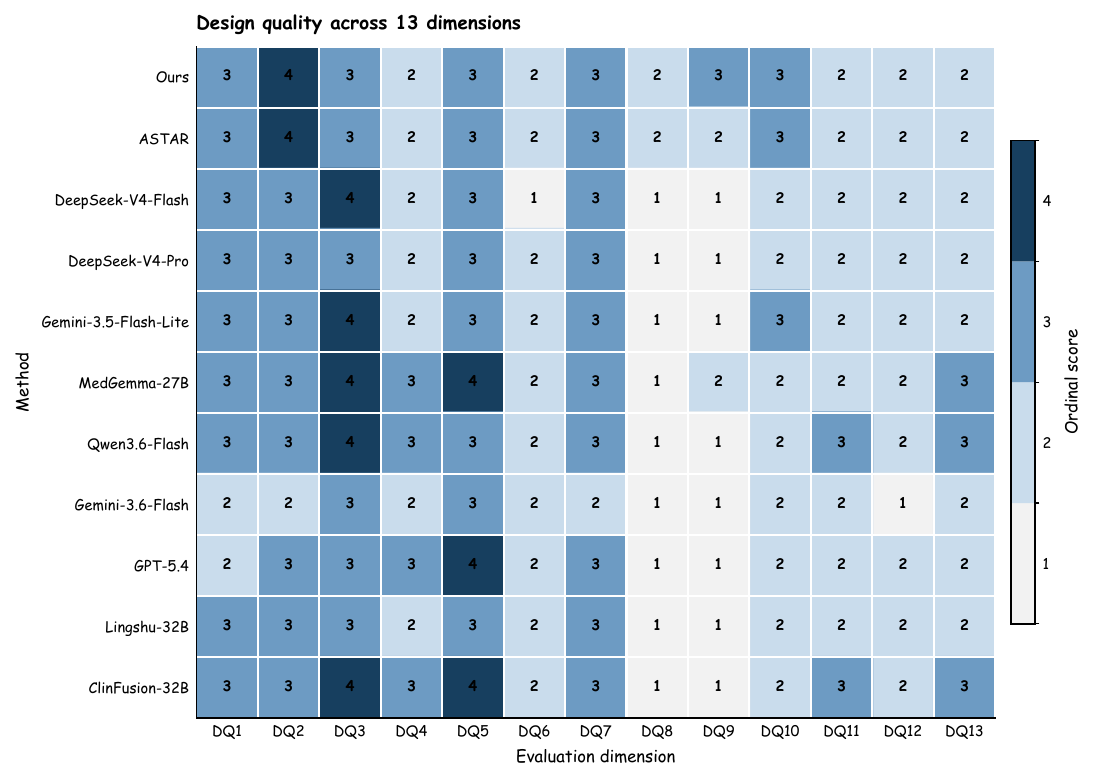}
    \caption{
    \textbf{Design Quality evaluation on the PETCT dataset.}
    LLM-as-judge scores of our method and baselines across the 13 design quality dimensions
    (DQ1--DQ13; Table~\ref{tab:template-design-quality}) of Template Quality.
    Darker colors indicate higher ordinal scores.
    }
    \label{fig:petct-1}
\end{figure}

\begin{figure}[h]
    \centering
    \includegraphics[width=\textwidth]{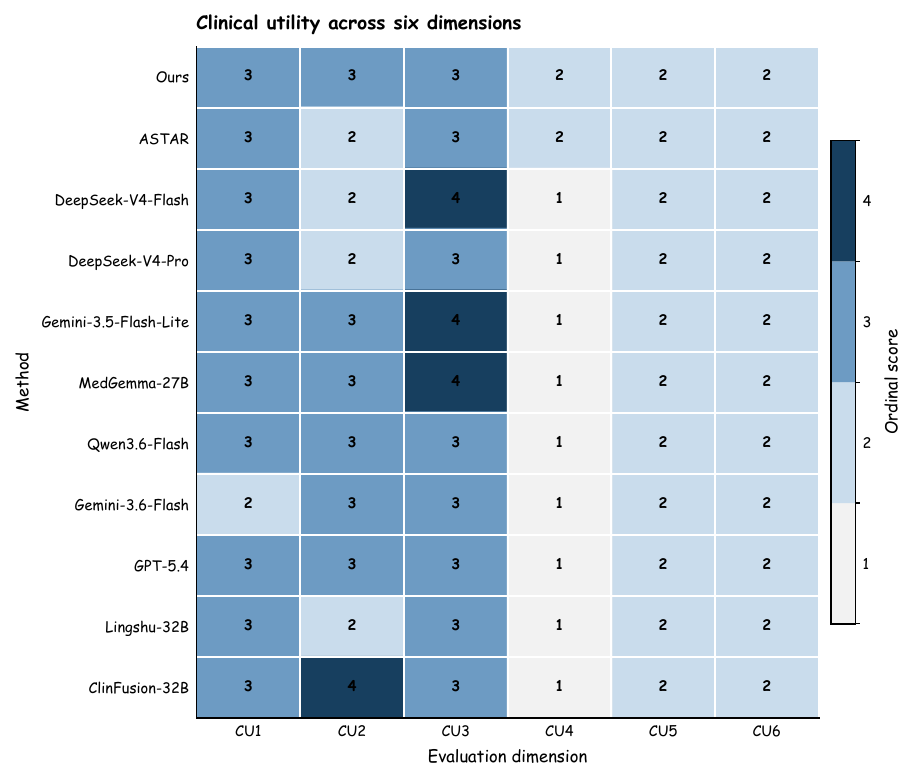}
    \caption{
\textbf{Clinical Utility evaluation on the PETCT dataset.} LLM-as-judge scores of our method and baselines across the 6 clinical utility dimensions (CU1–CU6; Table~\ref{tab:clinical-utility}) of Template Quality. Darker colors indicate higher ordinal scores.
    }
    \label{fig:petct-2}
\end{figure}

\begin{figure}[h]
    \centering
    \includegraphics[width=\textwidth]{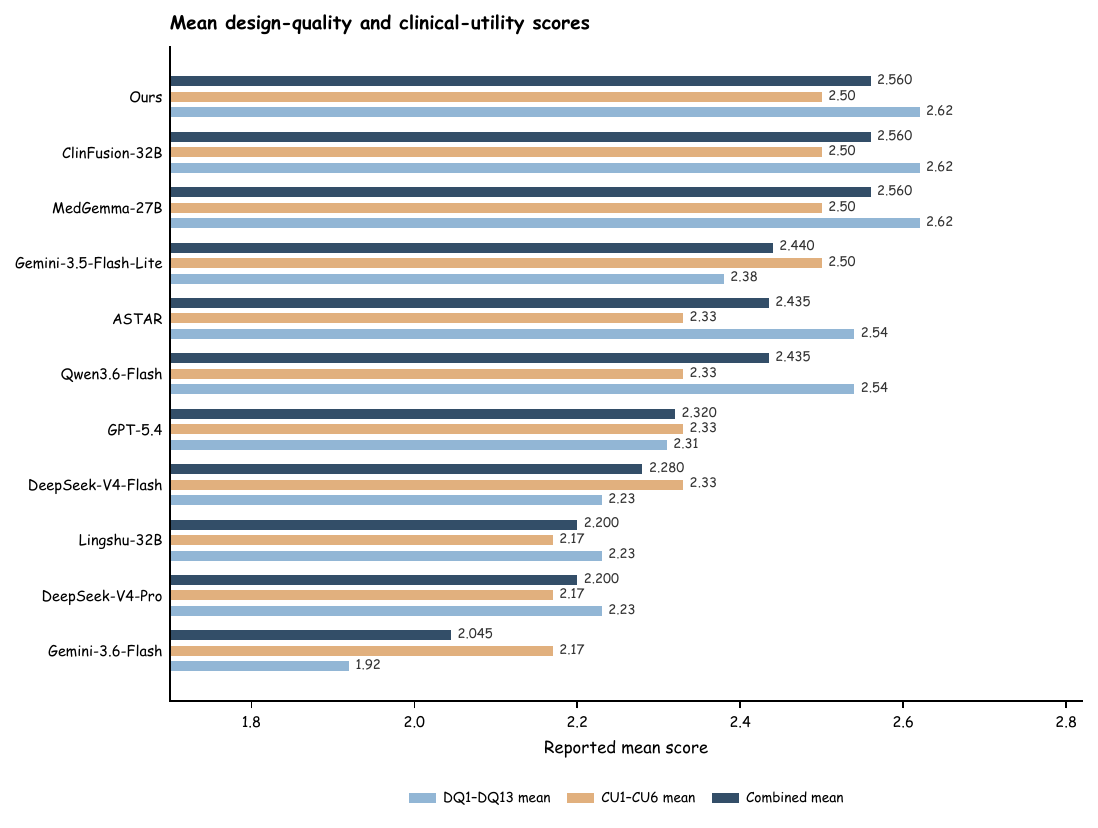}
\caption{
\textbf{LLM-as-judge Template Quality evaluation on the PETCT dataset.}
The figure reports the mean Design Quality score, mean Clinical Utility score, and their combined Template Quality score for our method and baselines. The combined score is computed as the average of the two aspect-level mean scores.
}
    \label{fig:petct-3}
\end{figure}

\begin{figure}[h]
    \centering
    \includegraphics[width=\textwidth]{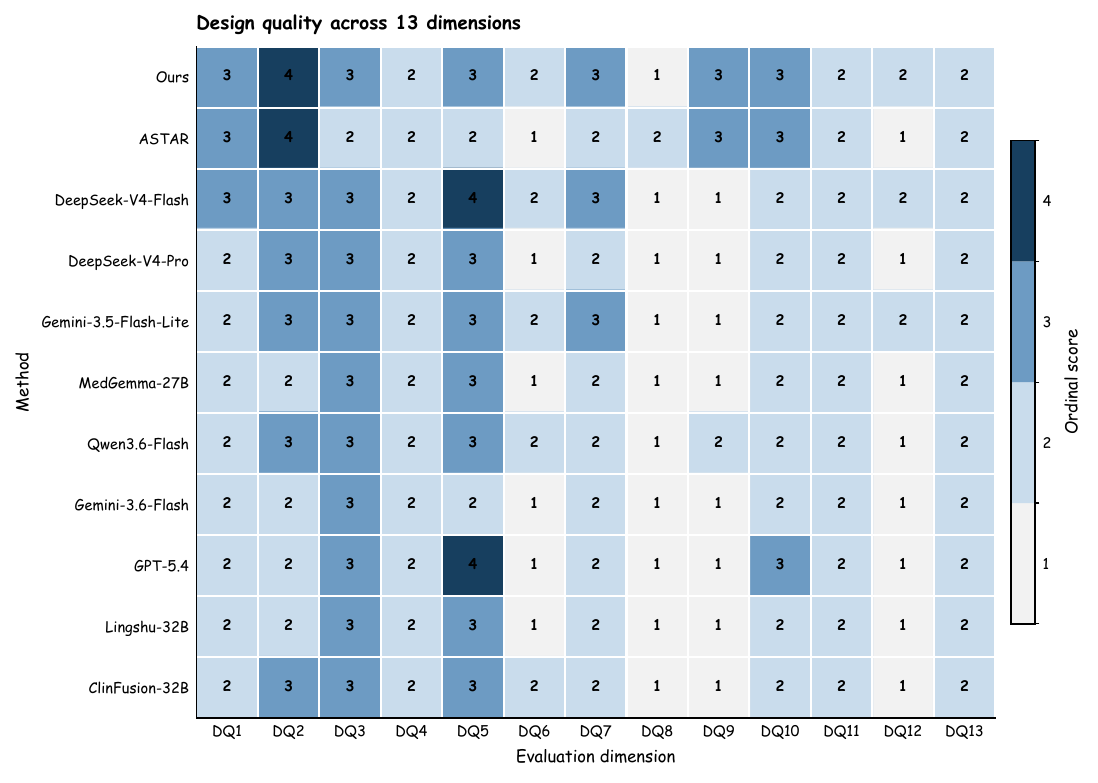}
    \caption{
    \textbf{Design Quality evaluation on the ViMed dataset.}
    LLM-as-judge scores of our method and baselines across the 13 design quality dimensions
    (DQ1--DQ13; Table~\ref{tab:template-design-quality}) of Template Quality.
    Darker colors indicate higher ordinal scores.
    }
    \label{fig:vimed-1}
\end{figure}

\begin{figure}[h]
    \centering
    \includegraphics[width=\textwidth]{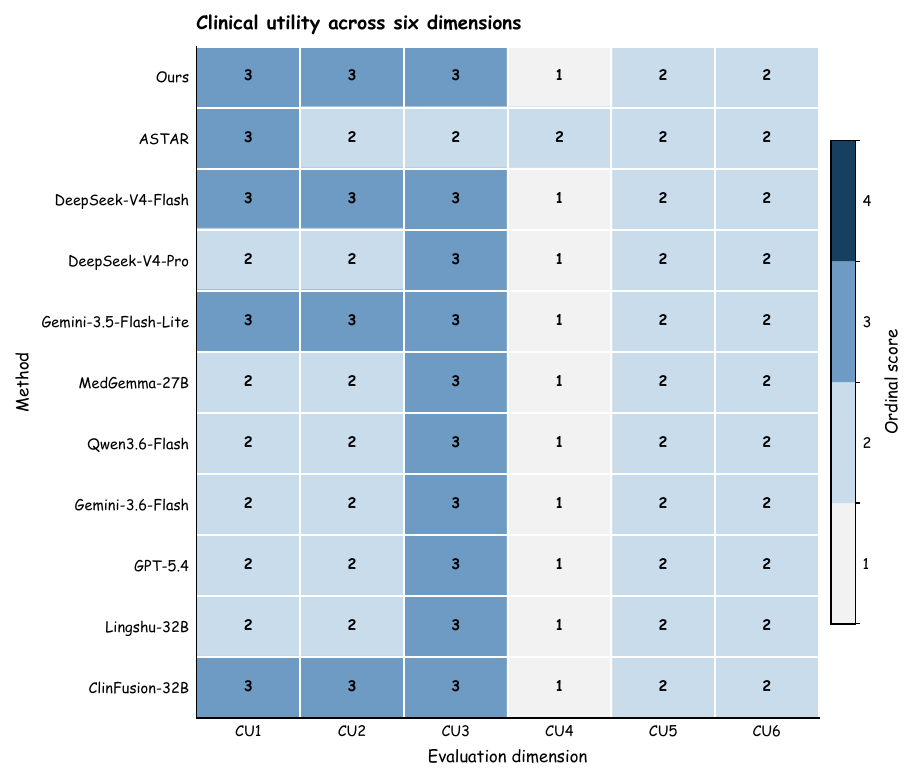}
    \caption{
\textbf{Clinical Utility evaluation on the ViMed dataset.} LLM-as-judge scores of our method and baselines across the 6 clinical utility dimensions (CU1–CU6; Table~\ref{tab:clinical-utility}) of Template Quality. Darker colors indicate higher ordinal scores.
    }
    \label{fig:vimed-2}
\end{figure}

\begin{figure}[h]
    \centering
    \includegraphics[width=\textwidth]{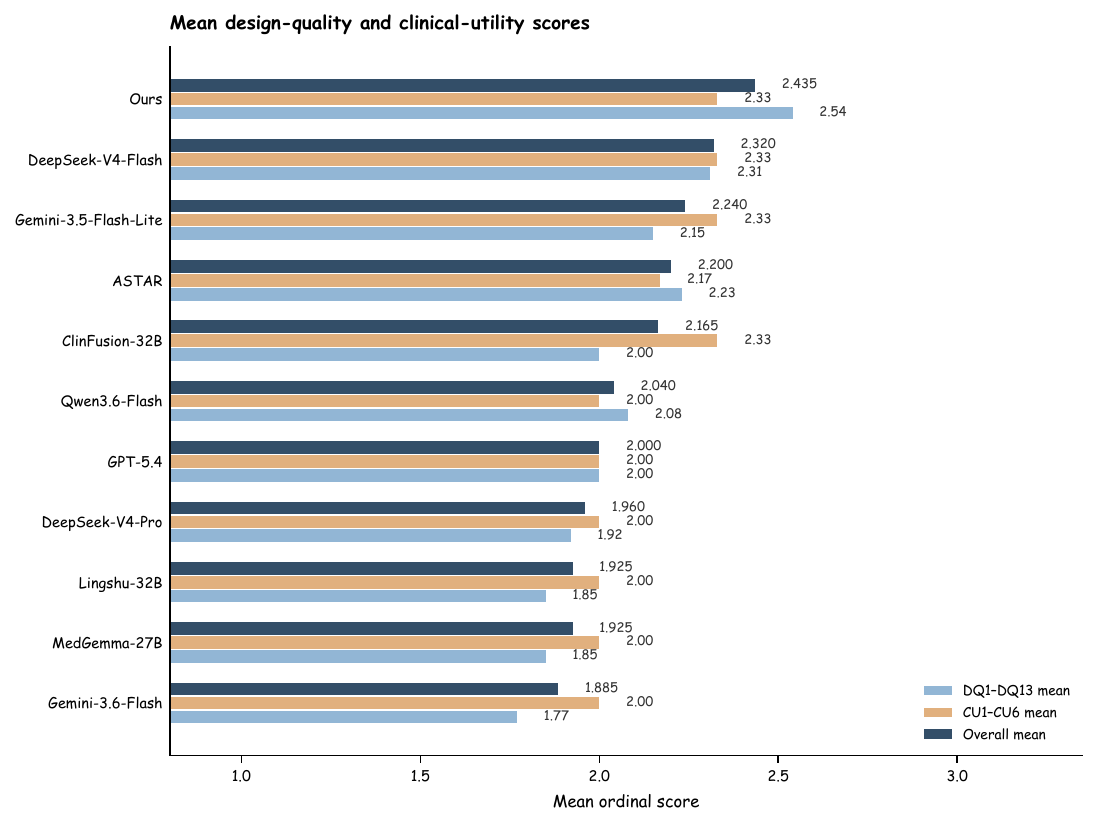}
\caption{
\textbf{LLM-as-judge Template Quality evaluation on the ViMed dataset.}
The figure reports the mean Design Quality score, mean Clinical Utility score, and their combined Template Quality score for our method and baselines. The combined score is computed as the average of the two aspect-level mean scores.
}
    \label{fig:vimed-3}
\end{figure}

\begin{figure}[h]
    \centering
    \includegraphics[width=\textwidth]{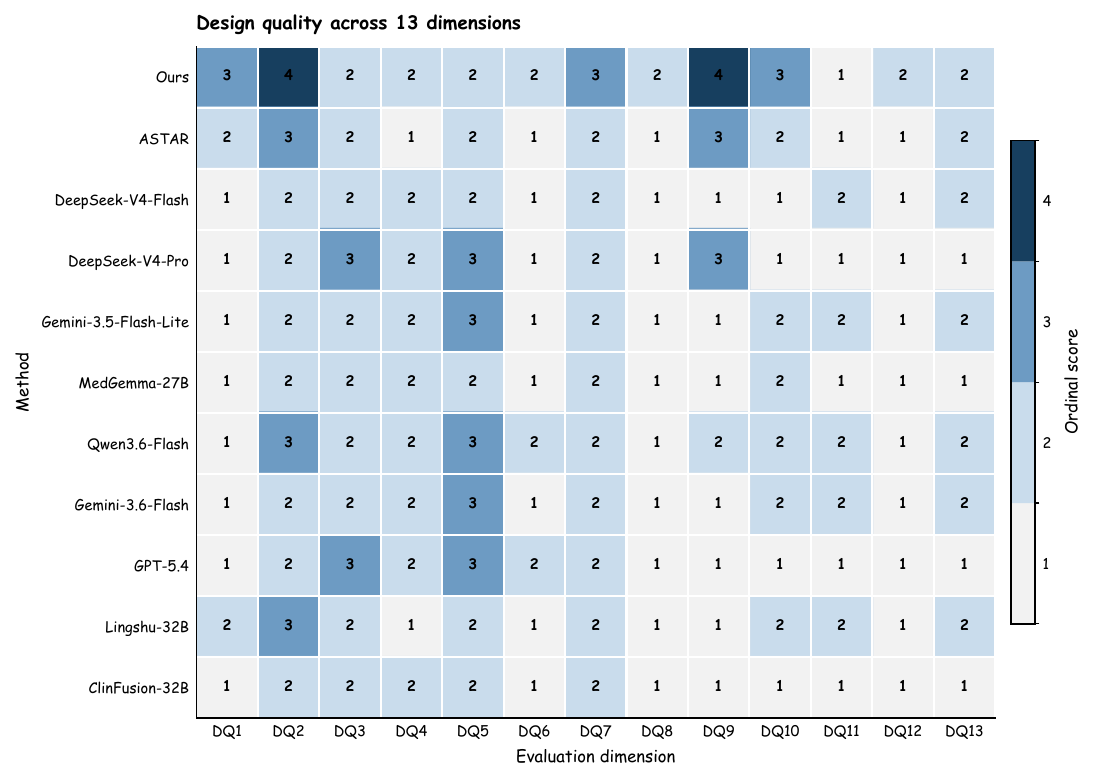}
    \caption{
    \textbf{Design Quality evaluation on the MERLIN dataset.}
    LLM-as-judge scores of our method and baselines across the 13 design quality dimensions
    (DQ1--DQ13; Table~\ref{tab:template-design-quality}) of Template Quality.
    Darker colors indicate higher ordinal scores.
    }
    \label{fig:merlin-1}
\end{figure}

\begin{figure}[h]
    \centering
    \includegraphics[width=\textwidth]{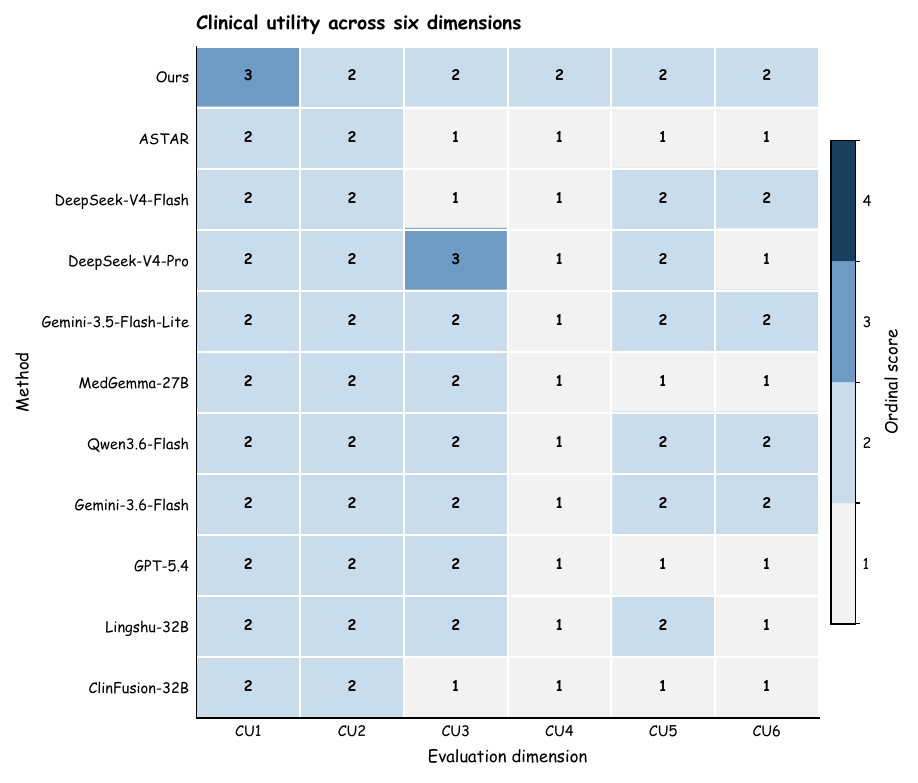}
    \caption{
\textbf{Clinical Utility evaluation on the MERLIN dataset.} LLM-as-judge scores of our method and baselines across the 6 clinical utility dimensions (CU1–CU6; Table~\ref{tab:clinical-utility}) of Template Quality. Darker colors indicate higher ordinal scores.
    }
    \label{fig:merlin-2}
\end{figure}

\begin{figure}[h]
    \centering
    \includegraphics[width=\textwidth]{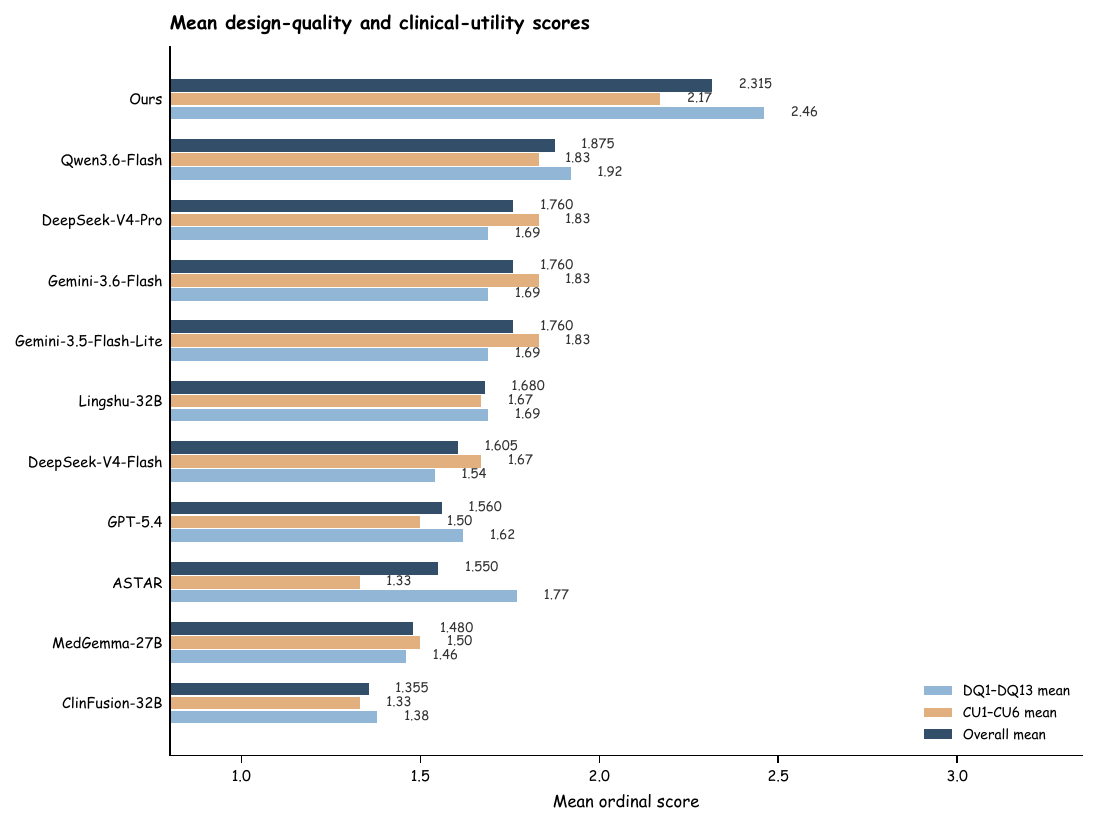}
\caption{
\textbf{LLM-as-judge Template Quality evaluation on the MERLIN dataset.}
The figure reports the mean Design Quality score, mean Clinical Utility score, and their combined Template Quality score for our method and baselines. The combined score is computed as the average of the two aspect-level mean scores.
}
    \label{fig:merlin-3}
\end{figure}

\begin{figure}[h]
    \centering
    \includegraphics[width=\textwidth]{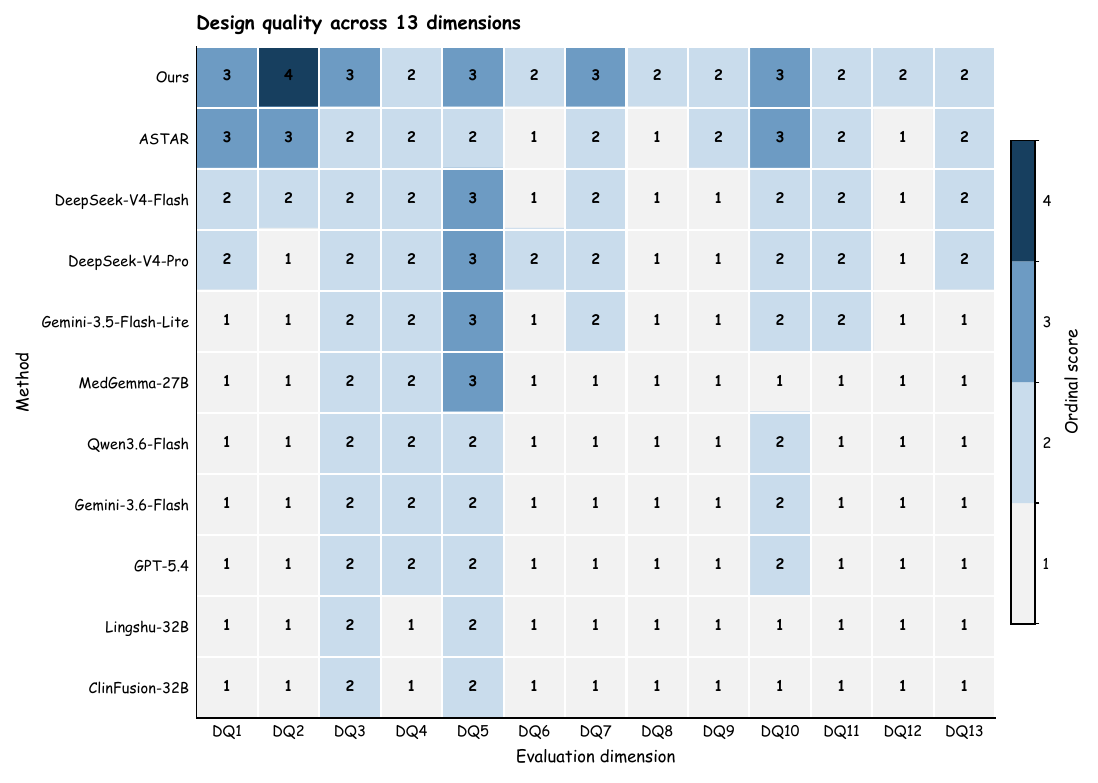}
    \caption{
    \textbf{Design Quality evaluation on the BDMAP dataset.}
    LLM-as-judge scores of our method and baselines across the 13 design quality dimensions
    (DQ1--DQ13; Table~\ref{tab:template-design-quality}) of Template Quality.
    Darker colors indicate higher ordinal scores.
    }
    \label{fig:bdmap-1}
\end{figure}

\begin{figure}[h]
    \centering
    \includegraphics[width=\textwidth]{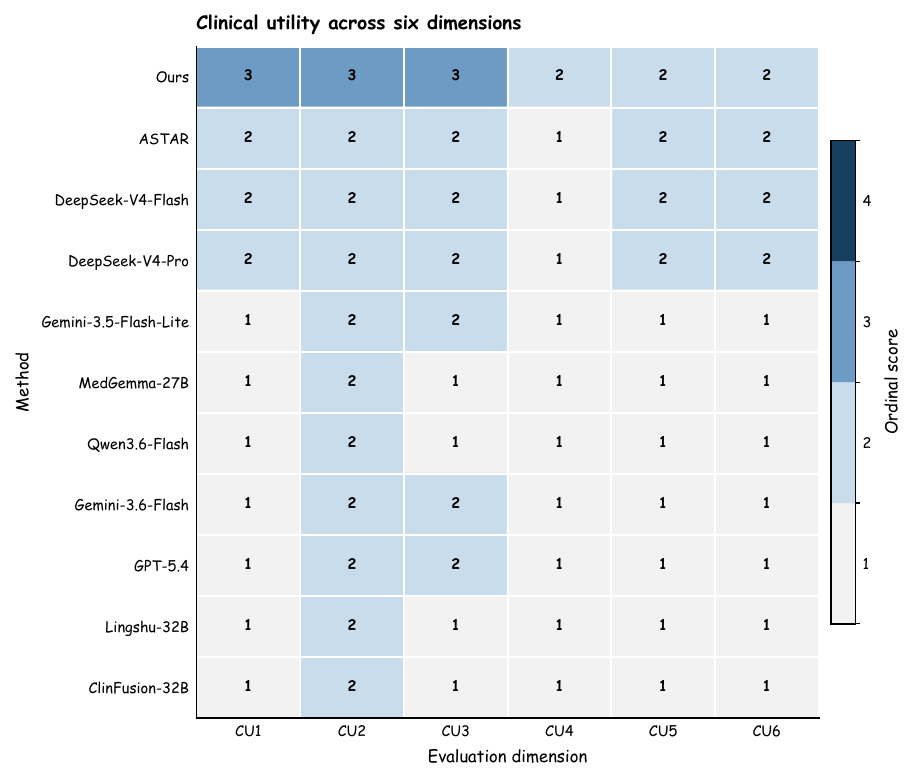}
    \caption{
    \textbf{Clinical Utility evaluation on the BDMAP dataset.} LLM-as-judge scores of our method and baselines across the 6 clinical utility dimensions (CU1–CU6; Table~\ref{tab:clinical-utility}) of Template Quality. Darker colors indicate higher ordinal scores.
    }
    \label{fig:bdmap-2}
\end{figure}

\begin{figure}[h]
    \centering
    \includegraphics[width=\textwidth]{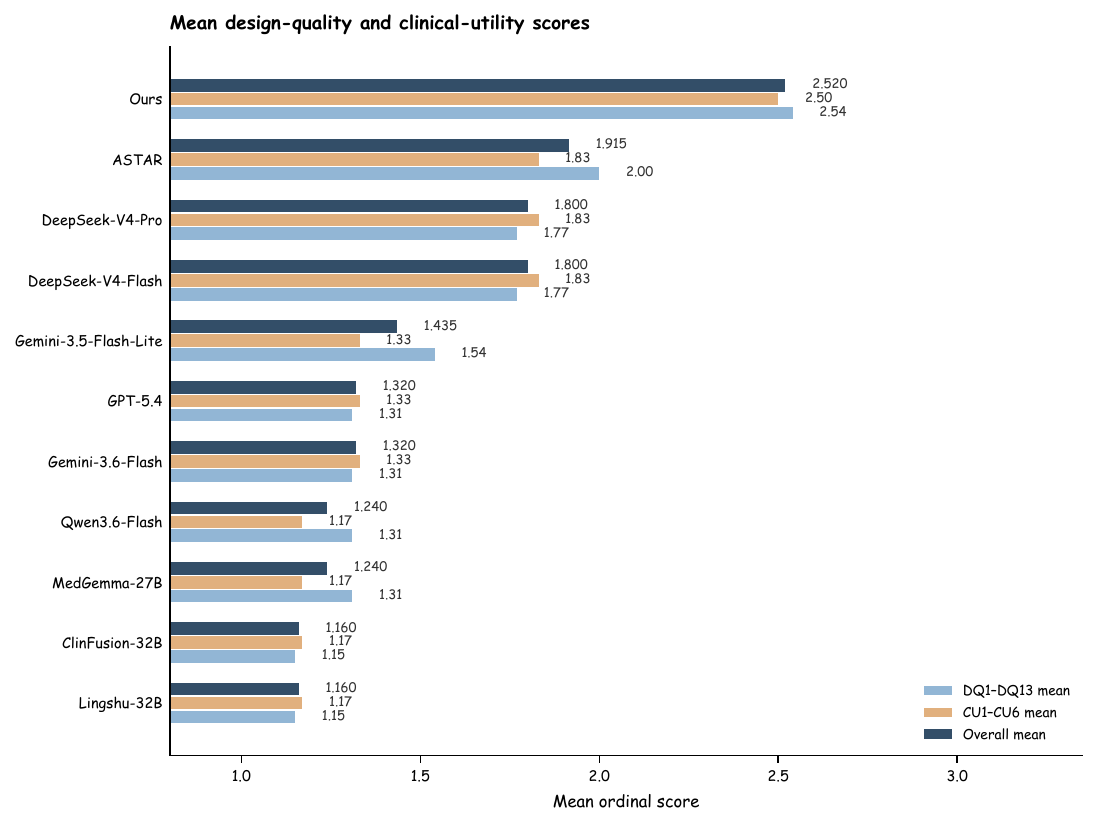}
\caption{
\textbf{LLM-as-judge Template Quality evaluation on the BDMAP dataset.}
The figure reports the mean Design Quality score, mean Clinical Utility score, and their combined Template Quality score for our method and baselines. The combined score is computed as the average of the two aspect-level mean scores.
}
    \label{fig:bdmap-3}
\end{figure}

The overall template-quality results compare our method and all baselines across the four datasets under both expert and LLM evaluation (Fig.~\ref{fig:template-quality-llm}). Beyond the aggregate template-quality scores reported in the main figure, this appendix presents dimension-level design-quality and clinical-utility ratings and their aggregates for PETCT (Figs.~\ref{fig:petct-1}--\ref{fig:petct-3}), ViMed (Figs.~\ref{fig:vimed-1}--\ref{fig:vimed-3}), MERLIN (Figs.~\ref{fig:merlin-1}--\ref{fig:merlin-3}), and BDMAP (Figs.~\ref{fig:bdmap-1}--\ref{fig:bdmap-3}).


\FloatBarrier

\subsection{Failure Analysis of Additional Baselines}
\label{app:failure-analysis}

Beyond the baseline results reported in the main text, we evaluated several additional models on the same template-induction task. These baselines were not included in the main comparison because they failed during template construction or the subsequent merge stage. The observed failures arose from insufficient context capacity, unstable long-form generation, malformed structured outputs, and repetition degeneration. We report these results to document the scope of the comparison and to distinguish recoverable formatting errors from cases in which no usable template was generated.

\subsubsection{Additional baseline models}
\label{app:baseline-screening}

\textbf{Doubao Seed series.}
We first removed endpoints intended for embedding, image, video, code, or role-playing applications, as well as models with a 32k-token context window. In the screening workload, the largest induction prompt contained 26,526 tokens and required approximately 13k additional tokens for the generated template, already exceeding a 32k-token window. We screened the remaining long-context candidates using a real induction prompt (Table~\ref{tab:doubao-screening}). The requests did not specify \texttt{max\_tokens}. Nevertheless, four models terminated with \texttt{finish\_reason=length} after producing between 11,541 and 21,623 output tokens, which indicates that an output-length limit was reached; the observed finish reason alone does not distinguish a service default budget from a model limit. The two Doubao 2.1 endpoints did not return within an acceptable time budget, despite responding normally to a trivial JSON request.

\begin{table*}[t]
    \centering
    \small
    \caption{Feasibility screening of the Doubao Seed series on a real induction prompt. No request-level output-token limit was specified.}
    \label{tab:doubao-screening}
    \begin{tabular}{lrrl}
        \toprule
        Model & Finish reason & Output tokens & Outcome \\
        \midrule
        \texttt{seed-2.0-pro}   & length & 21,623 & truncated JSON \\
        \texttt{seed-1.8}       & length & 18,836 & truncated JSON \\
        \texttt{seed-2.0-mini}  & length & 18,634 & truncated JSON \\
        \texttt{seed-2.0-lite}  & length & 11,541 & truncated JSON \\
        \texttt{seed-1.6}       & stop   &  6,982 & valid JSON in screening \\
        \texttt{seed-2.1-pro}   & --     & --      & no response after approximately 40 min \\
        \texttt{seed-2.1-turbo} & --     & --      & timeout at 180 s and 1,800 s \\
        \bottomrule
    \end{tabular}
\end{table*}

The apparently successful \texttt{seed-1.6} result was not stable under repeated evaluation. With an identical prompt and temperature set to zero, two consecutive calls produced 10,099 and 10,098 output tokens, respectively. The first completed normally, whereas the second terminated with \texttt{finish\_reason=length} and five unmatched braces. In the full induction run, only 6 of 22 batches (27\%) succeeded without retrying. Seventeen retry rounds were required to obtain outputs for all 22 batches, after which the merge call was still truncated at 14,514 output tokens. By comparison, successful merge outputs from the other models contained approximately 24k--44k tokens. We therefore exclude the Doubao models from the primary comparison because they cannot reliably execute the fixed protocol, rather than because of their downstream template quality.

\textbf{Local medical models.}
For local models, the configured context window provides an initial screening criterion. We measured peak prompt and template lengths for each dataset to characterize context demand (Table~\ref{tab:context-demand}). The sums of the separately reported input and output peaks range from 38,204 to 43,624 tokens; these sums are screening estimates rather than measurements of a single request. \texttt{HuatuoGPT-Vision-34B} has a 4,096-token window without RoPE scaling and cannot accommodate even the prompt skeleton plus one report and the required output budget. \texttt{Hulu-Med-32B} has an effective 32,768-token window. Although its configuration contains \texttt{sliding\_window=131072}, sliding-window attention is disabled. The observed merge workload also included PET/CT inputs exceeding 63k tokens and a MERLIN output of 65,067 tokens. Both models were excluded from the main comparison under the evaluated protocol. These observations do not establish infeasibility for every alternative batching or hierarchical-merging schedule.

\begin{table}[t]
    \centering
    \small
    \caption{Observed peak input and output lengths for induction, measured with the Qwen tokenizer on completed runs. The final column sums the two peaks, which need not occur in the same request.}
    \label{tab:context-demand}
    \begin{tabular}{lrrr}
        \toprule
        Dataset & Peak input & Peak output & Sum of peaks \\
        \midrule
        PET/CT  & 24,872 & 13,332 & 38,204 \\
        ViMed   & 29,387 &  8,964 & 38,351 \\
        BDMAP   & 39,461 &  3,558 & 43,019 \\
        MERLIN  & 30,039 & 13,585 & 43,624 \\
        \bottomrule
    \end{tabular}
\end{table}

The remaining long-context candidates were \texttt{Lingshu-32B} (128k), \texttt{MedGemma-27B-IT} (131,072), \texttt{Hulu-Med-30A3} (262,144), and \texttt{ClinFusion-32B} (262,144). \texttt{Hulu-Med-30A3} was subsequently excluded from the primary comparison. In two of its 22 raw outputs, the model remained inside an unclosed \texttt{<think>} block, entered a repetitive loop, and terminated without producing a single opening brace. This behavior is a semantic generation failure: no template object exists for a parser to recover.

\subsubsection{Failure taxonomy for structured outputs}
\label{app:json-failures}

We analyzed 409 raw outputs from the local-model runs. The strict extraction path, which removes Markdown fences and reasoning blocks before applying ordinary JSON parsing, accepted 294 outputs (71.9\%). The remaining 115 outputs (28.1\%) failed strict parsing. Parser exceptions alone were not a reliable description of the underlying failure. For example, a restart that occurs in the middle of a string is reported as an invalid control character, although the relevant event is that the model abandoned one JSON attempt and began another. We therefore classify failures by their generated structure rather than by
the exception message. Truncation accounts for the majority of strict-parsing failures in this structural taxonomy (Table~\ref{tab:failure-taxonomy}).

\begin{table*}[t]
    \centering
    \small
    \caption{Structural causes of the 115 outputs rejected by strict JSON parsing. Bracket accounting ignores delimiters that occur inside strings.}
    \label{tab:failure-taxonomy}
    \begin{tabular}{lrl}
        \toprule
        Cause & Count & Structural signature \\
        \midrule
        Truncation & 78 & 1--59 missing \texttt{\}} and up to 14 missing \texttt{]} \\
        Generation restart & 17 & a second JSON fence or a second top-level \texttt{desc} \\
        Excess closing delimiter & 17 & an otherwise complete object followed by extra delimiters \\
        Balanced but invalid syntax & 3 & e.g., an unambiguous trailing comma \\
        \bottomrule
    \end{tabular}
\end{table*}

\subsubsection{Fallback recovery}
\label{app:fallback-parser}

When strict JSON parsing fails, we apply a conservative fallback that retains only complete structures already present in the output. Depending on the failure type, it removes an incomplete suffix, discards extra closing delimiters, or separates multiple JSON attempts produced by a generation restart. The recovered object must still satisfy the template schema, and the procedure never introduces new slots, keys, or values. Outputs without an extractable template remain failures.

Fallback recovery prevented 12 call-level failures, all in local-model runs, while the evaluated hosted-API outputs passed strict parsing. Without this step, a malformed batch output containing a recoverable template could be discarded before merging. The fallback therefore prevents usable outputs from being discarded because of local formatting errors without changing their semantic content.

\subsubsection{Repetition degeneration and canonicalization}
\label{app:degeneration}

Structural recovery exposes a second failure mode: some models emit long sequences of repeated but individually well-formed components before truncation. In one \texttt{MedGemma-27B-IT} output, fallback recovered a slot whose candidate-value list $\Omega$ contained 1,719 entries but only six unique values. One 130-character value was repeated 1,714 times. Removing exact duplicates reduced this slot from 226,751 to 635 characters without deleting any unique value.

Repetition was strongly concentrated in outputs that required fallback. The duplicate rate within $\Omega$ was 5.0\% for strictly parsed local outputs and 80.7\% for fallback-recovered outputs. This association suggests a common generation process: repetition expands the sequence, the expanded sequence reaches an output limit, and the resulting truncation leaves an invalid JSON suffix. The association does not by itself establish that repetition caused every truncation, but it explains why malformed outputs contain substantially more duplicated content than directly parsed outputs.

We observed two distinct forms of degeneration. \texttt{MedGemma-27B-IT}, \texttt{Lingshu-32B}, and \texttt{Hulu-Med-30A3} primarily repeated values within the same $\Omega$ list. \texttt{ClinFusion-32B}, in contrast, repeated entire slots. A representative \texttt{ClinFusion-32B} output contained 509 recovered slots but only 32 unique slots, while each individual $\Omega$ list contained two non-duplicated values. We therefore canonicalize recovered templates at two levels: exact duplicate values are removed within each $\Omega$, and slots with the same canonical key are merged while preserving the union of their unique values. Across 114 recovered templates, value-level deduplication removed 80.7\% of the recovered characters, and slot-level merging increased the total reduction to 84.2\%. The operation is content-preserving with respect to the set of unique generated keys and values.

Together, these analyses explain why the additional baselines were absent from the main results and how malformed local-model outputs were handled. Models that could not complete template induction and merging were excluded, recoverable formatting errors were repaired without adding semantic content, and outputs containing no usable template remained explicit generation failures.


\providecommand{\methodname}{\textsc{TEMPLAR}}
\section{Representative Examples from Final Templates}
\label{app:template_examples}

The final templates are hierarchical clinical specifications
organized by anatomical paths. We present 50 representative examples
from the final templates constructed for PETCT, ViMed, MERLIN, and BDMAP (Table~\ref{tab:template-example-distribution}).
These examples cover organ status, lesion morphology, metabolic
activity, quantitative measurements, temporal change, post-surgical
findings, and anatomical relationships. Each example presents its
anatomical path, canonical field key $k$, value type $\tau$, allowed
values $\Omega$, and clinical note. All fields are taken directly from
the corresponding final template (Tables~\ref{tab:template-examples-petct}--\ref{tab:template-examples-bdmap}).

\begin{table}[t]
\centering
\small
\begin{tabular}{lrrr}
\toprule
Dataset & Examples & Categorical & Numerical \\
\midrule
PETCT & 12 & 11 & 1 \\
ViMed & 12 & 11 & 1 \\
MERLIN & 14 & 13 & 1 \\
BDMAP & 12 & 8 & 4 \\
\midrule
Total & 50 & 43 & 7 \\
\bottomrule
\end{tabular}
\caption{Distribution of representative examples from the final templates presented in this appendix.}
\label{tab:template-example-distribution}
\end{table}

\subsection{PETCT}
\label{app:template-examples-petct}

The PETCT examples span the brain, thyroid, lungs, mediastinal lymph nodes, liver, kidneys, female pelvis, and skeletal system. They show how morphology and FDG activity are represented by separate but anatomically organized template fields.

\begin{slotcatalog}{PETCT}{PETCT template specification}
\slotentry
  {head\_and\_neck/brain}
  {brain\_fdg\_metabolism\_status}
  {categorical}
  {[normal, increased, decreased, absent, asymmetric, focal\_hypermetabolism]}
  {Normal: symmetrical cortical \& subcortical uptake; abnormal: focal hypermetabolism (tumor/inflammation), global hypometabolism (neurodegeneration), asymmetry, or absent uptake.}
\slotentry
  {head\_and\_neck/thyroid}
  {thyroid\_overall\_status}
  {categorical}
  {[normal, enlarged, atrophic, small, heterogeneous, irregular, asymmetric, nodular, diffuse, residual]}
  {Normal: homogeneous, smooth, SUVmax \textless{} liver; abnormal: focal/diffuse uptake \textgreater{} liver, enlarged, heterogeneous, nodular, or irregular margins.}
\slotentry
  {chest\_and\_thorax/lungs}
  {lung\_fdg\_metabolism}
  {categorical}
  {[normal, increased, decreased, focal, diffuse, abnormal]}
  {Normal: homogeneous, low-grade FDG uptake (SUVmax \textless{}2.5). Focal/diffuse/increased uptake suggests inflammation, infection, or malignancy.}
\slotentry
  {chest\_and\_thorax/lungs}
  {lung\_fdg\_suvmax}
  {numerical}
  {[\textless{}num\textgreater{}]}
  {Normal lung parenchyma SUVmax \textasciitilde{}0.5-1.0; suspicious lesion usually \textgreater{}2.5, but inflammatory uptake can overlap. Use visual comparison with liver blood pool.}
\slotentry
  {chest\_and\_thorax/lungs/lung\_lesion\_margin}
  {lung\_lesion\_margin}
  {categorical}
  {[clear, indistinct, spiculated, lobulated, smooth, irregular]}
  {Smooth/lobulated margins often benign; spiculated/irregular margins suggest malignancy. Indistinct margin may indicate inflammation or infiltration.}
\slotentry
  {chest\_and\_thorax/lungs/nodules}
  {nodule\_type}
  {categorical}
  {[solid, part\_solid, ground\_glass, calcified, plaque]}
  {Solid: homogeneous opacity; part-solid: ground-glass + solid component; ground-glass: hazy, no solid; calcified: benign. Compare with prior CT.}
\slotentry
  {chest\_and\_thorax/mediastinum\_and\_hilar\_lymph\_nodes}
  {hilar\_mediastinal\_lymph\_node\_status}
  {categorical}
  {[normal, enlarged, absent, multiple]}
  {Normal: no FDG-avid hilar/mediastinal nodes above background. Abnormal: focal uptake \textgreater{}= mediastinal blood pool, with or without enlargement (\textgreater{}=10mm short axis).}
\slotentry
  {abdomen\_and\_pelvis/liver}
  {liver\_focal\_lesions}
  {categorical}
  {[absent, present, single, multiple, cystic, solid, mixed, low\_density, high\_density]}
  {Normal: no focal uptake \textgreater{} adjacent liver background. Solid/mixed lesions with SUVmax \textgreater{}3.0 or \textgreater{}liver suggest malignancy. Cysts: typically no uptake.}
\slotentry
  {abdomen\_and\_pelvis/liver}
  {liver\_fdg\_metabolism}
  {categorical}
  {[normal, increased, decreased, focal, diffuse, absent, uneven]}
  {Normal liver FDG uptake is homogeneous and similar to blood pool; suspicious if focal/diffuse increased uptake exceeds blood pool or is heterogeneous.}
\slotentry
  {abdomen\_and\_pelvis/kidneys\_and\_urinary\_tract}
  {kidney\_morphology\_and\_size}
  {categorical}
  {[normal, enlarged, atrophic, irregular, asymmetric]}
  {Normal: smooth contour, \textless{}12 cm length; enlarged \textgreater{}12 cm (e.g., obstruction, mass); atrophic \textless{}8 cm (e.g., scarring); irregular/asymmetric suggests focal pathology.}
\slotentry
  {abdomen\_and\_pelvis/uterus\_and\_cervix}
  {uterus\_morphology\_size\_and\_contour}
  {categorical}
  {[normal, enlarged, atrophic, deformed, irregular, nodular, absent, not visualized]}
  {Normal: anteverted, smooth contour, homogeneous; enlarged: \textgreater{}10 cm length or \textgreater{}5 cm AP diameter; abnormal: irregular contour, nodularity, or focal deformity suggests leiomyoma or malignancy.}
\slotentry
  {musculoskeletal\_and\_spine/systemic\_bones}
  {skeletal\_fdg\_metabolism\_status}
  {categorical}
  {[normal, abnormal, increased, decreased, absent]}
  {Normal: homogeneous low-grade uptake \textasciitilde{}SUVmax 1.5-2.5 (\textless{} liver); abnormal: focal or diffuse increased uptake, often SUVmax \textgreater{} liver, suggesting malignancy or inflammation.}
\end{slotcatalog}
\captionof{table}{\textbf{Representative examples from the final PETCT template.}}
\label{tab:template-examples-petct}

\subsection{ViMed}
\label{app:template-examples-vimed}

The ViMed examples emphasize physiological versus suspicious FDG patterns, nodal morphology and station labels, organ-level uptake, quantitative SUVmax, and post-surgical imaging findings.

\begin{slotcatalog}{ViMed}{ViMed template specification}
\slotentry
  {head\_and\_neck}
  {head\_and\_neck\_fdg\_characteristics}
  {categorical}
  {[absent, physiological, mild, focal, diffuse, asymmetric, increased]}
  {Normal: symmetric, mild FDG uptake in Waldeyer's ring, salivary glands, and muscles; suspicious: focal, asymmetric, or intense uptake (SUVmax \textgreater{} 4-5) suggests malignancy.}
\slotentry
  {head\_and\_neck/oral\_cavity\_and\_oropharynx}
  {tonsil\_fdg\_uptake}
  {categorical}
  {[absent, mild, moderate, severe]}
  {Normal: symmetric, mild-to-moderate FDG uptake (SUVmax typically 2-4). Abnormal: marked asymmetry, focal intense uptake, or SUVmax \textgreater{}5-6, suspicious for malignancy or infection.}
\slotentry
  {head\_and\_neck/cervical\_and\_supraclavicular\_lymph\_nodes}
  {cervical\_nodal\_level}
  {categorical}
  {[iia, iib, iii, iv, v, vi, iia\_iib\_iii]}
  {Normal: no FDG-avid lymph nodes in any cervical level. Suspicious: focal FDG uptake \textgreater{} liver activity, especially in levels II-IV, with irregular margins or size \textgreater{}1 cm short axis.}
\slotentry
  {thorax\_and\_chest/lungs\_and\_pulmonary\_parenchyma}
  {lung\_fdg\_uptake\_status}
  {categorical}
  {[absent, normal, increased, decreased, focal, diffuse, physiological]}
  {Normal: homogeneous, diffuse low-grade uptake (SUVmax \textless{}2.5). Abnormal: focal or increased uptake \textgreater{}2.5 suggests malignancy; diffuse increased uptake may indicate inflammation.}
\slotentry
  {thorax\_and\_chest/axillary\_lymph\_nodes}
  {axillary\_lymph\_node\_fdg\_status}
  {categorical}
  {[absent, normal, mild, increased]}
  {Normal: absent/low-level FDG uptake (similar to blood pool). Abnormal: focal/asymmetric increased uptake (SUVmax \textgreater{}2.5) suggests metastasis or reactive inflammation.}
\slotentry
  {abdomen\_and\_pelvis/liver\_and\_intrahepatic\_biliary\_tree}
  {liver\_fdg\_uptake\_intensity}
  {categorical}
  {[normal, increased, reduced, absent]}
  {Characterizes the overall intensity of hepatic FDG uptake.}

\slotentry
  {abdomen\_and\_pelvis/liver\_and\_intrahepatic\_biliary\_tree}
  {liver\_fdg\_uptake\_pattern}
  {categorical}
  {[homogeneous, heterogeneous, focal, diffuse]}
  {Characterizes the spatial distribution pattern of hepatic FDG uptake.}

\slotentry
  {abdomen\_and\_pelvis/large\_intestine\_and\_rectum}
  {colon\_fdg\_uptake\_status}
  {categorical}
  {[physiological, inflammatory, focal, diffuse, heterogeneous, increased, mildly\_increased]}
  {Normal: diffuse, mild, homogeneous uptake (grade 1-2); suspicious: focal, intense, or segmental uptake exceeding liver activity, or irregular/mass-like.}
\slotentry
  {abdomen\_and\_pelvis/spleen}
  {spleen\_fdg\_uptake}
  {categorical}
  {[physiological, pathological, increased, decreased, focal, diffuse, heterogeneous]}
  {Normal spleen: diffuse, homogeneous uptake \textless{}= liver; pathological: focal, heterogeneous, or diffuse \textgreater{} liver uptake (e.g., lymphoma, metastasis).}
\slotentry
  {lymph\_nodes}
  {lymph\_node\_morphology}
  {categorical}
  {[normal, enlarged, reactive, calcified, necrotic, preserved\_fatty\_hila, enhancing]}
  {Normal: homogeneous, smooth margins, preserved fatty hila, SUVmax \textless{}= mediastinal blood pool; abnormal: focal/heterogeneous uptake, irregular margins, loss of fatty hilum, necrosis, calcification.}
\slotentry
  {suvmax\_measurements}
  {fdg\_suvmax}
  {numerical}
  {[\textless{}num\textgreater{}]}
  {Normal SUVmax varies by organ; for liver \textasciitilde{}2.0-3.0, blood pool \textasciitilde{}1.2-2.0; lesion SUVmax \textgreater{} liver often suggests malignancy.}
\slotentry
  {musculoskeletal\_and\_other/surgical\_sites\_and\_postoperative\_changes}
  {surgical\_site\_imaging\_findings}
  {categorical}
  {[absent, normal, mild, diffuse, diffuse\_increased, increased, heterogeneous\_density, fdg\_avid, stable, new, recurrence, resected, soft\_tissue\_lesion, postoperative\_inflammation]}
  {Post-surgical site: normal = linear/mild SUV uptake (\textless{}= adjacent background); suspicious = increasing, heterogeneous, or FDG-avid focal uptake (SUVmax \textgreater{} liver or time-dependent rise).}
\end{slotcatalog}
\captionof{table}{\textbf{Representative examples from the final ViMed template.}}
\label{tab:template-examples-vimed}

\subsection{MERLIN}
\label{app:template-examples-merlin}

The MERLIN examples cover gastrointestinal caliber and obstruction, hepatobiliary findings, peritoneal abnormalities, renal morphology and function, vascular patency, osseous lesions, post-surgical change, longitudinal comparison, and quantitative measurement.

\begin{slotcatalog}{MERLIN}{MERLIN template specification}
\slotentry
  {gastrointestinal/gastric\_intestinal\_caliber}
  {gastric\_intestinal\_caliber}
  {categorical}
  {[normal, distended, collapsed, narrowed, absent]}
  {Normal: thin-walled, minimally distended; abnormal: marked dilation (\textgreater{}6 cm small bowel, \textgreater{}9 cm colon), wall thickening, or abrupt transition indicating obstruction.}
\slotentry
  {gastrointestinal/small\_bowel}
  {small\_bowel\_dilation\_status}
  {categorical}
  {[normal, dilated, focally\_dilated, collapsed, decompressed, narrow]}
  {Normal caliber is \textless{}=3 cm; dilation \textgreater{}3 cm with proximal distension indicates obstruction, while collapse denotes \textless{}2 cm.}
\slotentry
  {hepatobiliary/bile\_ducts}
  {biliary\_dilation\_severity}
  {categorical}
  {[absent, mild, moderate, severe]}
  {Normal CBD \textless{}=6 mm; mild 7-10 mm, moderate 11-15 mm, severe \textgreater{}15 mm. Intrahepatic ducts normally \textless{}1/3 adjacent portal vein diameter.}
\slotentry
  {hepatobiliary/liver}
  {lesion\_characterization}
  {categorical}
  {[cyst, hemangioma, metastasis, HCC, cholangiocarcinoma, abscess, granuloma\_calcification, neuroendocrine\_tumor, benign, malignant, indeterminate, too\_small\_to\_characterize]}
  {Normal: homogeneous parenchyma. Characterize: cysts non-enhancing water-density; hemangiomas peripheral nodular fill-in; HCC/metastases arterial hyperenhancement; abscesses rim enhancement.}
\slotentry
  {peritoneum\_mesenteric\_and\_abdominal\_wall/peritoneum}
  {pneumoperitoneum\_free\_air}
  {categorical}
  {[absent, trace, small, moderate, large, multiple\_foci, loculated]}
  {Normal: absent. Abnormal: -1000 HU gas anterior to liver/spleen or outlining bowel; loculation or large volume indicates perforation or contained leak.}
\slotentry
  {peritoneum\_mesenteric\_and\_abdominal\_wall/peritoneum}
  {ascites\_free\_fluid}
  {categorical}
  {[absent, trace, small, moderate, large, simple\_fluid]}
  {Normal: absent; trace/small shows water-density fluid (\textasciitilde{}0-20 HU) in dependent spaces, while moderate/large extends superiorly separating bowel and appears homogeneous if simple.}
\slotentry
  {genitourinary\_and\_adrenal/kidneys}
  {kidney\_morphology\_and\_appearance}
  {categorical}
  {[normal, abnormal, lobulated, irregular, homogeneous, heterogeneous, symmetric]}
  {Normal kidneys show smooth contours and homogeneous enhancement; abnormal morphology features irregular margins, heterogeneous attenuation, or asymmetry.}
\slotentry
  {genitourinary\_and\_adrenal/kidneys}
  {nephrolithiasis\_status}
  {categorical}
  {[absent, present, obstructing, nonobstructing, punctate, single, multiple, calcific\_density]}
  {Normal: absent. Present: hyperdense (\textgreater{}100 HU) collecting system foci; obstructing causes hydronephrosis and perinephric stranding, whereas nonobstructing shows no dilation.}
\slotentry
  {genitourinary\_and\_adrenal/kidneys}
  {kidney\_excretion}
  {categorical}
  {[normal, delayed, absent, symmetric, prompt, compromised, retained\_contrast]}
  {Normal: prompt symmetric opacification of collecting systems and ureters; abnormal: delayed/asymmetric excretion or retained parenchymal contrast suggests obstruction or impaired function.}
\slotentry
  {vascular\_and\_lymph\_nodes/vessels}
  {vessel\_patency}
  {categorical}
  {[patent, partially\_occluded, occluded, attenuated]}
  {Normal: continuous homogeneous contrast opacification without filling defects; abnormal: partial occlusion shows focal luminal narrowing or intraluminal thrombus, occlusion shows absent opacification.}
\slotentry
  {musculoskeletal\_and\_spine/bones}
  {bone\_lesion\_status}
  {categorical}
  {[absent, present, diffuse, multifocal, progressive, contusion, erosion, sclerosis]}
  {Normal: intact cortex, uniform trabeculae; suspicious: focal lytic (hypodense) or sclerotic (hyperdense) lesions, cortical breach, periosteal reaction, or soft-tissue mass.}
\slotentry
  {other/postsurgical\_changes}
  {postsurgical\_changes}
  {categorical}
  {[absent, present, normal, stable, unchanged, organized\_fluid\_collection]}
  {Normal: surgical clips, mild fat stranding, expected resection defect; suspicious: new/enlarging rim-enhancing fluid, abscess, recurrent mass, or obstruction.}
\slotentry
  {other/temporal\_stability\_and\_change}
  {temporal\_stability\_2}
  {categorical}
  {[stable, improved, progressed, new, resolved]}
  {Compare to prior CT. Stable: \textless{}20\% size change, unchanged enhancement. Progressed: \textgreater{}=20\% growth or new lesions. Improved/Resolved: decreased size or resolution.}
\slotentry
  {measurements}
  {size\_measurement}
  {numerical}
  {[\textless{}num\textgreater{} cm]}
  {Size interpretation is anatomy-specific; e.g., lymph nodes \textgreater{}1 cm short axis denote adenopathy, while solid organ lesions require correlation with attenuation and enhancement.}
\end{slotcatalog}
\captionof{table}{\textbf{Representative examples from the final MERLIN template.}}
\label{tab:template-examples-merlin}

\subsection{BDMAP}
\label{app:template-examples-bdmap}

The BDMAP examples focus on liver, spleen, pancreas, kidneys, and peripancreatic vessels. They highlight CT attenuation, morphology, lesion measurement, enhancement, and adjacent-structure relations.

\begin{slotcatalog}{BDMAP}{BDMAP template specification}
\slotentry
  {liver\_and\_biliary/liver}
  {liver\_status}
  {categorical}
  {[normal, abnormal]}
  {Normal: homogeneous attenuation, smooth contour, uniform enhancement; Abnormal: heterogeneous density, nodular margins, focal hypo/hyperattenuating lesions, or altered enhancement.}
\slotentry
  {liver\_and\_biliary/liver}
  {liver\_attenuation\_hu}
  {numerical}
  {[\textless{}num\textgreater{} HU]}
  {Normal hepatic attenuation is 40-60 HU; values below 40 HU indicate steatosis, whereas above 60-70 HU suggest iron overload.}
\slotentry
  {liver\_and\_biliary/liver}
  {liver\_attenuation}
  {categorical}
  {[normal, decreased, increased, steatosis, heterogeneous]}
  {Normal parenchyma is homogeneous (\textasciitilde{}50-60 HU); decreased/steatosis appears hypodense vs spleen (\textless{}40 HU), increased suggests iron/glycogen (\textgreater{}70 HU), and heterogeneous implies cirrhosis.}
\slotentry
  {spleen/splenic\_parenchyma}
  {spleen\_density}
  {numerical}
  {[\textless{}num\textgreater{} HU]}
  {Normal spleen measures 40-60 HU on portal venous phase; focal or diffuse attenuation outside this range suggests cysts, infarcts, or infiltrative disease.}
\slotentry
  {spleen/splenic\_parenchyma}
  {spleen\_parenchyma\_appearance}
  {categorical}
  {[normal, homogeneous, heterogeneous, coarse, abnormal, decreased, increased, splenomegaly, hypersplenism]}
  {Normal: homogeneous attenuation, smooth margins; abnormal: coarse/heterogeneous texture, focal hypoattenuating lesions, or splenomegaly (\textgreater{}13 cm).}
\slotentry
  {pancreas\_and\_peripancreatic\_vessels/pancreas}
  {pancreas\_normality}
  {categorical}
  {[normal, unremarkable, within normal limits]}
  {Normal: homogeneous enhancement, smooth margins, non-dilated duct; abnormal: focal hypoattenuating mass, irregular borders, ductal dilatation, or peripancreatic stranding.}
\slotentry
  {pancreas\_and\_peripancreatic\_vessels/pancreas}
  {pancreas\_size\_morphology}
  {categorical}
  {[normal, enlarged, mildly\_enlarged, focal\_enlargement, atrophic]}
  {Normal: homogeneous parenchyma, smooth contour, body \textless{}=3 cm; abnormal: diffuse enlargement \textgreater{}3 cm, focal mass, or atrophy with lobularity and fat infiltration.}
\slotentry
  {pancreas\_and\_peripancreatic\_vessels/pancreas}
  {pancreas\_enhancement\_pattern}
  {categorical}
  {[normal, homogeneous, heterogeneous, hypoenhancing, hyperattenuating, hypoattenuating, isoattenuating, calcified]}
  {Normal shows homogeneous enhancement; focal hypoattenuation/hypoenhancement suggests carcinoma, while heterogeneity, calcification, or hyperenhancement indicate chronic pancreatitis or tumors.}
\slotentry
  {pancreas\_and\_peripancreatic\_vessels/pancreas/pancreas\_mean\_hu}
  {pancreas\_mean\_hu}
  {numerical}
  {[\textless{}num\textgreater{} HU]}
  {Normal parenchyma measures 30-50 HU non-contrast; deviations below 30 HU suggest steatosis, while elevations above 50 HU indicate fibrosis or inflammation.}
\slotentry
  {pancreas\_and\_peripancreatic\_vessels/pancreatic\_lesions}
  {pancreas\_lesion\_size}
  {numerical}
  {[\textless{}num\textgreater{} cm]}
  {Normal: no focal mass; abnormal: solid lesion \textgreater{}=2 cm or cyst \textgreater{}=3 cm typically warrants further evaluation.}
\slotentry
  {kidneys\_and\_adrenals/kidneys/kidney\_lesion\_attenuation}
  {renal\_lesion\_attenuation}
  {categorical}
  {[hypoattenuating, hyperattenuating, isoattenuating, normal, mixed, varying]}
  {Hypoattenuating suggests simple cysts (\textless{}10 HU); hyperattenuating indicates hemorrhage or calcification (\textgreater{}20 HU), while isoattenuating blends with cortex and mixed denotes complex solid components.}
\slotentry
  {pancreas\_and\_peripancreatic\_vessels/peripancreatic\_vessels/vascular\_involvement}
  {vascular\_involvement}
  {categorical}
  {[absent, present, contact, encased]}
  {Normal: intact fat plane, smooth vessel wall; abnormal: fat plane loss, luminal narrowing, irregular contour, or circumferential contact indicates invasion.}
\end{slotcatalog}
\captionof{table}{\textbf{Representative examples from the final BDMAP template.}}
\label{tab:template-examples-bdmap}

\section{Representative Examples from Knowledge Graph}
\label{app:knowledge-graph-examples}
\begin{figure*}[t]
    \centering
    \includegraphics[
        width=\textwidth
    ]{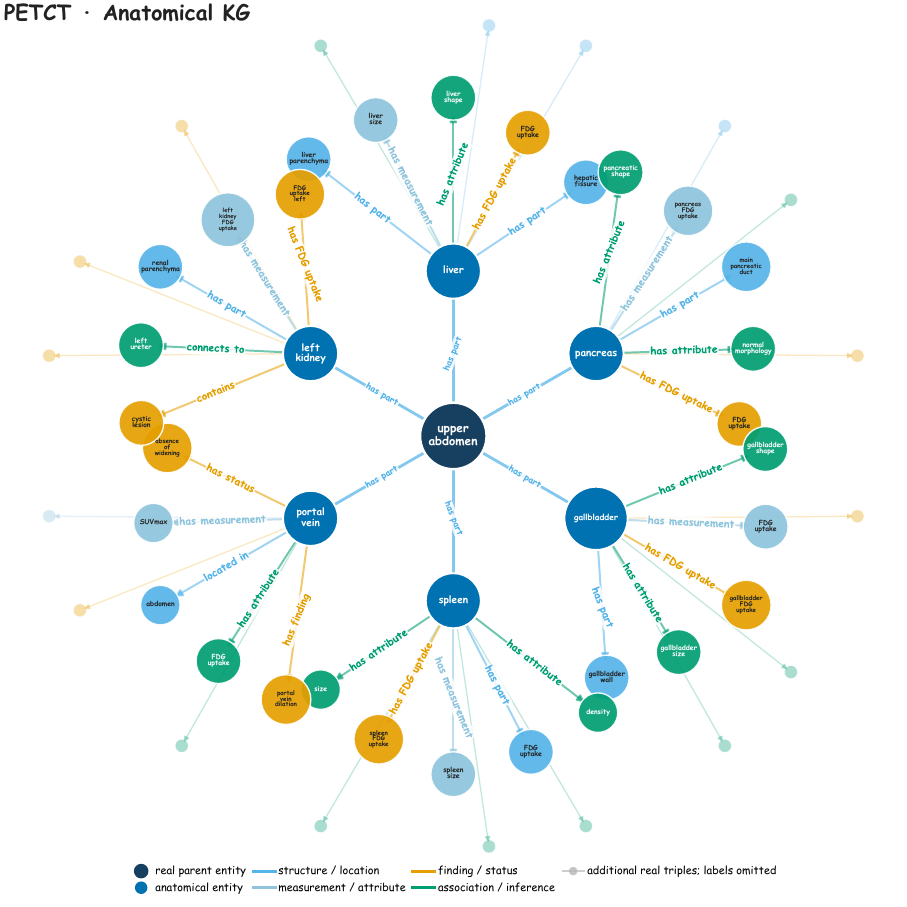}
    \caption{
        \textbf{Visualization of the PETCT anatomical knowledge graph.}
        The displayed subgraph organizes representative anatomical entities
        under their parent anatomical region and connects them to structural,
        measurement, attribute, finding, and status concepts.
        Node and edge colors distinguish entity and relation classes,
        respectively, as defined in the legend.
        Faint unlabeled branches indicate additional observed triples included
        to preserve the local graph structure.
    }
    \label{fig:kg-petct-anatomical}
\end{figure*}

\begin{figure*}[t]
    \centering
    \includegraphics[
        width=\textwidth
    ]{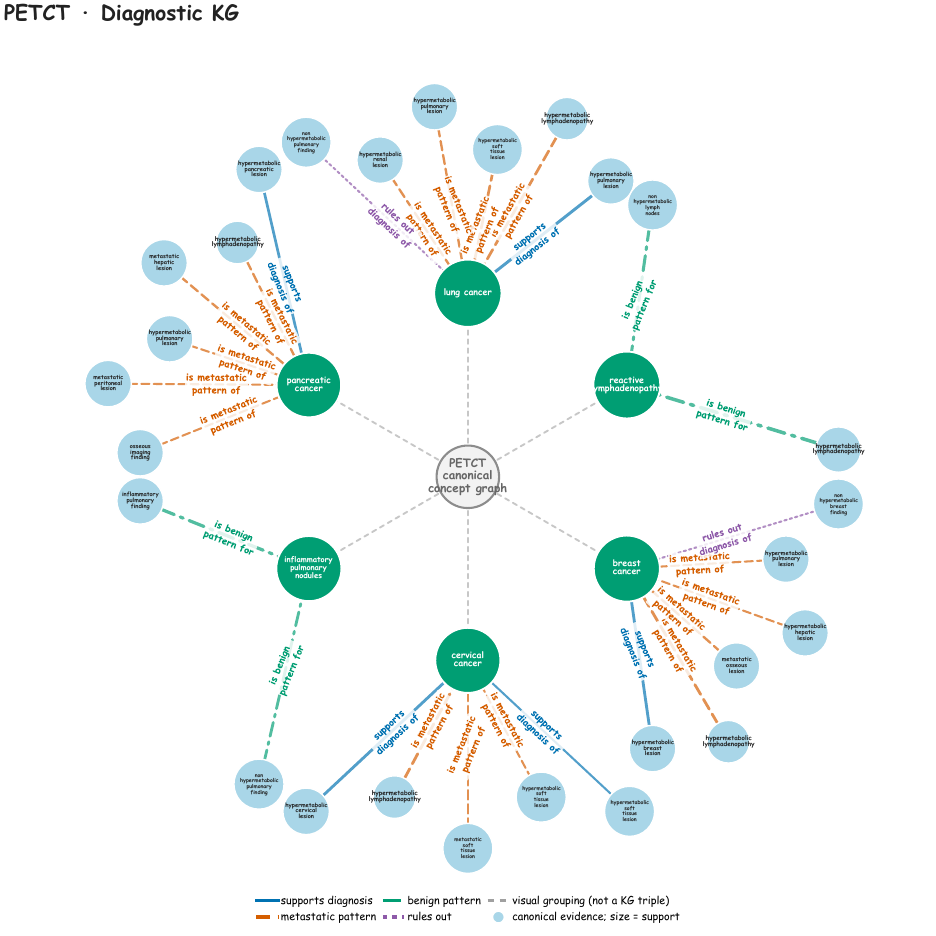}
    \caption{
        \textbf{Concept-level diagnostic knowledge graph for PETCT.}
        Representative canonical imaging-evidence concepts are connected to
        diagnostic targets through the extracted diagnostic relation types.
        Node area is proportional to the supporting raw-triple count, while
        arrow labels specify the relation between each evidence concept and
        diagnostic target.
        Gray dashed links indicate visual grouping and are not knowledge-graph edges.
    }
    \label{fig:kg-petct-diagnostic}
\end{figure*}

\begin{figure*}[t]
    \centering
    \includegraphics[
        width=\textwidth
    ]{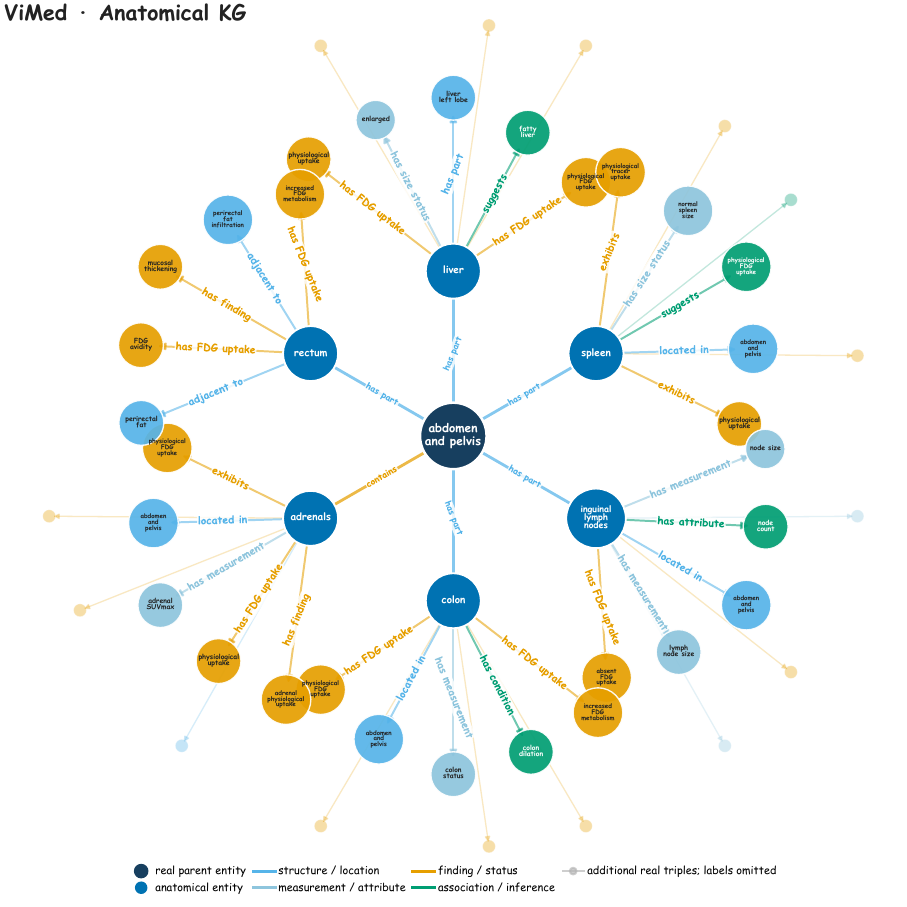}
    \caption{
        \textbf{Visualization of the ViMed anatomical knowledge graph.}
        The displayed subgraph organizes representative anatomical entities
        under their parent anatomical region and connects them to structural,
        measurement, attribute, finding, and status concepts.
        Node and edge colors distinguish entity and relation classes,
        respectively, as defined in the legend.
        Faint unlabeled branches indicate additional observed triples included
        to preserve the local graph structure.
    }
    \label{fig:kg-vimed-anatomical}
\end{figure*}

\begin{figure*}[t]
    \centering
    \includegraphics[
        width=\textwidth
    ]{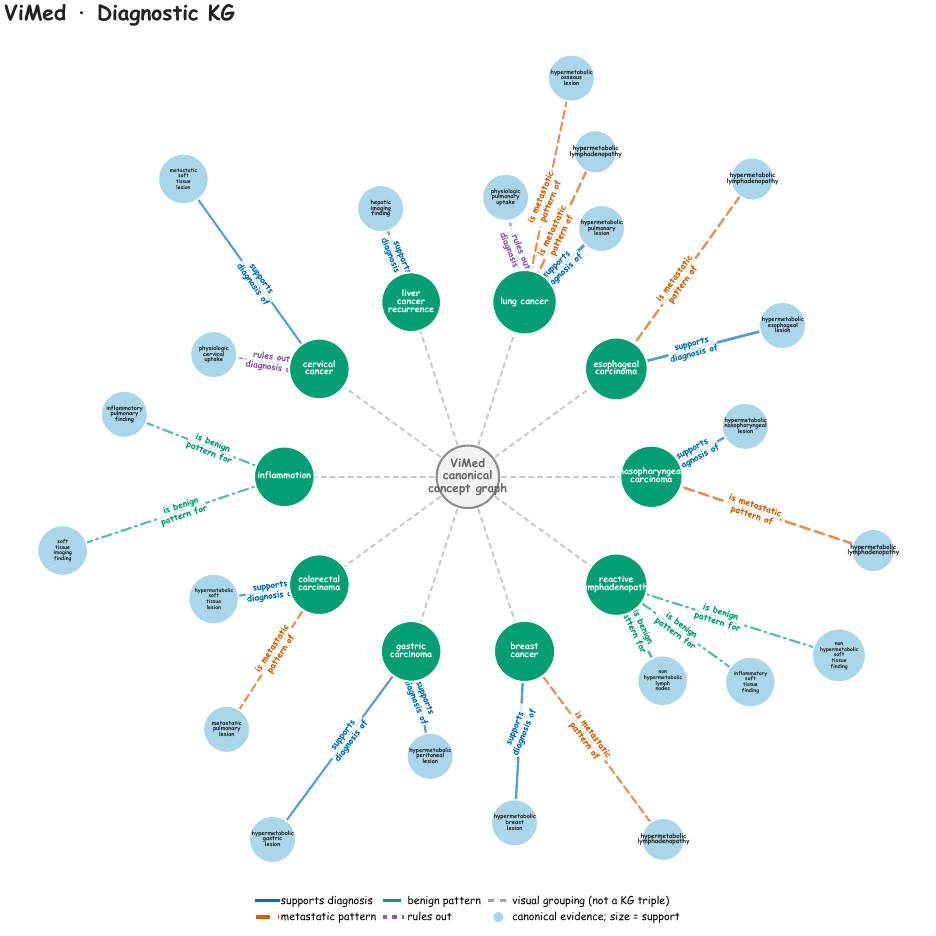}
    \caption{
        \textbf{Concept-level diagnostic knowledge graph for ViMed.}
        The visualization connects representative canonical imaging-evidence
        concepts to diagnostic targets through the extracted diagnostic
        relation types.
        Node area is proportional to the supporting raw-triple count, while
        arrow labels specify the relation between each evidence concept and
        diagnostic target.
        Gray dashed links indicate visual grouping and are not knowledge-graph edges.
    }
    \label{fig:kg-vimed-diagnostic}
\end{figure*}

\begin{figure*}[t]
    \centering
    \includegraphics[
        width=\textwidth
    ]{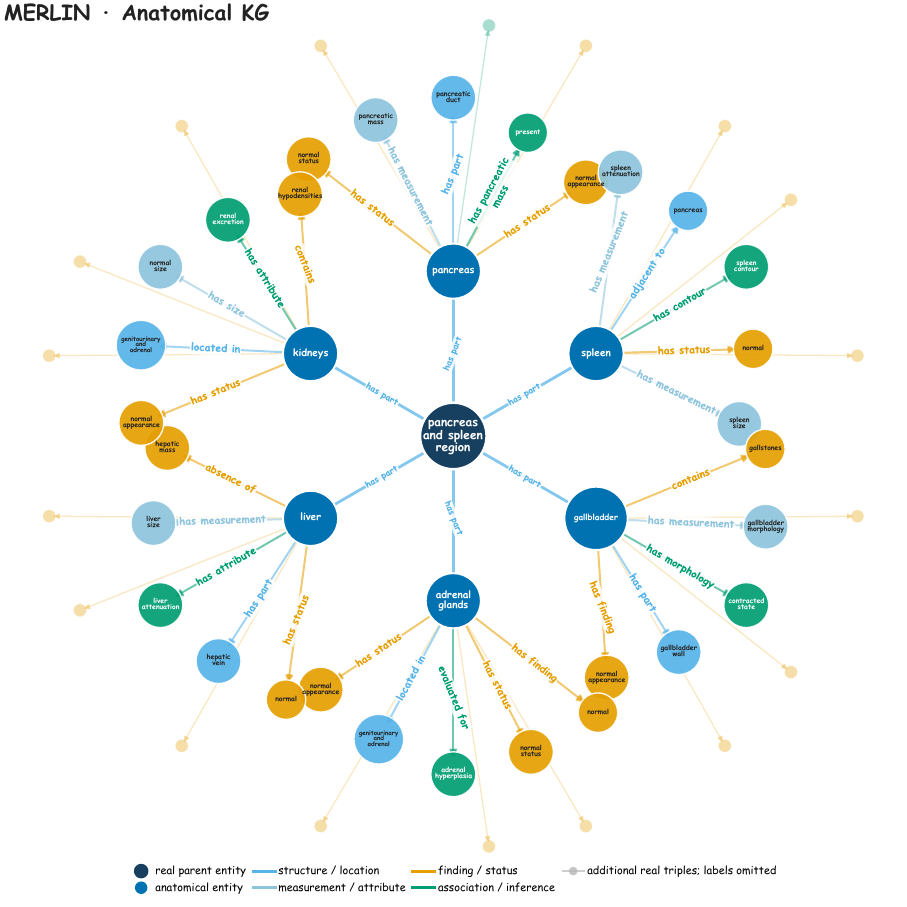}
    \caption{
        \textbf{Visualization of the MERLIN anatomical knowledge graph.}
        The displayed subgraph organizes representative anatomical entities
        under their parent anatomical region and connects them to structural,
        measurement, attribute, finding, and status concepts.
        Node and edge colors distinguish entity and relation classes,
        respectively, as defined in the legend.
        Faint unlabeled branches indicate additional observed triples included
        to preserve the local graph structure.
    }
    \label{fig:kg-merlin-anatomical}
\end{figure*}

\begin{figure*}[t]
    \centering
    \includegraphics[
        width=\textwidth
    ]{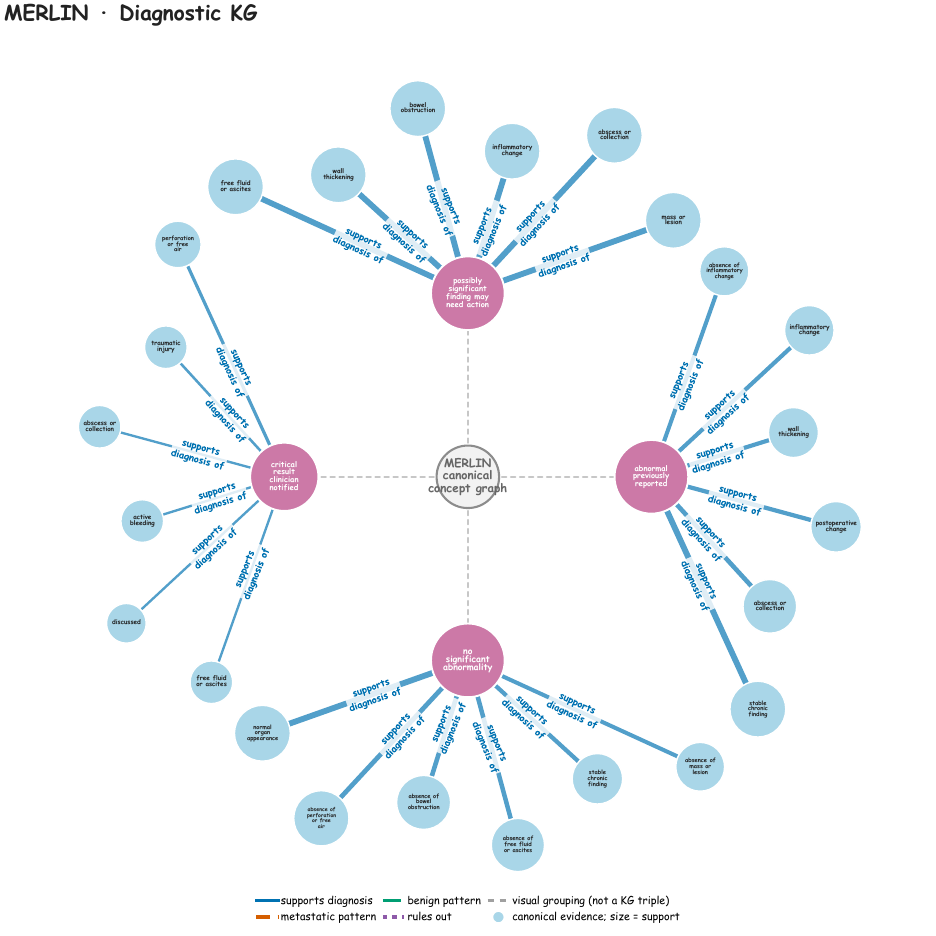}
    \caption{
        \textbf{Concept-level clinical-significance knowledge graph for MERLIN.}
        Representative canonical report-evidence concepts are connected to
        clinical-significance outcomes that characterize the reported
        abnormality and its potential need for clinical action.
        Node area is proportional to the supporting raw-triple count, while
        arrow labels specify the extracted evidence-to-outcome relation.
        Gray dashed links indicate visual grouping and are not knowledge-graph edges.
    }
    \label{fig:kg-merlin-clinical-significance}
\end{figure*}

\begin{figure*}[t]
    \centering
    \includegraphics[
        width=\textwidth
    ]{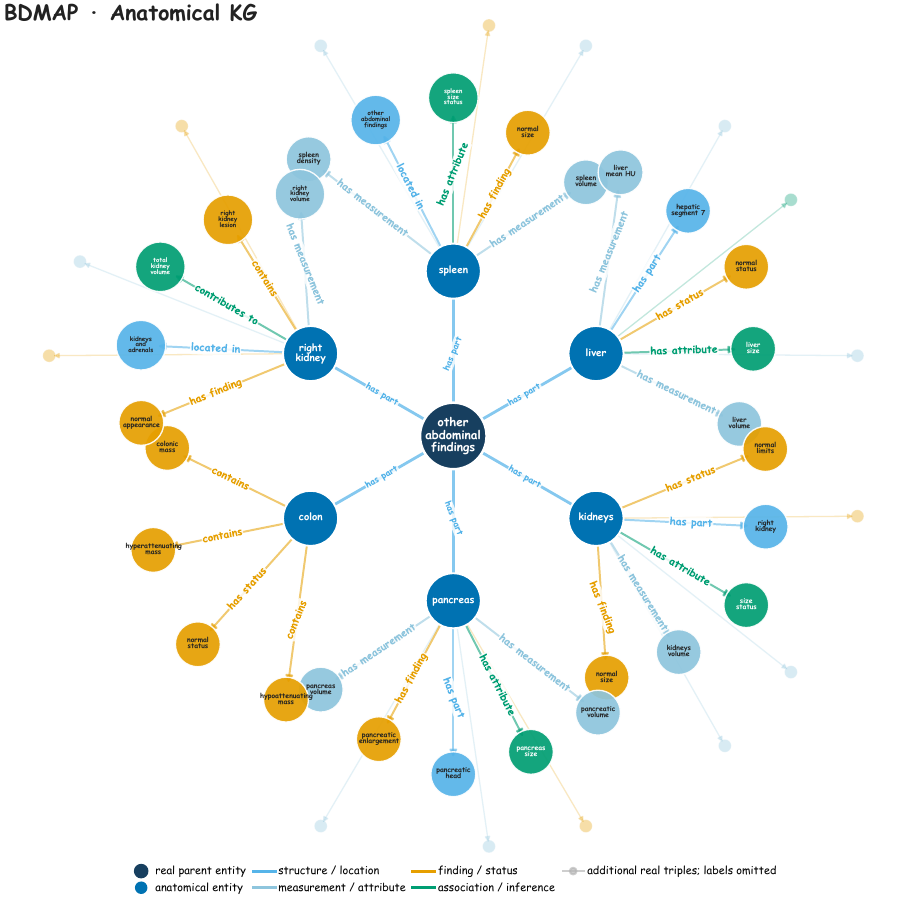}
    \caption{
        \textbf{Visualization of the BDMAP anatomical knowledge graph.}
        The displayed subgraph organizes representative anatomical entities
        under their parent anatomical region and connects them to structural,
        measurement, attribute, finding, and status concepts.
        Node and edge colors distinguish entity and relation classes,
        respectively, as defined in the legend.
        Faint unlabeled branches indicate additional observed triples included
        to preserve the local graph structure.
    }
    \label{fig:kg-bdmap-anatomical}
\end{figure*}

\begin{figure*}[t]
    \centering
    \includegraphics[
        width=\textwidth
    ]{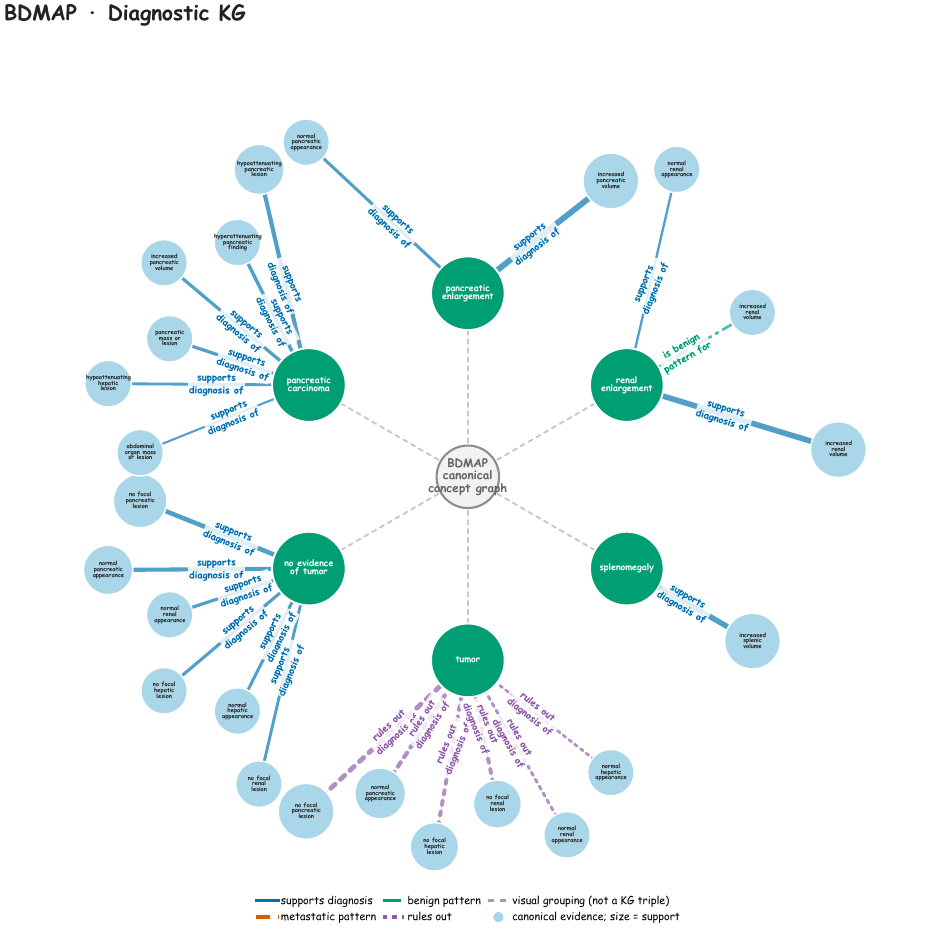}
    \caption{
        \textbf{Concept-level diagnostic knowledge graph for BDMAP.}
        Representative canonical imaging-evidence concepts are connected to
        diagnostic and clinical-status targets through the extracted diagnostic
        relation types.
        Node area is proportional to the supporting raw-triple count, while
        arrow labels specify the relation between each evidence concept and
        target.
        Gray dashed links indicate visual grouping and are not knowledge-graph edges.
    }
    \label{fig:kg-bdmap-diagnostic}
\end{figure*}

The PETCT examples illustrate anatomical organization and finding--diagnosis relations (Figs.~\ref{fig:kg-petct-anatomical} and~\ref{fig:kg-petct-diagnostic}). The ViMed examples show the corresponding anatomical and diagnostic structures (Figs.~\ref{fig:kg-vimed-anatomical} and~\ref{fig:kg-vimed-diagnostic}). The MERLIN examples pair anatomical relations with report-evidence links to clinical-significance outcomes (Figs.~\ref{fig:kg-merlin-anatomical} and~\ref{fig:kg-merlin-clinical-significance}). The BDMAP examples show anatomical relations and diagnostic or clinical-status targets (Figs.~\ref{fig:kg-bdmap-anatomical} and~\ref{fig:kg-bdmap-diagnostic}).

\FloatBarrier
\section{Prompt Specifications}
\label{app:prompts}

Baseline prompts specify batch-level schema induction and template merging (Figs.~\ref{fig:prompt-batch-template-induction} and~\ref{fig:prompt-template-merge}).

Induction prompts separate reports into ordered anatomical segments (Fig.~\ref{fig:prompt-anatomy-segmentation}), extract and align clinical spans and relations (Figs.~\ref{fig:prompt-span-kg-extraction} and~\ref{fig:prompt-span-relation-alignment}), and induce community-level slots (Fig.~\ref{fig:prompt-community-slot-induction}). Further prompts consolidate slots sharing a key (Fig.~\ref{fig:prompt-slot-consolidation}) and normalize relations within and across communities (Figs.~\ref{fig:prompt-relation-normalization} and~\ref{fig:prompt-anatomical-relation-consolidation}). The PET/CT diagnostic-graph prompts define cancer concepts and relation predicates before extracting diagnostic triples (Figs.~\ref{fig:prompt-cancer-taxonomy-audit}--\ref{fig:prompt-diagnostic-triple-extraction}).

Evolution prompts induce an anatomical backbone (Fig.~\ref{fig:prompt-anatomical-backbone-induction}), assign and reassign slot paths (Figs.~\ref{fig:prompt-first-round-slot-assignment} and~\ref{fig:prompt-global-path-reassignment}), and reorganize slots within a node (Fig.~\ref{fig:prompt-intra-node-reorganization}). Evidence-grounding prompts generate retrieval queries and synthesize reference notes (Figs.~\ref{fig:prompt-evidence-query-generation} and~\ref{fig:prompt-evidence-note-synthesis}).

Clinical prompts guide region-conditioned structuring, report reconstruction, and diagnostic reasoning (Figs.~\ref{fig:prompt-region-conditioned-structuring}--\ref{fig:prompt-diagnostic-reasoning}). Coverage-evaluation prompts extract report concepts and match them to template slots (Figs.~\ref{fig:prompt-clinical-concept-extraction} and~\ref{fig:prompt-concept-slot-matching}).

\begin{promptcardcompact}[title={Template Induction -- Batch Schema}]
You are a radiology informatics expert building a structured slot schema for whole-body 18F-FDG PET/CT reports.

You will be shown $n_reports report DESCRIPTION sections (imaging findings only, not diagnoses). Induce ONE template that captures the fields these reports have in common: which anatomical regions are described, and for each, which attributes are reported and what values those attributes take.

### Output format

Return a single JSON object, matching the schema of the example below exactly.

Top level is "desc".

Under it, each key is a region group (e.g. head_and_neck) containing organ keys, OR a flat section holding _slots_core directly. Both forms are valid; use whichever fits the content.

_slots_core holds slots for attributes present in MOST reports of the batch.

_slots_optional holds slots for attributes present in only SOME. Use this key only when such attributes exist.

Each slot has exactly these fields:

slot_id -- short unique identifier, snake_case

k -- English attribute name, snake_case

k_zh -- the Chinese gloss of k

tau -- one of categorical, numerical, freetext

omega -- a list of the values actually observed in this batch. For numerical, give value PATTERNS using <NUM> as the number placeholder (e.g. "<NUM> cm", "<NUM> x <NUM> cm", "SUVmax=<NUM>"), never concrete numbers. For categorical, list the distinct observed categories. For freetext, list representative descriptor values.

### Example (format reference only)

The following is a deliberately SIMPLIFIED example. It shows the required shape, not the required coverage -- it contains only a handful of regions and slots, whereas a real batch supports many more. Follow its structure; do not copy its content, and do not stop at its level of detail.

$seed_template

### Requirements

Cover every anatomical region the batch actually describes, not just those in the example. The example is a scaffold, not a ceiling.

Use the batch's own vocabulary for omega values. Do not invent categories the reports never use.

During induction, normalize numerical values to <NUM> patterns; numerical omega therefore stores patterns rather than concrete measurements.

Give every slot an accurate k_zh. It is a translation of k, not a restatement of the region name.

Put an attribute in _slots_core only if most reports in the batch have it; otherwise _slots_optional.

### Batch

This is batch $batch_index of $n_batches.

$reports

Output the JSON object alone. No preamble, no commentary, no code fences.
\end{promptcardcompact}

\captionof{figure}{
\textbf{System prompt for batch-level template induction.}
The prompt induces a structured slot schema from a batch of PET/CT report descriptions.
}
\label{fig:prompt-batch-template-induction}

\begin{promptcardcompact}[title={Template Merge -- Schema Consolidation}]
You are a radiology informatics expert consolidating slot schemas for whole-body 18F-FDG PET/CT reports.

Below are $n_templates templates. Each was induced independently from a different batch of reports drawn from the same corpus, so they describe the same underlying report structure as seen through different samples. Merge them into ONE canonical template.

### Merge rules, in priority order

Union of coverage. Every region, organ, and slot present in ANY input template must survive into the output. Dropping coverage is the primary failure mode of this step. When in doubt, keep it.

Deduplicate by meaning, not by string. Slots describing the same attribute under different k or slot_id names are ONE slot. Merge them, keeping the clearest name. Likewise for organs and regions.

Union the omega sets. When merging slots, the result's omega is the union of the inputs' values, deduplicated.

Core vs optional by prevalence. A slot belongs in _slots_core if it appeared in _slots_core of MOST input templates; otherwise put it in _slots_optional. This distinction is the main thing merging adds over simple concatenation -- do not put everything in core.

Reconcile tau conflicts. If the same attribute is numerical in some templates and freetext in others, choose the type that fits the omega values: use numerical when values are <NUM> patterns, categorical when they are a small closed set, freetext otherwise.

Anatomical ordering. Order regions head-to-toe. Keep each input's internal organ grouping where the templates agree.

Fill in k_zh. Every slot in the output needs an accurate Chinese gloss, including slots merged from inputs that disagreed on wording.

### Templates

$templates

Output the merged JSON object alone, in the same schema as the inputs. No preamble, no commentary, no code fences.
\end{promptcardcompact}

\captionof{figure}{
\textbf{System prompt for template consolidation.}
The prompt merges independently induced batch-level schemas into one canonical PET/CT reporting template.
}
\label{fig:prompt-template-merge}


\begin{promptcard}[title={Induction Agent -- Anatomy Segmentation}]
You are a PET/CT report structuring expert. Cut the report description into SEGMENTS. A segment is one or more consecutive sentences describing ONE anatomical region. Start a new segment when the subject anatomy changes.

For each segment output:
- anatomy_focus : English name of the segment's main anatomy
- evidence_text : the segment's text, verbatim from the report

OUTPUT JSON:
{"segments": [ {"anatomy_focus": "...", "evidence_text": "..."} ]}
\end{promptcard}

\captionof{figure}{
\textbf{System prompt for anatomy-guided report segmentation.}
The prompt divides a PET/CT report into consecutive segments according to changes in anatomical focus.
}
\label{fig:prompt-anatomy-segmentation}

\begin{promptcardcompact}[title={Induction Agent -- Span and KG Extraction}]
You are a PET/CT report structuring expert. Given one segment of a report (describing one anatomical region), perform two independent extraction tasks and emit them as two separate JSON blocks in a single response.

Task 1 -- Spans: Enumerate every distinct finding, measurement, or status assertion in the segment as a span.

Task 2 -- KG: Build a knowledge graph for the segment consisting of entities (nodes) and relations (edges).

Output format -- exactly two JSON blocks, each preceded by its marker line. Do not wrap them in a parent object, and do not add anything outside these blocks (no prose and no code fences).

===Spans_JSON===
{
  "spans": [
    {
      "a_struct": "should be exactly the Anatomy focus value passed in this turn (e.g., if Anatomy focus is 'head_and_neck', a_struct must be 'head_and_neck').",
      "a_sub": "FINER anatomical location below a_struct. E.g., for a right maxillary sinus finding when a_struct='head_and_neck', a_sub='maxillary_sinus_right'.",
      "a_attr": "attribute name (e.g., mass, nodule, density, size)",
      "a_type": "value type (categorical/numerical/freetext)",
      "a_val": "induction-normalized value (replace numbers with <NUM>, e.g., <NUM> mm or <NUM> cm)",
      "a_role": "clinical role (finding/measurement/diagnosis)",
      "a_sec": "report section (desc)"
    }
  ]
}

===KG_JSON===
{
  "entities": [
    {
      "name": "snake_case English entity name, e.g., liver, right_lobe, hepatic_mass",
      "type": "entity type, e.g., anatomy, lesion, measurement"
    }
  ],
  "relations": [
    {
      "subject": "entity name from entities, e.g., liver",
      "predicate": "relation predicate, e.g., has_part, has_measurement, adjacent_to, suggests, located_in",
      "object": "entity name from entities, e.g., right_lobe"
    }
  ]
}

Rules -- non-negotiable:

- The marker lines ===Spans_JSON=== and ===KG_JSON=== must appear exactly as shown, each on its own line before the corresponding JSON block.

- Emit both blocks every time.

- Every subject and object in relations must also appear as a name in entities. If you write {"subject":"liver",...}, then {"name":"liver",...} must be present in entities.

- Minimums: spans >= 1, entities >= 2, relations >= 1.
\end{promptcardcompact}

\captionof{figure}{
\textbf{System prompt for segment-level span and knowledge-graph extraction.}
The prompt independently extracts clinical spans and entity--relation structures from an anatomy-specific report segment.
}
\label{fig:prompt-span-kg-extraction}
\begin{promptcardcompact}[title={Induction Agent -- Span--Relation Alignment}]
You are a PET/CT report structuring expert. You are given:

- spans: a list of distinct findings, measurements, or status assertions in the segment.
- kg: the segment-level knowledge graph containing {entities, relations}.

Task:

Emit exactly one atom for each span, producing exactly len(spans) atoms in the same order as spans.

For each span:

1. Assign exactly one relation from kg.relations.
2. Select the most semantically relevant relation for that span.
3. If no direct match exists, use the most anatomically compatible relation from the same segment.
4. Copy the subject and object entities referenced by the selected relation from kg.entities into that atom's entities[].

Coverage:

- Try to use every relation in kg.relations at least once across the assignments whenever possible.
- A relation may be reused by multiple spans when appropriate.

Output format:

{
  "atoms": [
    {
      "span_tuple": {...},
      "entities": [
        {"name": "...", "type": "..."},
        {"name": "...", "type": "..."}
      ],
      "relations": [
        {"subject": "...", "predicate": "...", "object": "..."}
      ]
    }
  ]
}

Hard rules:

- Emit exactly one atom per span, preserving the original span content and order.
- Each atom must contain exactly one relation copied from kg.relations.
- entities[] must contain exactly the subject and object entities of the selected relation, copied from kg.entities.
- Do not modify or invent any span, entity, or relation.

Example -- 4 spans, 2 relations

Input:

Anatomy focus: thyroid

spans: [
  {
    "a_struct": "thyroid",
    "a_sub": "bilateral_lobes",
    "a_attr": "fdg_uptake_status",
    "a_type": "categorical",
    "a_val": "increased",
    "a_role": "finding",
    "a_sec": "desc"
  },
  {
    "a_struct": "thyroid",
    "a_sub": "bilateral_lobes",
    "a_attr": "fdg_uptake_pattern",
    "a_type": "categorical",
    "a_val": "diffuse_mild",
    "a_role": "finding",
    "a_sec": "desc"
  },
  {
    "a_struct": "thyroid",
    "a_sub": "bilateral_lobes",
    "a_attr": "suvmax",
    "a_type": "numerical",
    "a_val": "<NUM>",
    "a_role": "measurement",
    "a_sec": "desc"
  },
  {
    "a_struct": "thyroid",
    "a_sub": "bilateral_lobes",
    "a_attr": "metabolic_activity",
    "a_type": "categorical",
    "a_val": "mildly_increased",
    "a_role": "finding",
    "a_sec": "desc"
  }
]

kg: {
  "entities": [
    {"name": "thyroid", "type": "anatomy"},
    {"name": "fdg_uptake", "type": "attribute"},
    {"name": "suvmax", "type": "measurement"}
  ],
  "relations": [
    {
      "subject": "thyroid",
      "predicate": "shows",
      "object": "fdg_uptake"
    },
    {
      "subject": "thyroid",
      "predicate": "has_suvmax",
      "object": "suvmax"
    }
  ]
}

Correct output:

{
  "atoms": [
    {
      "span_tuple": {
        "a_struct": "thyroid",
        "a_sub": "bilateral_lobes",
        "a_attr": "fdg_uptake_status",
        "a_type": "categorical",
        "a_val": "increased",
        "a_role": "finding",
        "a_sec": "desc"
      },
      "entities": [
        {"name": "thyroid", "type": "anatomy"},
        {"name": "fdg_uptake", "type": "attribute"}
      ],
      "relations": [
        {
          "subject": "thyroid",
          "predicate": "shows",
          "object": "fdg_uptake"
        }
      ]
    },
    {
      "span_tuple": {
        "a_struct": "thyroid",
        "a_sub": "bilateral_lobes",
        "a_attr": "fdg_uptake_pattern",
        "a_type": "categorical",
        "a_val": "diffuse_mild",
        "a_role": "finding",
        "a_sec": "desc"
      },
      "entities": [
        {"name": "thyroid", "type": "anatomy"},
        {"name": "fdg_uptake", "type": "attribute"}
      ],
      "relations": [
        {
          "subject": "thyroid",
          "predicate": "shows",
          "object": "fdg_uptake"
        }
      ]
    },
    {
      "span_tuple": {
        "a_struct": "thyroid",
        "a_sub": "bilateral_lobes",
        "a_attr": "suvmax",
        "a_type": "numerical",
        "a_val": "<NUM>",
        "a_role": "measurement",
        "a_sec": "desc"
      },
      "entities": [
        {"name": "thyroid", "type": "anatomy"},
        {"name": "suvmax", "type": "measurement"}
      ],
      "relations": [
        {
          "subject": "thyroid",
          "predicate": "has_suvmax",
          "object": "suvmax"
        }
      ]
    },
    {
      "span_tuple": {
        "a_struct": "thyroid",
        "a_sub": "bilateral_lobes",
        "a_attr": "metabolic_activity",
        "a_type": "categorical",
        "a_val": "mildly_increased",
        "a_role": "finding",
        "a_sec": "desc"
      },
      "entities": [
        {"name": "thyroid", "type": "anatomy"},
        {"name": "fdg_uptake", "type": "attribute"}
      ],
      "relations": [
        {
          "subject": "thyroid",
          "predicate": "shows",
          "object": "fdg_uptake"
        }
      ]
    }
  ]
}

Output the JSON object alone. No preamble, no commentary, no code fences.
\end{promptcardcompact}

\captionof{figure}{
\textbf{System prompt for span--relation alignment.}
The prompt aligns each extracted clinical span with one segment-level knowledge-graph relation while preserving span order and entity provenance.
}
\label{fig:prompt-span-relation-alignment}
\begin{promptcard}[title={Induction Agent -- Community-to-Slot Induction}]
You are a medical report structuring expert.

You are given spans from one community. These spans all come from the same anatomy bucket and have already been semantically clustered.

Your job:

1. Subdivide the spans into semantically consistent subgroups.
2. Induce one normalized slot (k, tau, omega) for each subgroup.

Output JSON format:

{
  "subgroups": [
    {
      "group_id": "g1",
      "span_indices": [0, 1, 2, ...],
      "slot": {
        "k": "standardized key name in English snake_case, e.g., liver_parenchymal_mass",
        "k_zh": "accurate Chinese translation of k",
        "tau": "categorical/numerical/freetext",
        "omega": ["allowed_value1", "allowed_value2"]
      },
      "description": "semantic description"
    }
  ]
}
\end{promptcard}

\captionof{figure}{
\textbf{System prompt for community-level slot induction.}
The prompt partitions anatomically constrained clinical spans into semantically consistent subgroups and induces a normalized slot for each subgroup.
}
\label{fig:prompt-community-slot-induction}

\begin{promptcard}[title={Induction Agent -- Relation Normalization}]
You are a medical knowledge graph expert. Given a list of (subject, predicate, object) triples extracted from one community of clinical atoms, your goals are:

1. Deduplicate near-synonymous triples. For example, (lung_nodule, has_measurement, suvmax) and (lung_nodule, has_attribute, suvmax) may refer to the same fact.

2. Unify surface forms using lowercase snake_case. For example, merge variants such as FDG_uptake, fdg_uptake, and fdg_metabolism into one canonical form.

Output JSON format:

{
  "canonical_triples": [
    {
      "triple_id": "t1",
      "subject": "canonical subject (snake_case)",
      "predicate": "canonical predicate (snake_case)",
      "object": "canonical object (snake_case)",
      "input_indices": [0, 3, 7],
      "description": "one-line description (optional)"
    }
  ]
}
\end{promptcard}

\captionof{figure}{
\textbf{System prompt for within-community relation normalization.}
The prompt deduplicates semantically equivalent clinical triples and consolidates their surface forms into canonical entity and predicate representations.
}
\label{fig:prompt-relation-normalization}

\begin{promptcard}[title={Induction Agent -- Slot Consolidation}]
You are a medical report structuring expert. Given a group of slots that all share the same standardized key name (k), merge them into one canonical slot.

Tasks:

1. Merge the allowed value sets (omega). Keep distinct canonical values and remove near-duplicates.

2. Unify the value type (tau) using majority voting across the member slots.

3. Produce one canonical Chinese name (k_zh) and a one-line semantic description.

Output JSON format:

{
  "slot_id": "unique identifier based on k",
  "k": "standardized key name",
  "k_zh": "Chinese key name",
  "tau": "categorical/numerical/freetext",
  "omega": ["merged allowed values"],
  "description": "semantic description"
}
\end{promptcard}

\captionof{figure}{
\textbf{System prompt for cross-community slot consolidation.}
The prompt merges slots sharing the same standardized key into a single canonical slot with a unified value type and candidate-value vocabulary.
}
\label{fig:prompt-slot-consolidation}

\begin{promptcardcompact}[title={Induction Agent -- Relation Consolidation}]
You are given medical triples from the same anatomical region. Merge triples with the same clinical meaning into canonical triples.

Canonicalization rules:

1. Merge semantically equivalent finding predicates into has_finding when they indicate the presence of a finding, e.g., shows, exhibits, and demonstrates.

Example:

vertebral_body_margin | shows   | osteophyte
vertebral_body_margin | exhibits | osteophyte

->

vertebral_body_margin | has_finding | osteophyte

2. Merge semantically equivalent FDG-related predicates into has_fdg_uptake, e.g., has_fdg_activity, shows_fdg_uptake, and exhibits_fdg_uptake.

3. Do not blindly merge measurement or attribute predicates such as has_size, has_diameter, has_ct_value, has_suvmax, has_density, or has_attenuation. Merge them only when their clinical meanings are clearly equivalent.

4. Merge clearly equivalent entity variants, including singular/plural forms, minor spelling differences, word-order variants, and redundant prefixes, when the anatomical meaning is unchanged.

5. Preserve laterality. Do not merge left/right, unilateral/bilateral, or other side-specific entities.

6. Preserve polarity and normality. Negative or normality predicates must not be merged with positive finding or FDG predicates. If uncertain, keep them separate.

Output JSONL, one canonical group per line:

{"subject":"<canonical subject>","predicate":"<canonical predicate>","object":"<canonical object>","merged_from_triple_ids":["<triple_id_1>","<triple_id_2>",...]}

Output rules:

- Every input triple_id must appear in exactly one group.
- Singleton groups are allowed.
- Use lowercase snake_case for the canonical subject, predicate, and object.
- Do not invent triple_ids.
- Output only JSONL lines, with no prose or code fences.
\end{promptcardcompact}


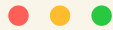
\captionof{figure}{
\textbf{System prompt for anatomy-constrained relation consolidation.}
The prompt merges semantically equivalent medical triples while preserving anatomical laterality, clinical polarity, measurement semantics, and source identifiers.
}
\label{fig:prompt-anatomical-relation-consolidation}

\begin{promptcardcompact}[title={Induction Agent -- Cancer Taxonomy Audit}]
You are an oncology-trained PET/CT auditor. You are given a frequency-ranked list of raw diagnosis strings from a training corpus.

Task:

Identify the cancer-type taxonomy represented in the corpus. Include primary cancer types that occur frequently enough to warrant separate labels, and collect their observed diagnostic expressions as aliases.

For each cancer type, output:

- label: canonical lowercase snake_case name.

- aliases: lowercase diagnostic phrases actually observed in the input that refer to this cancer type.

- proportion_hint: approximate number of input diagnoses belonging to this type, used for downstream stratified sampling.

Prefer the following canonical labels when supported by the corpus:

lung_cancer, liver_cancer, hepatocellular_carcinoma, rectal_cancer, colon_cancer, breast_cancer, cervical_cancer, esophageal_cancer, pancreatic_cancer, gastric_cancer, renal_cancer, nasopharyngeal_cancer, prostate_cancer, ovarian_cancer, thyroid_cancer, bladder_cancer.

Guidelines:

- Prefer between 8 and 20 cancer types when supported by the corpus. Do not create labels solely to reach this range.

- Derive aliases from the provided diagnosis strings rather than inventing unseen variants.

- Merge synonymous expressions into one canonical label.

- When a specific subtype is explicitly stated, prefer the more specific label and avoid double-counting it under a broader parent category.

Output:

{
  "cancer_types": [
    {
      "label": "lung_cancer",
      "aliases": [
        "lung cancer",
        "pulmonary carcinoma"
      ],
      "proportion_hint": 25
    }
  ],
  "reasoning": "One brief sentence explaining the induced taxonomy."
}
\end{promptcardcompact}

\captionof{figure}{
\textbf{System prompt for cancer-taxonomy auditing.}
The prompt identifies corpus-supported primary cancer categories and consolidates their observed diagnostic expressions into canonical aliases.
}
\label{fig:prompt-cancer-taxonomy-audit}

\begin{promptcardcompact}[title={Induction Agent -- Diagnostic Predicate Schema}]
You are an oncology-trained PET/CT radiologist. You are given a set of real (description, diagnosis) pairs sampled from the training corpus.

Task:

Induce a compact predicate schema for extracting diagnostic triples (subject, predicate, object) that describe how imaging findings support, characterize, or argue against a diagnosis.

For each predicate, output:

- name: lowercase snake_case predicate name.

- description: one-sentence description of its clinical meaning.

- subject_type: expected subject type, e.g., finding_constellation, metastasis_finding, counterevidence_finding, or benign_finding.

- object_type: expected object type, e.g., primary_cancer, cancer_type, or benign_disease.

Guidelines:

- Prefer a compact schema containing between 3 and 6 predicates.

- When applicable, prefer the following canonical predicates:

  supports_diagnosis_of

  is_metastatic_pattern_of

  rules_out_diagnosis_of

  is_benign_pattern_for

- Introduce a new predicate only when the observed diagnostic relation cannot be represented by these predicates.

- Predicates should encode diagnostic evidence rather than pure anatomical structure or imaging parameters.

Output:

{
  "predicates": [
    {
      "name": "supports_diagnosis_of",
      "description": "...",
      "subject_type": "finding_constellation",
      "object_type": "primary_cancer"
    }
  ],
  "reasoning": "One brief sentence explaining the induced schema."
}
\end{promptcardcompact}

\captionof{figure}{
\textbf{System prompt for oncology-specific diagnostic predicate induction.}
The prompt induces a compact relation schema describing how PET/CT findings support, characterize, or argue against oncological diagnoses.
}
\label{fig:prompt-oncology-diagnostic-predicate-schema}

\begin{promptcardcompact}[title={Induction Agent -- Diagnostic Triple Extraction}]
You are an oncology-trained PET/CT radiologist building a diagnostic knowledge graph.

You are given one PET/CT case consisting of an imaging description and the radiologist's diagnosis. Extract diagnostic triples (subject, predicate, object) that capture how imaging findings support, characterize, or argue against the diagnosis.

Predicate types:

{{PREDICATE_DEFINITIONS_LEARNED_FROM_TRAINING_DATA}}

For each triple, also output:

- anatomy_bucket: one of {{ANATOMY_BUCKET_ENUM}}.

- subject_keywords: between 3 and 6 anatomical or morphological keywords for downstream matching.

Extraction rules:

- Use lowercase snake_case for subjects, predicates, and objects.

- Primary cancer objects must come from the diagnosis text.

- If multiple primary cancers are stated, extract triples for each relevant cancer.

- For metastatic findings, the object should be the corresponding primary cancer, not the metastatic site.

- Exclude pure anatomy, scan parameters, and other non-diagnostic information.

- Findings must be grounded in the description, and diagnostic objects must be grounded in the diagnosis text.

- Prefer a small set of high-confidence triples, typically between 3 and 8 per case.

Output:

{
  "primary_cancers": [
    "<snake_case cancer 1>",
    "..."
  ],
  "triples": [
    {
      "subject": "<finding pattern, snake_case>",
      "predicate": "{{PREDICATE_ENUM}}",
      "object": "<primary cancer or benign disease, snake_case>",
      "anatomy_bucket": "<one of the buckets>",
      "subject_keywords": [
        "tok1",
        "tok2",
        "..."
      ]
    }
  ]
}

Output the JSON object only. No prose or code fences.
\end{promptcardcompact}

\captionof{figure}{
\textbf{System prompt for diagnostic-triple extraction.}
The prompt extracts description-grounded finding--diagnosis relations using predicate and anatomical schemas induced from the training corpus.
}
\label{fig:prompt-diagnostic-triple-extraction}

\begin{promptcard}[title={Evolution Agent -- Anatomical Backbone Induction}]
You are a medical report structuring expert. Based on the given medical report description samples, induce a high-level structural skeleton.

Output JSON format as a nested tree:

{
  "desc": {
    "chest": {
      "lung": {},
      "pleura": {},
      "heart": {},
      "breast": {},
      "..."
    },
    "abdomen_pelvis": {
      "liver": {},
      "gallbladder": {},
      "kidney": {},
      "pelvic_cavity": {},
      "..."
    }
  }
}

Rules:

1. The top level must contain only desc.

2. Organize the hierarchy according to anatomical structures or clinically meaningful logical categories.

3. Every leaf node must be represented as an empty object {}.
\end{promptcard}

\captionof{figure}{
\textbf{System prompt for anatomical backbone induction.}
The prompt derives a high-level hierarchical skeleton from representative medical report descriptions.
}
\label{fig:prompt-anatomical-backbone-induction}

\begin{promptcard}[title={Evolution Agent -- First-Round Slot Assignment}]
You are a medical report structuring expert. Given a slot and a hierarchical skeleton, decide where the slot should be attached.

Output JSON:

{
  "decision": "attach/propose_new/drop",
  "tier": "core/extended/drop",
  "path": "desc.abdomen_pelvis.liver",
  "new_branch": {
    "parent": "desc.chest",
    "name": "mediastinal_lymph_node"
  },
  "reason": "explanation"
}

Rules:

- attach: attach the slot to an existing node.

- propose_new: propose a new branch when no suitable node exists.

- drop: discard the slot.

- tier: use core for essential slots, extended for optional slots, and drop for discarded slots.
\end{promptcard}

\captionof{figure}{
\textbf{System prompt for first-round slot assignment.}
The prompt assigns an induced slot to an existing anatomical path, proposes a new branch when necessary, or removes the slot.
}
\label{fig:prompt-first-round-slot-assignment}

\begin{promptcard}[title={Evolution Agent -- Global Path Reassignment}]
You are a medical report structuring expert. Given a slot and the updated hierarchical skeleton, decide its final path.

Output JSON:

{
  "tier": "core/extended/drop",
  "path": "desc.abdomen_pelvis.liver",
  "reason": "explanation"
}

Rules:

- You cannot propose new nodes.

- The selected path must already exist in the updated skeleton.

- Use core for essential slots, extended for optional slots, and drop for discarded slots.
\end{promptcard}

\captionof{figure}{
\textbf{System prompt for global path reassignment.}
The prompt performs the final slot-placement decision using the complete set of paths in the updated anatomical hierarchy.
}
\label{fig:prompt-global-path-reassignment}

\begin{promptcard}[title={Evolution Agent -- Intra-Node Reorganization}]
You are a medical report structuring expert. Given a list of slots attached to one hierarchical node, perform the following tasks:

1. Merge slots only when they represent the same clinical attribute or measurement target and have compatible anatomical location, laterality, and polarity; substantial overlap in allowed-value sets may serve as supporting evidence but is not sufficient on its own.


2. Harmonize their value types using majority voting.

Output JSON:

{
  "slots": [
    {
      "slot_id": "id",
      "k": "key name",
      "k_zh": "Chinese name",
      "tau": "type",
      "omega": ["values"],
      "merged_from": ["original ids"]
    }
  ]
}
\end{promptcard}

\captionof{figure}{
\textbf{System prompt for intra-node semantic reorganization.}
The prompt consolidates slots with substantially overlapping candidate-value sets within one hierarchical node and harmonizes their value types.
}
\label{fig:prompt-intra-node-reorganization}

\begin{promptcard}[title={Evolution Agent -- Evidence Query Generation}]
You generate a single concise English web-search query to retrieve authoritative clinical reference knowledge for one PET/CT report slot.

Return strictly:

{"query": "..."}

Guidance by value type:

- For a numerical slot, search for reference values and clinically used abnormal thresholds for the measurement when available.

  Example: "normal spleen SUVmax reference range PET/CT"

- For a categorical or freetext slot, search for clinical interpretation criteria or features that distinguish relevant states.

  Example: "thyroid nodule FDG uptake benign malignant PET/CT"

Keep the query short, English, and specific to the anatomy, finding, and PET/CT context. Do not use Chinese.
\end{promptcard}

\captionof{figure}{
\textbf{System prompt for external-evidence query generation.}
The prompt produces a concise slot-specific search query for retrieving authoritative PET/CT reference knowledge.
}
\label{fig:prompt-evidence-query-generation}

\begin{promptcard}[title={Evolution Agent -- Evidence Note Synthesis}]
You are a senior radiologist writing one concise clinical reference note for a PET/CT report slot. The note should summarize clinically relevant reference knowledge that may help interpret the finding during reporting.

You are given the slot definition and retrieved evidence snippets from clinical papers and web sources. Use the provided evidence as grounding.

Return strictly:

{
  "note": "...",
  "reason": "one short sentence explaining the evidence basis or why the note is empty"
}

Rules:

- The note must contain no more than 200 characters. Write it in English using one or two concise clauses.

- For a numerical slot, summarize reference values or clinically used abnormal thresholds when supported by the retrieved evidence.

- For a categorical or freetext slot, summarize the main clinical features or criteria that distinguish relevant states or interpretations.

- Prefer evidence from higher-authority clinical sources when the retrieved sources disagree.

- Do not introduce precise numerical values, thresholds, or clinical criteria that are not supported by the provided evidence.

- If the retrieved evidence is insufficient to support a reliable reference note, return an empty note using "note": "", and briefly explain why in reason.
\end{promptcard}

\captionof{figure}{
\textbf{System prompt for external-evidence note synthesis.}
The prompt converts retrieved clinical evidence into a concise, source-grounded reference note for one PET/CT reporting slot.
}
\label{fig:prompt-evidence-note-synthesis}


\begin{promptcardcompact}[title={Clinical Agent -- Region-Conditioned Structuring}]
You are a medical report structuring expert. Extract structured information from a PET/CT report for one anatomical region.

Target region:

{region_name}

Slots to fill:

{slots_section}

These are all available fields for this region. Do not create additional keys.

Reference anatomical relations:

{hints_section}

Use these relations only as reminders of clinically relevant concepts that may appear in the report. They must not be treated as patient-specific evidence.

Full report:

{report}

Extraction rules:

- Only output keys defined in the Slots to fill section.

- Use null for slots not supported by the report.

- For categorical slots, select an allowed_values entry when it adequately represents the reported finding.

- For numerical slots, preserve the reported numerical value and unit.

- Use the reference anatomical relations to identify potentially relevant findings, but fill a slot only when the current report provides supporting evidence.

- Some slots may include linguistic_patterns, which provide slot-specific guidance for preserving clinically meaningful detail. When such patterns are provided and the corresponding details are explicitly present in the report, prioritize preserving those details rather than reducing the finding to a generic categorical value.

- Preserve relevant details indicated by the linguistic patterns, such as measurements and units, anatomical location, laterality, morphology, uptake characteristics, and other explicitly reported attributes.

- All extracted values must be grounded in the current report. Do not infer, complete, or introduce patient-specific information that is not supported by the source text.

Output JSON only, using exactly the following format:

{output_format}
\end{promptcardcompact}

\captionof{figure}{
\textbf{System prompt for region-conditioned report structuring.}
The prompt fills the predefined slots of one anatomical region using only patient-specific evidence contained in the current PET/CT report.
}
\label{fig:prompt-region-conditioned-structuring}

\begin{promptcard}[title={Clinical Agent -- Report Reconstruction}]
You are a medical report writing expert. Based on the given structured data, reconstruct a clinically faithful PET/CT report description.

Structured data:

{structured}

Reference sentences from real PET/CT reports:

{references}

Use the reference sentences only as stylistic and organizational guidance for phrasing, sentence rhythm, anatomical ordering, and terminology. They must not be used as a source of patient-specific clinical facts.

Requirements:

- Write in standard medical report style using plain prose, with no markdown, section headers, or bullet points.

- Include all clinically relevant information contained in the structured data.

- Treat the structured data as the sole source of patient-specific facts.

- Do not introduce, infer, or complete any finding, measurement, laterality, anatomical detail, or other clinical information that is absent from the structured data.

- Do not copy patient-specific facts from the reference sentences.

- Use the reference sentences only to guide wording and organization.

Output the reconstructed description text directly, without JSON or any other format.
\end{promptcard}

\captionof{figure}{
\textbf{System prompt for slot-conditioned report reconstruction.}
The prompt reconstructs a clinically faithful PET/CT description from structured data while using retrieved report sentences only as stylistic guidance.
}
\label{fig:prompt-report-reconstruction}

\begin{promptcard}[title={Clinical Agent -- Diagnostic Reasoning}]
You are a radiologist. Infer the diagnosis based on the following structured examination data.

Structured data:

{structured}

{diag_hints_section}

Output JSON only, using exactly the following format:

{
  "primary_diagnosis": "main diagnosis",
  "key_findings": [
    "finding1",
    "finding2"
  ],
  "recommendations": "recommendations"
}
\end{promptcard}

\captionof{figure}{
\textbf{System prompt for graph-assisted diagnostic reasoning.}
The prompt infers a primary diagnosis, supporting findings, and recommendations from the structured examination data and optional diagnostic-graph hints.
}
\label{fig:prompt-diagnostic-reasoning}

\begin{promptcard}[title={Evaluation Agent -- Clinical Concept Extraction}]
You are a medical report structuring expert. Extract all clinical concept keys from the given medical report.

Each concept key must use the entity+attribute form.

Examples:

liver_parenchymal_mass

liver_lesion_fdg_uptake

pelvic_soft_tissue_nodule

thyroid_diffuse_fdg_uptake

gallbladder_wall_thickening

renal_cystic_lesion

Output JSON only, using exactly the following format:

{
  "keys": [
    "concept_key_1",
    "concept_key_2",
    "..."
  ]
}
\end{promptcard}

\captionof{figure}{
\textbf{System prompt for clinical concept extraction.}
The prompt extracts normalized entity--attribute concept keys from a held-out medical report for template-coverage evaluation.
}
\label{fig:prompt-clinical-concept-extraction}

\begin{promptcard}[title={Evaluation Agent -- Concept-to-Slot Matching}]
Determine whether the report concept key semantically matches any key in the template.

Report concept key:

{report_key}

Template key set:

{template_keys}

Output JSON only, using exactly the following format:

{
  "match": true,
  "matched_key": "matched template key"
}

If no semantic match exists, return:

{
  "match": false,
  "matched_key": ""
}
\end{promptcard}

\captionof{figure}{
\textbf{System prompt for semantic concept-to-slot matching.}
The prompt determines whether a clinical concept extracted from a held-out report is semantically covered by a canonical template key.
}
\label{fig:prompt-concept-slot-matching}

\begin{promptcard}[title={Evaluation Agent -- Diagnostic Fidelity Judging}]
You are a medical report quality-evaluation expert. Compare the inferred diagnosis with the ground-truth diagnosis and assign scores according to the following metrics.

Metric definitions:

PDA is rated on a 0--5 scale: 0, absent or wholly incorrect;
1, nonspecific but compatible with the reference;
2, partially correct information but an incorrect primary diagnosis;
3, correct primary diagnosis with a clinically important qualifier
missing or incorrect;
4, correct diagnosis with only minor discrepancies;
5, clinically equivalent to the reference, including essential
qualifiers and diagnostic uncertainty. Important qualifiers include
anatomical site, laterality, and severity or stage when specified
in the reference.

KFP is the proportion of reference key findings correctly preserved:
$\mathrm{KFP}=M/K$, where $K$ is the number of distinct
diagnostically relevant reference findings and $M$ is the number
correctly retained. The reference findings and their essential
qualifiers are fixed before inspecting any method's output and
shared across methods. Semantically equivalent expressions count
as matches; omissions, contradictions, and missing essential
qualifiers do not. Unsupported additions receive no credit.
Cases with $K=0$ are excluded from KFP aggregation, with the
effective sample size reported.

CA measures whether the output provides sufficient supported
information for a reference-consistent clinical interpretation:
0, uninterpretable;
1, misleading;
2, usable only after major correction;
3, useful for a broad interpretation but missing important context;
4, adequate with minor omissions;
5, clear, sufficiently specific, and fully supported.
Fabricated findings or unsupported recommendations that could
materially change clinical interpretation or management cap CA
at 1. Recommendations are not required when the references
provide no basis for them.

Inferred diagnosis:

{inferred}

Ground-truth diagnosis:

{ground_truth}

Ground-truth description:

{ground_truth_desc}

Output JSON only, with no additional content:

{
  "pda": <integer from 0 to 5>,
  "kfp": <decimal from 0 to 1>,
  "ca": <integer from 0 to 5>,
  "reason": "scoring rationale"
}
\end{promptcard}

\captionof{figure}{
\textbf{System prompt for diagnostic-fidelity evaluation.}
The prompt compares an inferred diagnosis against the withheld reference diagnosis and description using primary-diagnosis accuracy, key-finding preservation, and clinical actionability.
}
\label{fig:prompt-diagnostic-fidelity-evaluation}














\end{document}